\documentclass[acmsmall,screen=true,review=false,anonymous=false,authorversion=false, timestamp=true]{acmart}

\usepackage{ifthen}
\newboolean{authnotes}
\setboolean{authnotes}{false}
\newboolean{majorReviewInRed}

\setboolean{majorReviewInRed}{false}
\newboolean{arxiv}
\newcommand{\arxivMode}{\setboolean{arxiv}{true}}
\setboolean{arxiv}{false}
\arxivMode
\ifthenelse{\boolean{arxiv}}
{
\setcopyright{none}
}
{
\copyrightyear{2020} 
\acmYear{2020} 
\received{March 2020}
\received[revised]{December 2020}
\received[accepted]{March 2021}
\setcopyright{licensedothergov}
\acmJournal{TIOT}
\acmPrice{xx.00}
\acmDOI{10.xxx}
\acmISBN{978-1-xxx}
}
\newcommand{\fakeparagraph}[1]{\vskip 3pt\noindent\textit{#1{.} }}

\usepackage[utf8]{inputenc}

\usepackage{url}
\usepackage{graphicx}
\usepackage{xspace}
\usepackage{paralist}
\usepackage{xcolor}
\usepackage{balance}
\usepackage{caption,subcaption}%

\usepackage{multirow}
\usepackage{booktabs}
\usepackage{widetable}
\usepackage{multirow}
\usepackage{units}
\usepackage{textcomp}
\usepackage{amsmath,amsfonts}
\usepackage{algorithmic}

\expandafter\def\expandafter\normalsize\expandafter{%
    \normalsize
    \setlength\abovedisplayskip{0pt}
    \setlength\belowdisplayskip{1pt}
    \setlength\abovedisplayshortskip{0pt}
    \setlength\belowdisplayshortskip{1pt}
}

\newcommand{\name}{BlueFlood\xspace}
\ifthenelse{\boolean{majorReviewInRed}}
{
\newcommand{\finalreview}[1]{\textcolor{red}{#1}}
}
{
\newcommand{\finalreview}[1]{#1}
}
\ifthenelse{\boolean{authnotes}}
{
\newcommand{\ba}[1]{\footnote{\textcolor{brown}{{\bf Beshr:}~{\em #1}}}}
\newcommand{\ol}[1]{\footnote{\textcolor{orange}{{\bf Olaf:}~{\em #1}}}}
\newcommand{\sd}[1]{\footnote{\textcolor{green}{{\bf Simon:}~{\em #1}}}}
\newcommand{\todo}[1]{\textcolor{blue}{{\bf Todo --\xspace}{\em #1}}}
\newcommand{\sketch}[1]{\textcolor{violet}{{\bf Sketch --\xspace}{\em #1}}}

\usepackage{soul}
\setstcolor{red}
\newcommand{\review}[1]{#1}
\newcommand{\newtext}[1]{#1}
\newcommand{\majorreview}[1]{#1}
}
{
\newcommand{\ba}[1]{}
\newcommand{\ol}[1]{}
\newcommand{\sd}[1]{}
\newcommand{\todo}[1]{}
\newcommand{\sketch}[1]{}
\newcommand{\st}[1]{}
\newcommand{\review}[1]{#1}
\newcommand{\newtext}[1]{#1}
\newcommand{\majorreview}[1]{#1}
}

\newcommand{\ie}{{\it i.e.},~}
\newcommand{\eg}{{\it e.g.},~}
\newcommand{\etal}{{\it et al.}\xspace}
\newcommand{\cf}{{\it c.f.},~}

\newcommand{\lsec}[1]{\label{sec:#1}}
\newcommand{\lfig}[1]{\label{fig:#1}}
\newcommand{\ltab}[1]{\label{tab:#1}}

\newcommand{\rsec}[1]{\S\ref{sec:#1}}
\newcommand{\rfig}[1]{Fig.~\ref{fig:#1}}
\newcommand{\rtab}[1]{Table~\ref{tab:#1}}

\newcommand{\equref}[1]{Equation~\emph{\ref{#1}}}

\newcommand{\capt}[1]{\mdseries{\emph{#1}}}

\begin{document}
\sloppy
\title[Concurrent Transmissions for Multi-Hop Bluetooth 5]{BlueFlood: Concurrent Transmissions for Multi-Hop Bluetooth 5 --- Modeling and Evaluation}
\ifthenelse{\boolean{arxiv}}
{
\titlenote{\textcopyright 2020 Copyrights are held by the authors. We grant arXiv.org a perpetual, non-exclusive license to distribute this article according to \url{http://arxiv.org/licenses/nonexclusive-distrib/1.0/}}
}{
}

\author{Beshr Al~Nahas}
\email{alnahas.beshr@gmail.com}
\affiliation{%
\institution{Chalmers University of Technology}
\city{Gothenburg}
\country{Sweden}}

\author{Antonio Escobar-Molero}
\email{antonio.escobar@infineon.com}
\affiliation{%
\institution{Infineon Technologies AG}
\city{Neubiberg}
\country{Germany}
}

\author{Jirka Klaue}
\email{jirka.klaue@airbus.com}
\affiliation{%
\institution{Airbus}
\city{Hamburg}
\country{Germany}
}

\author{Simon Duquennoy}
\email{simon.duquennoy@wittra.se}
\affiliation{%
\institution{Wittra Sweden AB}
\city{Stockholm}
\country{Sweden}
}

\author{Olaf Landsiedel}
\email{ol@informatik.uni-kiel.de}
\affiliation{%
\institution{Kiel University}
\city{Kiel}
\country{Germany}}
\affiliation{%
\institution{Chalmers University of Technology}
\city{Gothenburg}
\country{Sweden}}

\renewcommand{\shortauthors}{B. Al~Nahas et al.}

\begin{abstract}
Bluetooth is an omnipresent technology, available on billions of devices today.
While it has been traditionally limited to peer-to-peer communication and star networks, the recent Bluetooth Mesh standard extends it to multi-hop networking.
In addition, the Bluetooth 5 standard introduces new modes to allow for increased reliability. 
In this paper, we evaluate the feasibility of concurrent transmissions (CT) in Bluetooth via modeling and controlled experiments and then devise an efficient network-wide data dissemination protocol, BlueFlood, based on CT for multi-hop Bluetooth networks.

First, we model and analyze how CT distorts the received waveform and characterize the Bit Error Rate of a Frequency-Shift Keying receiver to show that CT is feasible over Bluetooth. 
Second, we verify our analytic results with a controlled experimental study of CT over Bluetooth PHY.
Third, we present BlueFlood, a fast and efficient network-wide data dissemination in multi-hop Bluetooth networks.

In our experimental evaluation, in two testbeds deployed in university buildings, we show that BlueFlood achieves 99.9\% end-to-end delivery ratio with a duty-cycle of 0.4\% for periodic dissemination of advertising packets of 38 bytes with 200 milliseconds intervals at 2 Mbps.
Moreover, we show that BlueFlood can be received by off-the-shelf devices such as smartphones, paving a seamless integration with existing technologies.

\end{abstract}


\maketitle

\section{Introduction}
\lsec{Introduction}

\fakeparagraph{Context}
Bluetooth is an omnipresent communication technology.
In 2020, the market volume of Bluetooth-enabled devices is expected to reach 8.4~Billion units, up from 1.3~Billion in 2013~\cite{ble-stats}. 
This makes Bluetooth predominant in our modern, connected society.
While Bluetooth has been available for many years, the release of Bluetooth Low Energy (BLE) in 2010 brought significant improvements in terms of energy efficiency for Bluetooth. 
Today, many wireless peripherals; \eg health, fitness and home automation use BLE as main communication technology.
With the recent release of Bluetooth~5 and Bluetooth~Mesh, the yearly growth of the deployment of Bluetooth devices is likely to further increase.
With its new transmission modes, Bluetooth~5 aims to offer a performance in terms of reliability, range, and energy efficiency that is on-par with IEEE~802.15.4~\cite{blev5-brief}.

\fakeparagraph{State of the Art}
In the past decade, the research community has designed a plethora of MAC, routing, and dissemination protocols for low-power wireless networking.
However, the focus for networking in low-power wireless has been nearly exclusively on IEEE~802.15.4.
For example, Glossy~\cite{ferrari11glossy} made a breakthrough in low-power wireless in disseminating information at network-scale quickly and efficiently.
It utilizes concurrent transmissions of tightly synchronized packets to realize flooding and synchronization services.
As of today, Glossy is practically limited to 802.15.4 in the \unit[2.4]{GHz} band and -- to a smaller degree -- ultra-wide band communication (UWB)~\cite{uwb16sensys, uwbranging18ewsn} and 802.15.4 in the sub-GHz band~\cite{dcoss13cx}.

Concurrent transmissions (CT) for Bluetooth, however, have been overlooked until today. 
It is, for example, not shown whether the concepts of concurrent transmissions are applicable to Bluetooth. 
The key differences between the Bluetooth physical layer (PHY) and IEEE~802.15.4 in the \unit[2.4]{GHz} band, are (i) the use of different modulation: Gaussian Frequency Shift Keying (GFSK) and Orthogonal Quadratic Phase Shift Keying (O-QPSK), respectively, (ii) the lack of Direct Sequence Spread Spectrum (DSSS) in Bluetooth and (iii) the support of four data rates in Bluetooth: \unit[125]{Kbps}, \unit[500]{Kbps}, \unit[1]{Mbps}, \unit[2]{Mbps} versus \unit[250]{Kbps} for 802.15.4.
This design makes Bluetooth less sophisticated in terms of physical layer features when compared to IEEE~802.15.4.
Moreover, analytic and experimental results indicate that the coding robustness provided by DSSS in 802.15.4 is essential to the reliability of Glossy~\cite{wilhelm14ct,lcn16ct}.
The recently adopted standard Bluetooth~5 provides convolutional encoding for the two long range modes with \unit[125]{Kbps} and \unit[500]{Kbps} bitrates, but still operates with GFSK modulation and without DSSS.
Thus, it is unclear how this coding scheme improves the robustness of concurrent transmissions.

\fakeparagraph{Approach and Impact}
We argue that adapting the concepts of concurrent transmissions to Bluetooth can open a variety of new application scenarios due to the ubiquitous availability of Bluetooth-enabled devices.
In this paper, we model and evaluate concurrent transmissions on top of Bluetooth PHY and exploit them in \name to provide network-wide flooding.
For example, in case of a fire in a building, we see the opportunity to use \name to disseminate a warning message with evacuation information as extended Bluetooth beacons.
As we show in this paper, such a CT-based flood of Bluetooth beacons is received, for example, by off-the-shelf smartphones.
Similarly, Bluetooth~Mesh extensively builds on network-wide flooding of messages which can benefit from concurrent transmissions to improve energy efficiency and reliability while reducing latency.

Finally, while Glossy was originally implemented on TelosB hardware utilizing the MSP430 MCU and a CC2420 radio, we now have modern SoCs with integrated radios available. 
We show in this article that these strongly simplify the design and implementation of protocols where transmissions need to be timed in the order of parts of a microsecond; \ie down to the individual ticks of the micro-controller, such as the case of concurrent transmissions.

\fakeparagraph{Contributions}
\majorreview{This article makes five key contributions}:
\begin{enumerate}[(i)]
    \item We model concurrent transmissions over FSK modulation and analyze the factors that affect its performance on the Bluetooth PHY.
	\item We demonstrate the feasibility of CT on the Bluetooth PHY through controlled experiments.
	\item We evaluate the performance trade-offs of the four transmission modes provided by Bluetooth~5 of 1 and \unit[2]{Mbps} and coded long range with 500 and \unit[125]{Kbps}, for CT. 
	\majorreview{\item We illustrate the applicability of CT over multi-hop networks with a simple protocol, \name: a low-power flooding protocol for Bluetooth PHY with improved performance, when compared to our original conference publication \cite{alnahas2019blueflood}. 
	We also demonstrate that \name is received by off-the-shelf receivers and does not need special software or hardware to receive on \eg smartphones.
	\item We evaluate \name in university buildings and show that \name achieves $99.9\%$ end-to-end delivery ratio in multi-hop networks with a duty cycle of 0.4\% for a periodic dissemination of advertising packets of \unit[38]{bytes} with \unit[200]{milliseconds} intervals.}
	Moreover, we show the fragility of CT over Bluetooth and the associated practical challenges.
	\name is available as open source\footnote{\url{https://github.com/iot-chalmers/BlueFlood}}. 
	This includes the code, the experimental data and the scripts needed to reproduce our results. 
\end{enumerate}%

\fakeparagraph{Article Structure and Outline}
To guide the reader through the article, we give an overview of the main parts of the article and outline its structure. 
This article is comprised of two main parts:
\begin{itemize}
    \item \emph{Analytical Modelling and Experimental Feasibility Study}:
    \rsec{feasibility} provides a discussion of the anticipated opportunities for applying CT and identifies challenges in the operation of CT over Bluetooth. 
    Then, we provide an analytic model of CT and show how the carrier frequency offset, transmission time offset and the number of transmitters affect its performance.
    Finally, we conclude this section with an experimental evaluation of CT over Bluetooth in a controlled setting and show how the different Bluetooth~5 modes perform under CT with different signal powers, transmission offsets and carrier frequency offsets.
    \item \emph{Design and Evaluation of \name}:
    The feasibility study demonstrates the potential of CT as a low-level primitive for building efficient protocols over Bluetooth. 
    \rsec{Design} illustrates the design of \name: A simplistic, yet efficient, multihop dissemination protocol.
\end{itemize}

The remainder of this article is structured as follows:
We review the related technical background about low power communication, Bluetooth and concurrent transmissions in \rsec{Background}, \finalreview{then we discuss the related state of the art research in \rsec{RelatedWork}.}
Later, we discuss, model and evaluate the feasibility of concurrent transmissions over Bluetooth in \rsec{feasibility}.
Next, we introduce the design of \name: A flooding protocol for network-wide dissemination and synchronization in \rsec{Design}. 
Finally, we evaluate \name in \rsec{evaluation} and conclude in \rsec{Conclusion}.

\section{Background}
\lsec{Background}

\newcommand{\figbeatingintro}{
    \begin{figure*}[t]
        \centering
            \begin{subfigure}[t]{0.45\columnwidth}
            \centering
            \includegraphics[width=0.8\columnwidth,keepaspectratio]{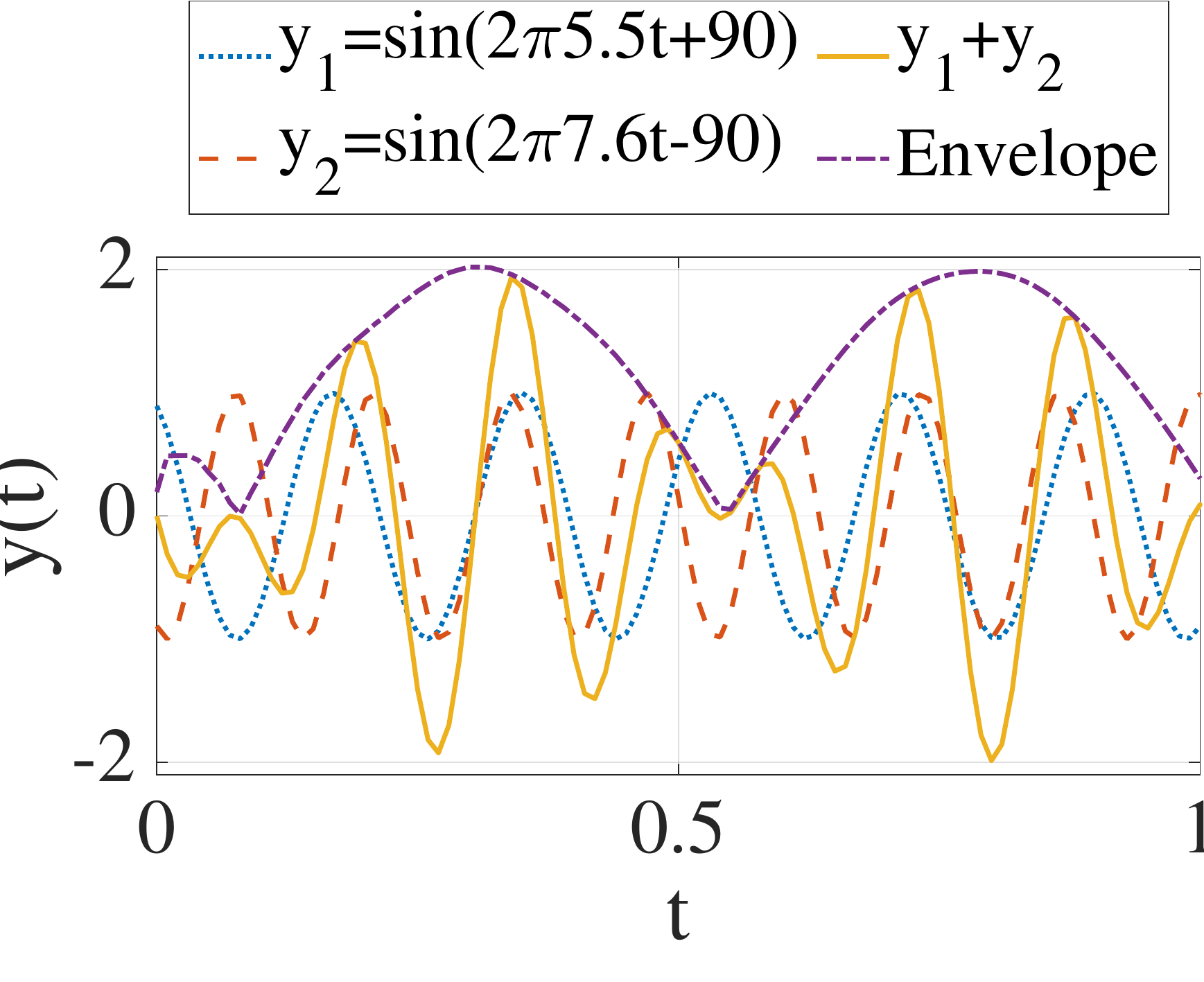}
            \caption{\review{Two concurrent} sinuous waves with different frequencies and phases result in a beating signal. Note that the two signals amplify and cancel each others periodically. See the sum of the signals at $t = 0$ to 0.1 and 0.3 to 0.4, for example.\lfig{beating-math}}
            \end{subfigure}%
            \hfill%
            \begin{subfigure}[t]{0.45\columnwidth}
            \centering
            \includegraphics[page=1,width=0.8\columnwidth,keepaspectratio]{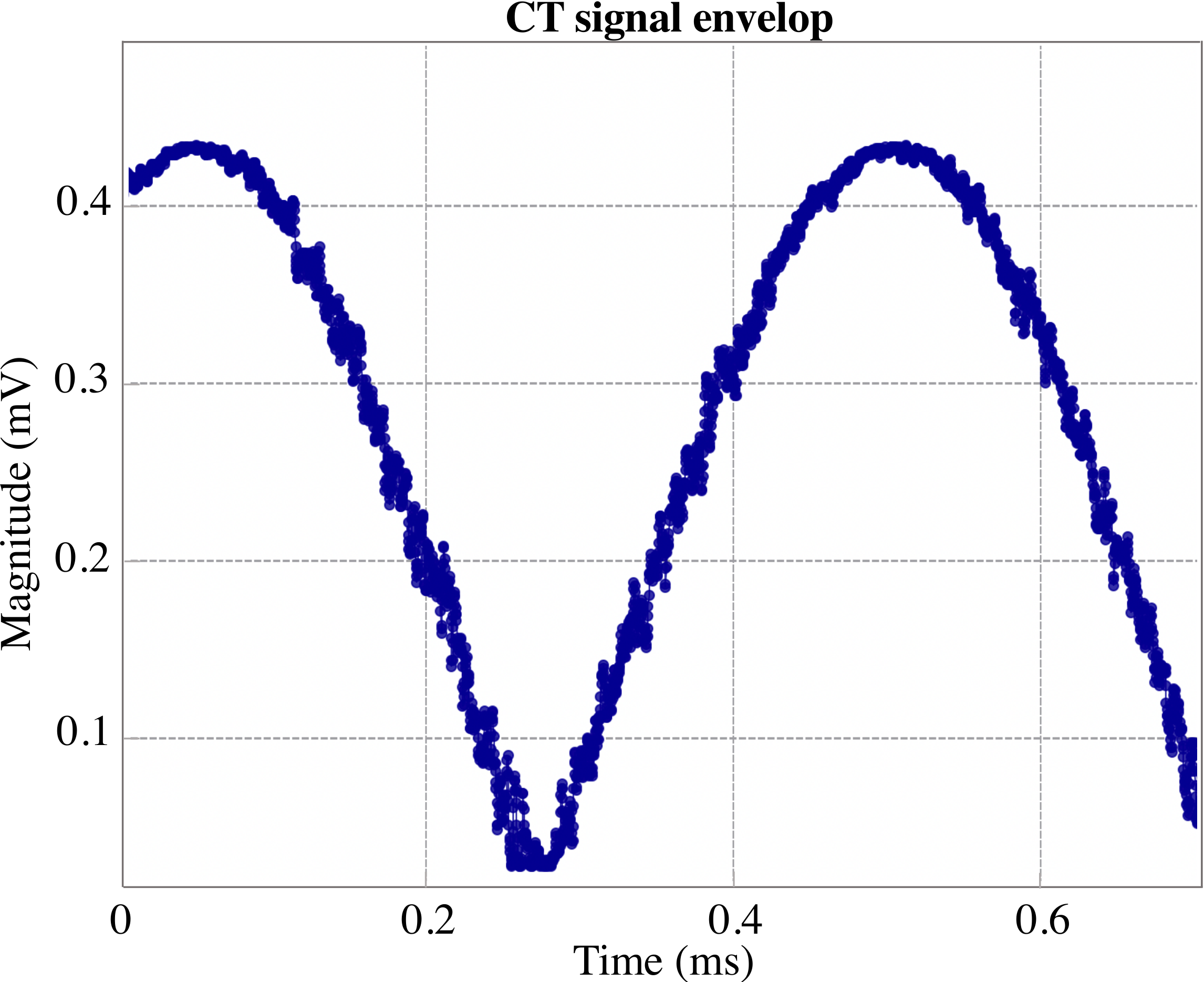}
            \caption{\todo{convert to a plot of the envelope of a (math)}Capturing the envelope of a beating carrier in CT reception from two transmitters using two nodes and a software defined radio. The figure shows the envelope of the signal in the baseband after removing the carrier, and shall be a constant line in the case of the optimal transmitter.\lfig{beating-pattern-sdr-bg}}
            \end{subfigure}%
        \caption[Concurrent transmissions result in carrier signal beating]{Concurrent transmissions lead to a beating radio signal instead of having a uniform magnitude. This is due to the frequency offset of the commercial transmitters from the nominal standard frequency. Therefore, CT might become destructive if the signal distortion is severe.\lfig{carrier_beating}}
    \end{figure*}
}

In this section, we provide the necessary technical background on Bluetooth and concurrent transmissions. 
Also, we relate to essential state of the art on both modeling and utilizing concurrent transmissions in low-power wireless. 
With these we identify the key challenges for concurrent transmissions on Bluetooth PHY. 
Later, \rsec{RelatedWork} provides a deeper discussion of the state of the art in the broader field of concurrent transmissions. 

\subsection{Low-Power Wireless Protocols: 802.15.4 vs. BLE}
\lsec{background:lpw}
\emph{ZigBee/IEEE 802.15.4 and Bluetooth Low Energy (BLE)} are today's widespread technologies for low-power wireless communication in the unlicensed \unit[2.4]{GHz} spectrum.
Each of them was initially designed for unique and distinct goals: 
While Bluetooth traditionally targets low-range single-hop communication with a bitrate suitable for \eg wearable and multimedia applications, ZigBee targets longer ranges and reliable multihop communication with a lower bitrate suitable for \eg home automation applications or industrial control. 
To this end, the IEEE~802.15.4 standard introduces a physical layer in the \unit[2.4]{GHz} band that utilizes O-QPSK modulation and DSSS for forward error correction (FEC): The PHY layer groups every 4~data-bits to make one PHY symbol and encodes it using 32~PHY signals or \emph{chips} --- a chip is the lower- or upper-half-sine wave that represents a logical 0 or 1.
With a chip rate of 2~M chips per second, it supports a bitrate of \unit[250]{Kbps} in 16~RF channels spaced \unit[5]{MHz} with a bandwidth of \unit[2]{MHz} each.
It offers packet sizes of up to \unit[127]{byte}.
On the other hand, both Bluetooth and 802.15.4 in sub-GHz use variants of FSK modulation, and both support \emph{uncoded detection} --- the physical layer symbols do not have redundancy and represent a one-to-one mapping to data bits.
BLE~4 uses GFSK and 802.15.4 in sub-GHz uses 2-FSK --- both modulation schemes represent bits 0 and 1 by using a $\pm \Delta f$ frequency shift from the central carrier frequency.
BLE~4 offers a bit-rate of \unit[1]{Mbps} in 40~channels with a bandwidth of \unit[2]{MHz} each and supports packets with PDUs up to 39~bytes.
Overall, the design choices of the narrower channels, a simpler modulation scheme and the lack of DSSS make Bluetooth the less robust communication scheme of the two. 
Next, we discuss how the recent Bluetooth~5 changes this.

\subsection{Bluetooth 5}
The recent Bluetooth~5 standard~\cite{blev5-brief} introduces (i) new long-range communication modes and (ii)~supports longer packets up to 255 bytes. 

The physical layer of Bluetooth~5 supports four PHY modes:
(i) two modes without forward error correction (FEC): \unit[2]{Mbps} mode in addition to the backward compatible \unit[1]{Mbps}, 
and (ii)~two long range modes that utilize FEC driven by a convolutional code: \unit[500]{Kbps} and \unit[125]{Kbps}, with up to $4\times$ longer range when compared to uncoded \unit[1]{Mbps}.
We note selected low-level details: 
(i) the different modes have different preamble lengths: one byte for \unit[1]{Mbps}, two bytes for \unit[2]{Mbps} and ten bytes for the coded modes \unit[500]{Kbps} and \unit[125]{Kbps},
(ii) the two coded modes \unit[500]{Kbps} and \unit[125]{Kbps} always transmit the header with FEC 1:8, and only afterwards change the coding rate to FEC 1:2 for the \unit[500]{Kbps} mode, and
(iii) all modes share a symbol rate of \unit[1]{M symbol/s} except for the \unit[2]{Mbps} mode which has \unit[2]{M symbol/s}.
\rtab{ble_modes} summarizes the operation modes.
When compared to 802.15.4, the physical layer of Bluetooth~5 still maintains the narrow channels of \unit[2]{MHz} and does not employ DSSS.
Nonetheless, the standard has the potential to be an enabler for IoT applications with a performance in terms of range, reliability, and energy-efficiency comparable to 802.15.4. %

\subsection{Bluetooth Mesh}
Bluetooth Mesh, part of the Bluetooth~4 standard, introduces multi-hop communication to Bluetooth:
Bluetooth Mesh follows a publish/subscribe paradigm where messages are flooded in the network so that all subscribers can receive them. 
Thus, Bluetooth Mesh does not employ routing nor does it maintain paths in the network. 
To reduce the burden on battery-powered devices, forwarding of messages in a Bluetooth Mesh is commonly handled by mains-powered devices. 
In recent studies with always-on, \ie mains-powered, nodes as backbone, Bluetooth Mesh reaches a reliability of above 99\% both in simulation~\cite{blemesh17ericsson} and experiments~\cite{blemesh18silabs}, and latency of 200 milliseconds, in networks of up to 6 hops with payloads of 16 bytes~\cite{blemesh18silabs}.

Because Bluetooth Mesh employs flooding, it differs strongly from established mesh and routing protocols in 802.15.4 such as CTP~\cite{ctp} or RPL~\cite{winter12rpl}.
We see the fact that Bluetooth Mesh is based on flooding is an additional motivation to evaluate the feasibility and performance of concurrent transmissions for network-wide flooding in Bluetooth~5.

\begin{figure}[tb]
	\centering
    \begin{minipage}{.65\textwidth}
		\includegraphics[width=1\columnwidth]{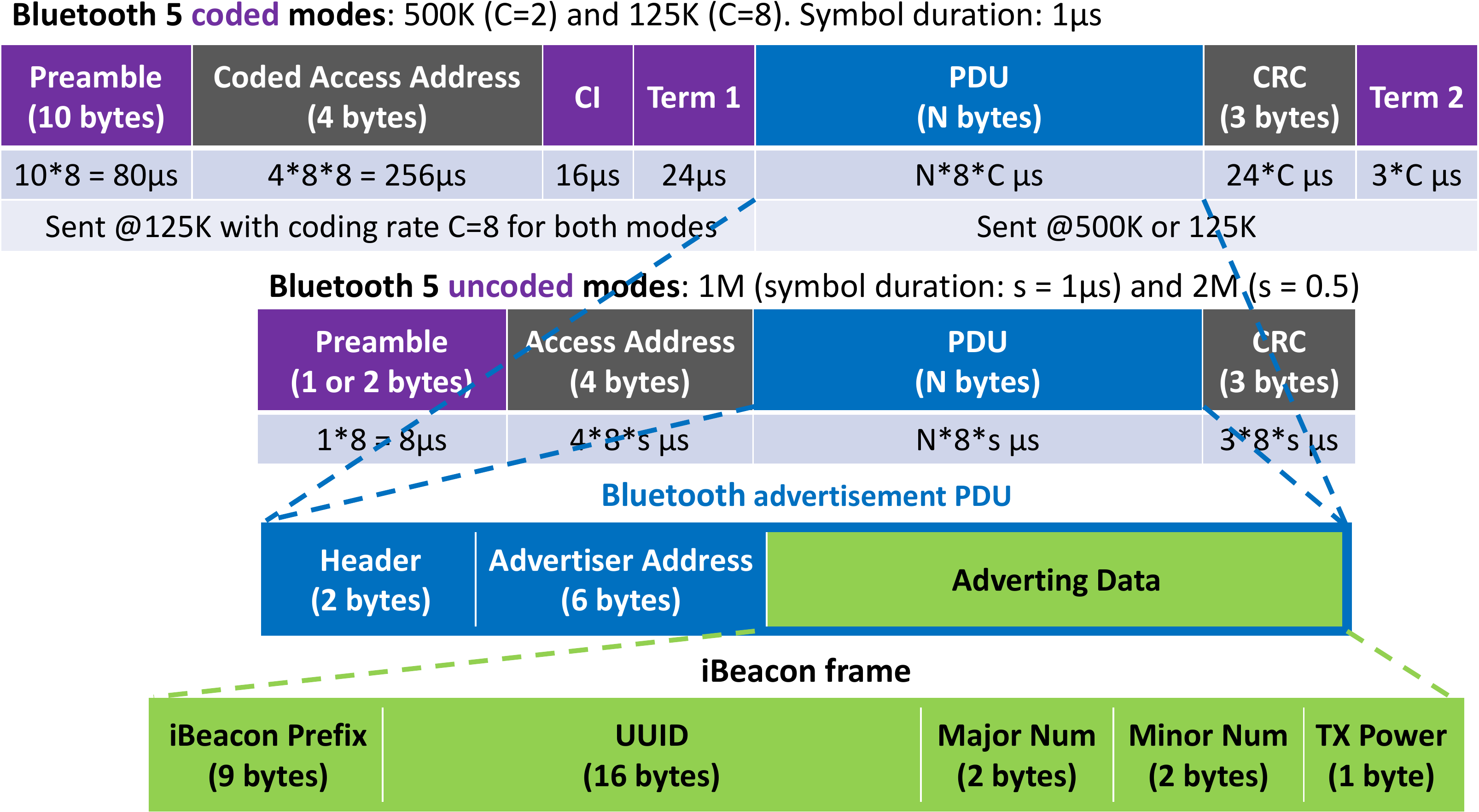}
		\captionof{figure}[Bluetooth Packet Structure]{Bluetooth packet structure for the coded and uncoded modes. \capt{Bluetooth advertisements formats are defined in defacto industrial standards such as iBeacon.}\lfig{bg:blepacket}}
    \end{minipage}
    \hfill
	\begin{minipage}{.32\textwidth}
    \includegraphics[width=1\columnwidth,keepaspectratio]{pics/sines.pdf}
    \vspace{-2.5em}
    \captionof{figure}{Two concurrent sinuous waves $y_1$ and $y_2$ with different frequencies and phases result in a beating signal $y_1 + y_2$ where they amplify and cancel each other periodically. 
    \lfig{carrier_beating}
    }
    \end{minipage}
\end{figure}
	
\subsection{Bluetooth Advertisements}
Traditionally, Bluetooth targets single-hop communication. 
For this, it operates in two modes: advertisement mode and connected mode.
In the advertisement mode, a Bluetooth device broadcasts short pieces of information.
This is commonly used by low-power devices such as, for example, temperature sensors to share their sensor readings, and localization beacons to announce their presence and their physical location. 
Moreover, this mode is used to advertise the availability of a device so that other devices can connect to it. 
The second mode --- the connected mode, establishes a connection between a master and a slave.
Here, a master and a slave communicate in time-synchronized \emph{connection events}.
In this paper, we focus on Bluetooth advertisements and refer the reader to Bluetooth core specifications~\cite{bluetooth5} for details about each mode. 

In this paper we use non-connectable beacons for lightweight flooding while staying compatible with off-the-shelf devices.
Bluetooth~5 extends this further by allowing a packet of up to 255~bytes versus the legacy 39~bytes limit.
Moreover, it allows advertising on any of the 40~channels instead of limiting it to three channels as in previous Bluetooth versions. 
While the Bluetooth specifications do not define the beacon payload format, there are several industrial standards, with two main formats~\cite{wiki:blebeacons}: 
(i) Apple's iBeacon (shown in \rfig{bg:blepacket}) and the open source alternative AltBeacon (by Radius Networks), and
(ii) Google's Eddystone.

\begin{table}[tb]
    \centering
    \caption[Bluetooth~5 and IEEE 802.15.4: PHY parameters and modes]{Bluetooth~5 and IEEE 802.15.4: PHY parameters and modes. 
    \capt{Note that: (i) in Bluetooth, each bit is encoded using 1, 2 or 8 symbols depending on FEC; (ii) Bluetooth coded modes \unit[500]{Kbps} and \unit[125]{Kbps} use the \unit[1]{Mbps} PHY mode beneath, and (iii) IEEE 802.15.4 uses a different terminology: one symbol represents 4~bits and is encoded using 32~chips --- a chip is the PHY layer signal that represents a logical 0 or 1. $\tau$ stands for period.}\ltab{ble_modes}
    }
    \footnotesize
    \begin{widetable}{\linewidth}{lcccccc}
    \toprule
    \emph{Bluetooth~5} & \emph{Bitrate} $[bps]$ & \emph{Symbol rate} [per sec] & \emph{Symbol $\tau$} [$\mu s$] & \emph{bit $\tau$} [$\mu s$]& \emph{FEC} & \emph{Preamble} [byte]\\
    \cmidrule(r){1-1}
    \cmidrule(lr){2-2}
    \cmidrule(lr){3-3}
    \cmidrule(lr){4-4}
    \cmidrule(lr){5-5}
    \cmidrule(lr){6-6}
    \cmidrule(l){7-7}
    GFSK & 2~M & 2~M & 0.5 & 0.5 & - & 2 \\
    GFSK & 1~M & 1~M & 1 & 1 & - & 1 \\
    GFSK & 500~K & 1~M & 1 & 2 & 1:2 & 10 \\
    GFSK & 125~K & 1~M & 1 & 8 & 1:8 & 10 \\
    \midrule
    \emph{IEEE 802.15.4} & \emph{Bitrate}  $[bps]$ & \emph{Chip rate} [per sec] & \emph{Chip $\tau$} [$\mu s$] & \emph{Symbol $\tau$} [$\mu s$] & \emph{FEC} & \emph{Preamble} [byte]\\
    \cmidrule(r){1-1}
    \cmidrule(lr){2-2}
    \cmidrule(lr){3-3}
    \cmidrule(lr){4-4}
    \cmidrule(lr){5-5}
    \cmidrule(lr){6-6}
    \cmidrule(l){7-7}
    O-QPSK & 250~K & 2~M & 0.5 & 16 & 1:8 & 4\\
    \bottomrule
    \end{widetable}
\end{table}

\subsection{Concurrent Transmissions and Capture}
\lsec{background:ct}

\fakeparagraph{Definitions}
In \emph{Concurrent Transmissions (CT)}, or \emph{Synchronous Transmissions}, multiple nodes synchronously transmit the data they want to share.
Nodes overhearing the concurrent transmissions receive one of them with a high probability, due to the capture effect~\cite{leentvaar76}, or non-destructive interference.
We shall note that we use both terms Concurrent Transmissions (CT) and Synchronous Transmissions interchangeably to refer to \emph{tightly synchronized simultaneous transmissions}.

\emph{Capture effect:} 
A receiving radio can capture one of the many colliding packets under specific conditions related to the used technology~\cite{201311LandsiedelChaos, leentvaar76}, which we briefly highlight next. 

\emph{Non-destructive interference:} 
If the colliding packets are tightly synchronized and have the same contents, then the resulting signal might be distorted, but it is highly probable that they do not destruct each other. 
Thus, the receiver can recover the contents with a high probability. 
\citeauthor{ferrari11glossy}~\cite{ferrari11glossy} present an in-depth evaluation of this effect on 802.15.4, but they assume it is \emph{constructive interference}.
Later work~\cite{wilhelm14ct, lcn16ct} has shown that is not constructive in practice, but not fully destructive either; \ie the receiver decodes the packet with a high probability, but the concurrent transmission link quality is lower than the best single-transmission link. 
We confirm this as well when studying CT over Bluetooth later in~\rsec{feasibility}.

\newtext{\fakeparagraph{Link-based Communications and CT Benefits}
Classic approaches to networking build on routing and link-based communications and utilize mechanisms to avoid packet collisions.
For example, IEEE 802.15.4 employs carrier sense multiple access (CSMA) to avoid sending when it senses energy in the channel, and utilizes acknowledgments (ACK) for re-transmission after a timeout when missing an acknowledgment.
A classic network stack uses a routing protocol to achieve multihop communication, \eg RPL in 6LoWPAN.
CT, however, embraces the broadcast nature of the wireless medium and synchronizes transmissions to enhance the probability of packet reception. 
CT enjoys the benefits of \emph{sender diversity}: the concurrent senders have independent links to the receiver.
More importantly, it is a simple yet efficient flooding primitive, that avoids the implementation and operation overhead of routing and link-based communications~\cite{zimmerling2020synchronous}.
Moreover, it has been demonstrated that CT-based protocols achieve enormous performance gains in terms of end-to-end reliability, latency and energy consumption~\cite{ferrari11glossy, alnahas17a2, dcoss13cx, LaneFlood, 201311LandsiedelChaos}, even under harsh interference conditions~\cite{Lim17Comp, Nahas16Comp, Nahas17Comp, Nahas18Comp, Escobar17Comp, ewsn17Comp, escobar2019competition, bigbangbus18comp}.
}

\fakeparagraph{CT with 802.15.4 radios in the \unit[2.4]{GHz} band}
Typically, in 802.15.4, the radio receives the stronger one of the concurrent transmissions if its signal is \unit[3]{dBm} stronger, the so-called \emph{co-channel rejection}, if they are synchronized within the preamble of 5~bytes, \ie \unit[160]{$\mu s$}~\cite{201311LandsiedelChaos}.
However, in the case of CT of the same data over 802.15.4, if the nodes transmit within \unit[0.5]{$\mu s$}, then no signal strength delta is necessary~\cite{ferrari11glossy}.
This is due to the use of DSSS:
802.15.4 radios in the \unit[2.4]{GHz} band utilize DSSS, where bits are encoded redundantly into \emph{chips} with a 1:8 FEC redundancy, \ie \unit[2]{M chip/s} encode a \unit[250]{Kbps} data stream, as highlighted in~\rsec{background:lpw}.
This encoding helps to recover bits from the distorted signal in both cases of CT of the same and different data.

\fakeparagraph{CT with 802.15.4 in the sub-GHz band}
In contrast to the \unit[2.4]{GHz} band, 802.15.4 in the sub-GHz band does not employ FEC. 
Per the study of Liao \etal~\cite{sensys16ctposter} on CT over 802.15.4 in the sub-GHz band, the most critical operation zone for CT is when both transmissions reach the receiver with the same power; \ie zero power delta.
The authors argue that in this case the timing offset needs to be smaller than $1~\mu s$.
In the case of \emph{different data} on the other hand, and under similar conditions, \majorreview{models suggest that there} needs to be a power delta of about 10~dB to have a packet reception rate of $20 - 30\%$, especially when \review{the packets are} not protected by FEC such as \majorreview{for 802.15.4 in the sub-GHz band~\cite{wilhelm14ct, lcn16ct}.
We later show that such threshold also exists for BLE, but it is in the area 6~to 8~dB.}
Moreover, these studies demonstrate a degraded receiver sensitivity and subsequently a declined reliability with CT when done over uncoded, non-DSSS communication, \ie without the protection of FEC.
On the other hand, the studies indicate that the use of FEC improves reliability and relaxes the conditions for successful reception.

\subsection{Glossy}
Glossy~\cite{ferrari11glossy} is a flooding protocol for network-wide time synchronization and data dissemination.
It established the design principle of concurrent transmissions of the same data in low-power wireless networks that are based on the IEEE 802.15.4 standard as it proved to be a highly reliable and efficient protocol.
Glossy operates in rounds, with a designated node, the initiator, that starts the concurrent flooding.
Nodes hearing the transmission synchronize to the network and join the flooding wave by repeating the packet.
The transmissions are tightly synchronized in order to achieve non-destructive CT.
Every node alternates between reception and transmission and repeat this multiple times to spread the information and achieve one-to-many data dissemination from the initiator to the rest of the network.

\section{Related Work}
\lsec{RelatedWork}
In this section, we discuss the state of the art of the broader field of concurrent transmissions, constructive interference, capture effect, and the protocols that base on these concepts in Wireless Sensor Networks (WSNs) and Internet of Things (IoT).
We provide the necessary technical background on Bluetooth and concurrent transmissions in \rsec{Background}.

\fakeparagraph{Understanding Concurrent Transmissions}
While the research into capture effect is not new and was first observed for FM transmitters~\cite{leentvaar76}, the capture effect in low-power wireless networking was first experimentally studied by Ringwald and R{\"o}mer~\cite{bitmac05} over On-Off-Keying (OOK) modulation where they design BitMAC, a MAC protocol that utilizes CT to implement simple in-network aggregates to provide collision-free communication. 
Later, Son~\etal~\cite{concurrent-son} evaluated CT over 802.15.4 compatible radios.
The success of concurrent transmissions in Glossy~\cite{ferrari11glossy} started a debate on how these work and what underlying physical phenomena enable it.
The authors of Glossy argue that the signals interfere constructively.
Later, Rao~\etal~\cite{murphy16infocom} demonstrated that through precise timing Glossy can achieve destructive interference to provide feedback.

In contrast, Yuan and Hollick~\cite{yuan13letstalk} show experimentally that frequency offsets between concurrent transmitters makes it hard to get constructive interference. 
Wilhelm~\etal~\cite{wilhelm14ct} introduce analytical models backed with experiments to parameterize concurrent transmissions and show that these are rather non-destructive interference.
Thus, they argue that the signals get degraded due to concurrent transmissions but can be decoded, nevertheless. 
Moreover, they argue that coding is essential to improve the reliability of concurrent transmissions. 
Similarly, Liao~\etal~\cite{lcn16ct} argue that DSSS coding helps CT survive beating. 
While the mentioned papers are limited to 802.15.4 in the \unit[2.4]{GHz} band, Liao~\etal~\cite{sensys16ctposter} have a limited study on CT over 802.15.4 in sub-GHz. 
Roest~\cite{Roest15ble} studies the capture effect and evaluates Chaos on BLE, 1~Mbps.
Schaper~\cite{schaper19ble} extends and repeats the feasibility study of CT over Bluetooth of our original conference paper on \name \cite{alnahas2019blueflood} in the controlled wireless environment of an Anechoic chamber and shows similar trends to our results.
To the best of our knowledge, no prior research has evaluated and utilized CT over Bluetooth~5 extensively. 

\fakeparagraph{Concurrent Transmissions Protocols}
BitMAC~\cite{bitmac05}, A-MAC~\cite{dutta10} and Glossy~\cite{ferrari11glossy} pioneered the field of concurrent transmissions in WSNs. 
LWB~\cite{ferrari12lwb}, Splash~\cite{2014DoddavenkatappaSplash} and Choco~\cite{Choco} base on Glossy to schedule individual floods to provide data collection while Crystal~\cite{Crystal} and its multichannel version~\cite{CrystalMC} reduce the number of Glossy floods by relying on data prediction. 
CXFS~\cite{dcoss13cx}, Sparkle~\cite{2014YuanSparkle} and others~\cite{PEASST, LaneFlood, RFT, Ripple} limit the number of concurrent transmitters in Glossy or LWB while Sleeping Beauty~\cite{sleepingbeauty16mass} combines both limiting the number of transmitters by putting them to sleep and scheduling Glossy floods to improve energy efficiency.
Baloo~\cite{baloo19} provides a framework for easing the development and implementation of Glossy-based synchronous transmissions protocols. 

SurePoint~\cite{uwb16sensys} builds an efficient concurrent network-wide flooding similar to Glossy in UWB and leverages it to provide a localization service while Corbal\'{a}n and Picco~\cite{uwbranging18ewsn} introduce concurrent ranging on UWB where a transceiver tag estimates the distance to anchors by exploiting the channel impulse response (CIR) estimation of the anchors' replies.
SnapLoc~\cite{snaploc19} \majorreview{exploits CT to localize tags with a single read operation of multiple anchors.}
Lobba~\etal~\cite{lobba2020uwbct} implements Glossy and evaluates CT over UWB, and shows how CT in UWB offers a higher energy efficiency and tighter synchronization when compared to narrow-band technologies, such as 802.15.4 and Bluetooth.

Chaos~\cite{201311LandsiedelChaos} on the other hand extends the design of Glossy to utilize the capture effect on 802.15.4 in the \unit[2.4]{GHz}-band to let nodes transmit different data and efficiently calculate network-wide aggregates by employing in-network data processing.
$A^2$~\cite{alnahas17a2} takes this further by introducing communication primitives for network-wide consensus. 
WirelessPaxos~\cite{Poirot19Paxos} builds on top of $A^2$ to achieve fault-tolerant eventual consistency. 
Mixer~\cite{herrmann2018mixer} and Codecast~\cite{codecast18} utilize network coding techniques for efficient many-to-many data sharing. 
However, since these protocols base on capture of different data rather than flooding the same data, they are more difficult to support on uncoded communication technologies such as the Bluetooth modes \unit[1 and 2]{Mbps}.

Overall, concurrent transmissions enable low-latency network-wide communication.
While none of the aforementioned protocols support Bluetooth, the concepts are generally extendable to other technologies given that concurrent transmissions are supported.
\name builds on these results to bring efficient network flooding to Bluetooth mesh networks.

\fakeparagraph{Low-power Channel Hopping}
Using frequency diversity techniques has proven to be effective for combating interference~\cite{watteyne10mitigating}.
It is wide-spread both in the established standards; such as Bluetooth~\cite{bluetooth5}, TSCH~\cite{stdieee802154-2015} and in the state of the art such as the top solutions in the dependability competition~\cite{ewsn17Comp} and BLEach~\cite{bleach}.
BLEach not only enables adaptive channels black-listing and adaptive duty-cycling to provide quality of service guarantees, but implements IPv6 over BLE as in RFC 7668~\cite{bleIpv6rfc7668}.
However, it only supports star networks as opposed to \name which supports multihop Bluetooth mesh networks.

\section{Feasibility of CT over Bluetooth}
\lsec{feasibility}

After providing the required background on both Bluetooth and concurrent transmissions, we set out to analyze and evaluate whether concurrent transmissions are practical on the Bluetooth physical layer. 
\majorreview{
Our goals are as follows: 
(i) show that CT is physically feasible on FSK modulation schemes. We want to address the concern that CT might be only accidentally working due to the implementation of the specific Bluetooth transceiver we use in the experimental sections of the paper; and
(ii) analyze what factors affect the performance of CT and how. 
}

\newtext{This section is structured as follows.
\begin{itemize}
    \item in \rsec{feasibility:possibilitiesAndChallenges}, we begin by outlining the lessons learned from the state-of-the-art, discuss why CT shall work, identify the challenges and discuss how they materialize for Bluetooth, then
    \item in \rsec{feasibility:CT_model} we present analytic models for concurrent transmissions over Frequency-Shift Keying modulation, which is similar to the modulation used in Bluetooth, albeit simpler.
    In this section we show how the various parameters of the concurrent transmissions affect its performance:
    \begin{itemize}
        \item we derive the analytic expression of the envelope of the CT signal and the resulting bit-error rate (BER) in~\rsec{feasibility:model:ct_envelope_ber},
        \item we show how the time delay of a concurrent transmitter and the power difference affects the BER of CT in~\rsec{feasibility:model:ct_time_shift_power_delta},
        \item we derive an upper bound of the number of concurrent transmissions in a dense deployment in~\rsec{feasibility:model:ct_scalability}, and
        \item we give a lower bound of the expected packet error rate (PER) with regard to the beating frequency in~\rsec{feasibility:model:beating_prr_analysis}, then
    \end{itemize}
    \item in \rsec{feasibility:CT_experimental}, we perform an experimental validation of the expected performance of CT \review{in controlled settings}:
        \begin{itemize}
        \item we show the actual carrier beating of sample transceivers in~\rsec{feasibility:carrier_beating_eval}, then
        \item we evaluate the PER of CT with transceivers that have different carrier frequency offsets in~\rsec{feasibility:cfo_eval}, later
        \item we evaluate how transmission power differences affect the packet reception ratio (PRR) of CT  in~\rsec{feasibility:txPower}, and
        \item finally, we evaluate how the transmission delay affects the PRR of CT in~\rsec{feasibility:txTimeDelta}.
    \end{itemize}
\end{itemize}
Note that we tag the figures in this section with [analytic] and [experiment], to help the reader distinguish the plots of the analytic expressions versus the experimental results.
}

\subsection{CT Opportunities and Challenges}
\lsec{feasibility:possibilitiesAndChallenges}
We begin by outlining why CT should fundamentally work over Bluetooth before discussing the practical challenges and limitations of achieving successful CT over Bluetooth. 
As discussed in \rsec{background:ct}, recent studies discuss CT over 802.15.4 both in the \unit[2.4]{GHz} and the sub-GHz bands. From these, we next draw lessons that are applicable to Bluetooth.

\begin{figure}[t]
  \centering
    \begin{subfigure}[t]{0.32\columnwidth}
    \centering
    \includegraphics[width=1\columnwidth]{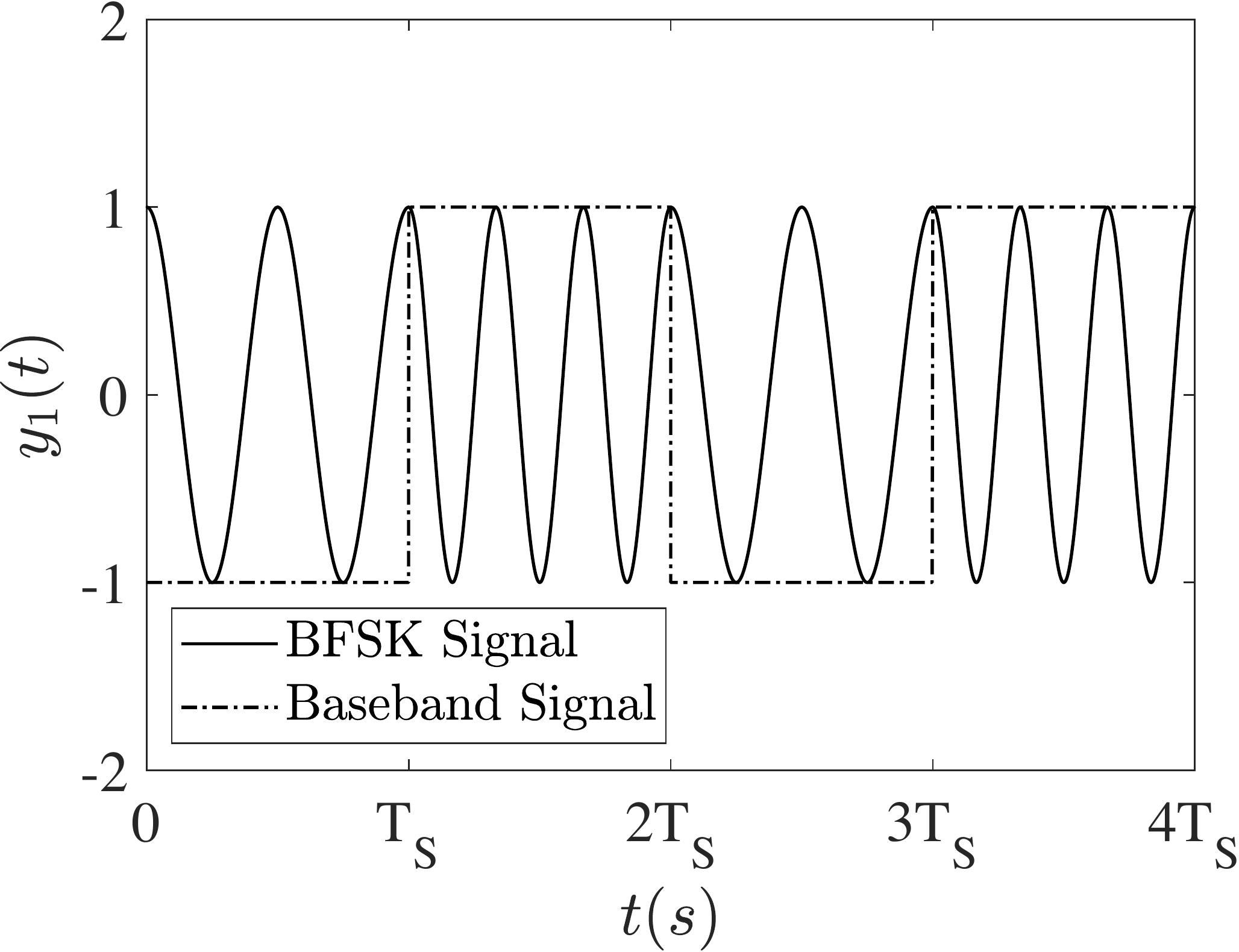}
    \caption{Modulated BFSK signal of a single transmitter.
    The signal frequency changes slightly to represent ones and zeros.  \label{fig:Modulated--signal-BFSK}}
    \end{subfigure}%
    \hfill%
    \begin{subfigure}[t]{0.32\columnwidth}
    \centering
    \includegraphics[width=1\columnwidth]{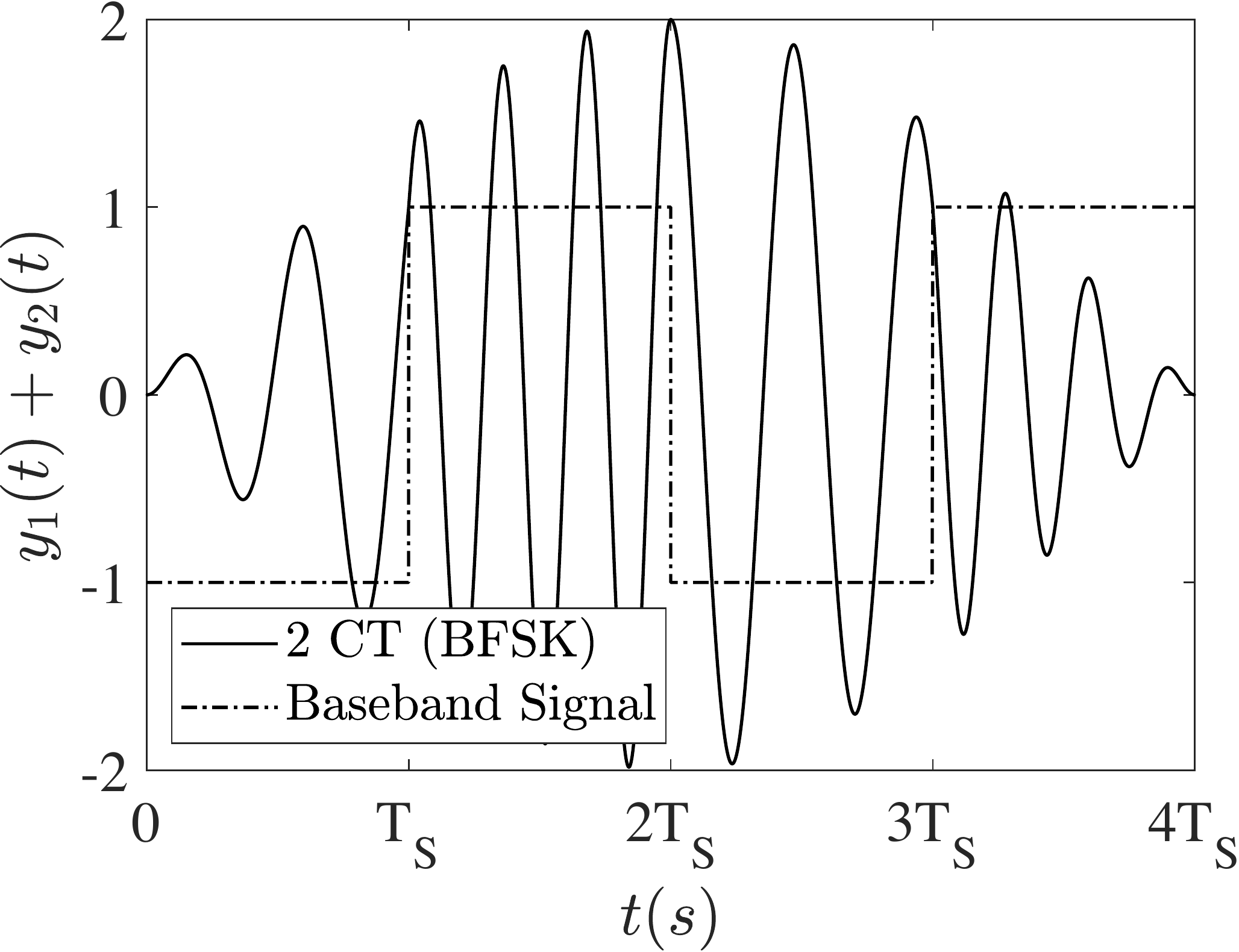}
    \caption{Beating at the receiver of two concurrent transmissions with the same magnitude and perfect time synchronization: $A_1 = A_2,\:\Delta t=0$.
    Beating happens because of the slightly different RF frequencies due to radio oscillator imperfections.
    In this example, the beating period is 4 symbols: $T_{beat} = 4T_S$.\label{fig:Interference-between-two-1-1}}
    \end{subfigure}
    \hfill%
    \begin{subfigure}[t]{0.32\columnwidth}
    \centering
    \includegraphics[width=1\columnwidth]{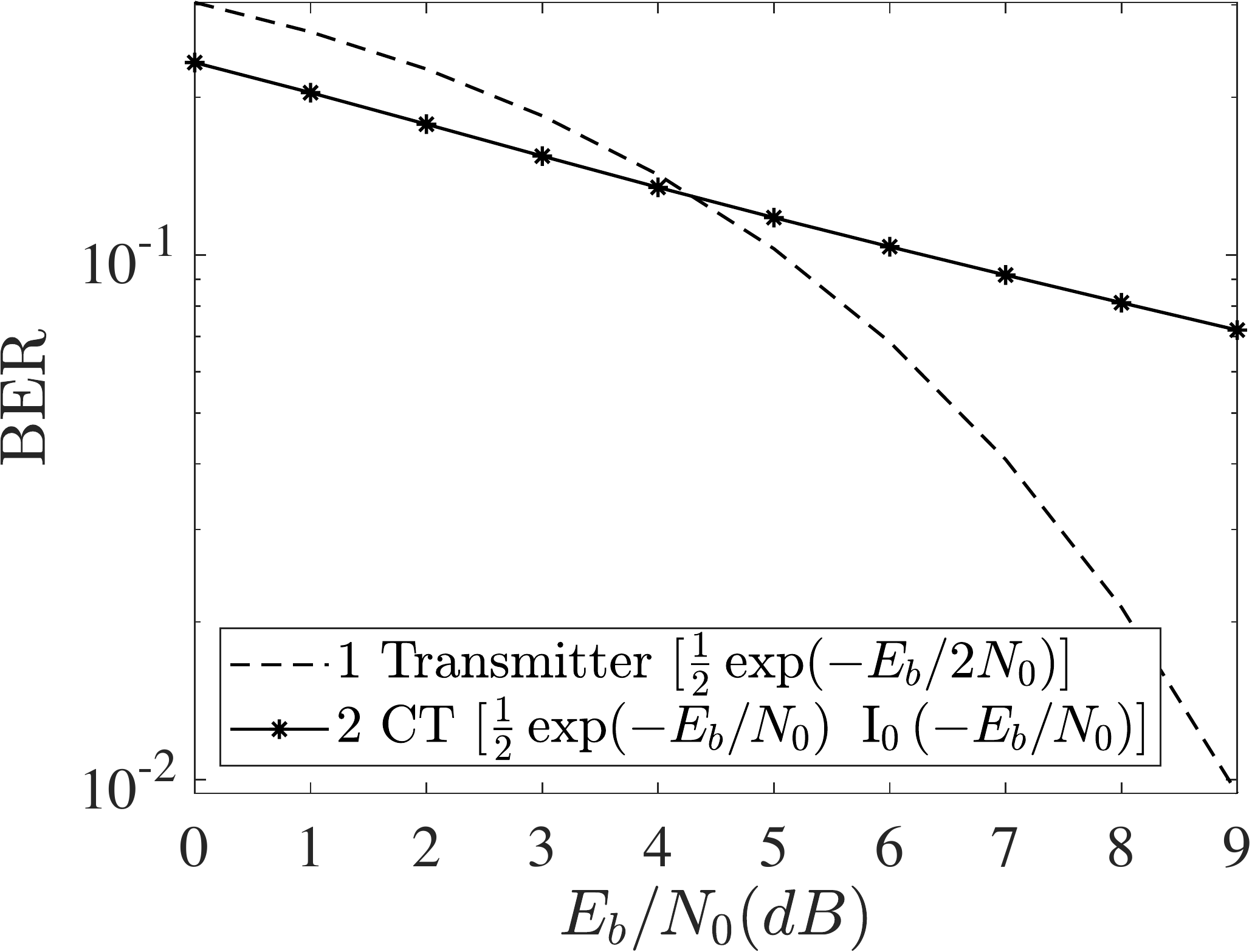}
    \caption{Analytical BER for one vs. two concurrent transmitters
    ($\Delta t=0$, $A_{1}$=$A_{2}$) at non-coherent receivers. We note the degraded BER of CT in \review{the region with the high energy-to-noise ratio, \ie the right-half of the x-axis ($E_b/N_0>\unit[4]{dB}$)}. \label{fig:-ber_ebn0-1-1-1-5-1-1}}
    \end{subfigure}%
\caption{Modeling Concurrent Transmissions in BFSK: Beating Effect and the Resulting BER. [analytic] \label{fig:big-fig-models}}
\end{figure}

Bluetooth uses Gaussian-filtered Frequency Shift Keying (GFSK).
We can describe it with a non-distorting simplification: in the base-band frequency spectrum of the modulated signal, bits 0 and 1 are shown as $\pm \Delta f$ frequency shifts from the central frequency, as shown in \rfig{Modulated--signal-BFSK}. 
In the case of \emph{ideal} synchronous concurrent transmissions (\ie no time, frequency or phase delta in the carrier band) of the same data, the two signals would overlay perfectly and a receiver would not notice a difference from the case of a single transmitter, except for a doubled magnitude.
On the other hand, with different data, the sum of the two signals of the two different bits need to be distinguishable from the center frequency, and lay on either $\pm \Delta f$ sides; \ie one signal needs to be sufficiently stronger than the other.
Nevertheless, real-life concurrent transmissions are not as simple: different transmitters have slightly different frequencies, drift independently and signals sum-up at the receiver with different phases, as we show later.

As discussed in \rsec{background:ct}, the performance and practical feasibility of CT depend on four factors~\cite{wilhelm14ct}: 
(i)~the time delta between the two packets, and 
(ii) the Received Signal Strength (RSS) delta. 
Moreover, both (iii) the choices of the radio technology (modulation and encoding), and 
(iv) whether the concurrently transmitted packets have an identical payload or not determine the range of the first two parameters for successful reception and the  robustness of the CT link. 

In practice, the carrier frequencies of the different transmitters are never equal, due to oscillator accuracy.
For example, an oscillator with a \unit[10]{ppm} accuracy results in a carrier frequency offset (CFO) of \unit[$\pm 24$]{KHz} for a \unit[2.4]{GHz} carrier, on average.
As a result, the concurrent transmission of the \emph{same data} leads to a \emph{beating} radio signal, where the signal magnitude alternates between peaks and valleys instead of being uniform, as illustrated in \rfig{carrier_beating}.
These variations in frequency and phase distort the signal; thus, CT might be destructive if the signal distortion is severe.
It shall be noted that the radios transmit preamble bytes to synchronize the frequency and phase of the receiver to that of the transmitter.
In the case of CT, the receiver synchronizes to the effective sum of the different preambles.
On the other hand, the concurrent transmission of \emph{different data} causes a destructive interference of the signal that is only recoverable when one received signal has a RSS sufficiently higher than the sum of the other transmissions as long as they are received within the duration of the preamble.

Wilhelm \etal~\cite{wilhelm14ct} suggest that the combination of the carrier-phase offset and the timing offset is detrimental to the reception of CT when sending the same data. 
Their paper gives bounds of the tolerable timing offset to be half of the symbol period; \ie $\tau/2$.
For Bluetooth, this translates to $0.25~\mu s$ for the \unit[2]{Mbps} mode and $0.5~\mu s$ for the other modes (as they share the same symbol rate).
On the other hand, the tolerable carrier-phase offset is estimated to be $0.4\pi$.
While we cannot control the phase offset in the low-energy commercial off-the-shelf (COTS) radios, we can synchronize the transmissions \review{timings} to be within the bounds noted above.
In addition, the signal preamble helps the receiver to synchronize the phase offset.
Thereby, a longer preamble can help a receiver to lock on a specific phase-offset and thereby improve the reception of this particular transmissions.

\fakeparagraph{Summary}
Based on the existing models presented in \rsec{background:ct} and our analysis above, we can summarize the status of CT over Bluetooth as follows:
(i) since Bluetooth employs non-DSSS communication, it is expected to suffer under CT when compared to, for example, 802.15.4 in the \unit[2.4]{GHz} band,
(ii)~the timing offset shall be kept under $0.25~\mu s$ for the \unit[2]{Mbps} mode and $0.5~\mu s$ for the other modes, 
(iii) the phase offset shall be below $\pm0.4\pi$, which we cannot control in COTS devices.
However, we argue that we can potentially increase the robustness by using Bluetooth transmission modes with longer preambles and thereby improve the synchronization of the receiver onto a specific phase-offset of a signal. 
(iv) The capture of CT of different data is not possible without a major signal power delta; especially without FEC (see also \rsec{background:ct}), and
(v) the use of FEC is expected to improve the performance, but obviously incurs a non-trivial overhead of 1:2 or 1:8 per the two modes \unit[500]{Kbps} and \unit[125]{Kbps}, respectively.

\majorreview{
While our analysis of the existing models gives indications, none of the models is specific to BLE. 
Moreover, each of the existing models gives partial results, for example, by focusing on either phase offset or time offsets. 
We argue that these need to be combined into a single model tailored specifically to BLE. 
Next, we address this by modeling}
CT over 2-FSK modulation, which is similar to the modulation used in Bluetooth \majorreview{to see in detail which of the above results apply also to BLE.}
Later, we experimentally evaluate CT performance over Bluetooth.
\subsection{Modeling CT in Frequency-Shift Keying (FSK) Systems}
\lsec{feasibility:CT_model}
In this section, we present analytic models for concurrent transmissions over Frequency-Shift Keying systems.
Bluetooth employs a compatible modulation --- Gaussian Frequency Shift Keying (GFSK).
\review{The difference is that the signal passes a Gaussian filter before being transmitted, to smooth the modulated signal transitions and limit the modulated spectrum width. 
Thus, it allows a narrower spacing between channels than FSK.}
For simplicity and without loss of generality, we consider the Binary Frequency-Shift Keying \review{(denoted as 2-FSK or BFSK), \ie without the Gaussian pulse shaping}.
\fakeparagraph{Scenario} We consider two concurrent Binary Frequency-Shift Keying (BFSK) transmitters and one receiver. 
Both transmitters are sending \textit{the same} pseudo-random bit-stream.

\fakeparagraph{Objectives} We (i) present an analytic expression of the bit error rate (BER) of the CT signal, with regard to the beating signal envelope;
(ii) simulate the bit error rate (BER) of the CT signal in presence of a time shift and a power difference;
(iii) discuss the scalability of CT in terms of achievable clock synchronization accuracy; and,
(iv) simulate the packet error rate (PER) under various beating conditions and power offsets.
We note that the simulations performed in this section are based on Monte Carlo sampling of the analytic expressions as finding a closed-form expression is \review{beyond the scope of this work}.

\fakeparagraph{Mathematical Representation} The transmitted signals, $y_1(t)$ and $y_2(t)$, have a relative temporal displacement, $\Delta t$, different amplitudes, $A_{1}$ and  $A_{2}$, and slightly different carrier frequencies, $f_{c1}$ and  $f_{c2}$. 
Both amplitude and frequency differences are unavoidable, since the signals take different paths and originate from different transmitters with different oscillators:

\begin{center}
\begin{equation}
\begin{aligned}
y_{1}(t)&=A_{1}\cos(2\pi(f_{c1}+n(\Delta f))t); \\
y_{2}(t)&=A_{2}\cos(2\pi(f_{c2}+n(\Delta f))(t-\Delta t));
n\in\{-1,+1\}
\end{aligned}
\end{equation}
\par\end{center}

At the receiver, the superposition of both signals, $y_1(t)+y_2(t)$, is a beating waveform due to summing sinuous waves with different frequencies.
The beating frequency is equal to the carrier frequency offset of one transmitter in regard to the other transmitter $f_{beat}=\left|f_{c1}-f_{c2}\right|$ \cite{escobar2019}.
Therefore, we base our discussion on the beating frequency $f_{beat}$ in the rest of this section.

\rfig{Modulated--signal-BFSK} shows the signal of one BFSK transmitter.
\rfig{Interference-between-two-1-1} shows the resulting beating waveform from the two concurrent transmitters.
As a first insight, we observe that beating causes a distortion of the amplitude and phase of the wave but luckily leaves the frequency shifts modulating the baseband signal visible. 

For illustrative purposes, we choose the following parameters:
$T_{S}$ as the symbol (bit) period and $h$ as the resulting modulation index.
This is the minimum modulation index for non-coherent orthogonal detection that is used in off-the-shelf IoT devices \cite{proakis2008digital}. 

\begin{center}
\begin{equation}
\Delta f=\frac{1}{2T_{S}};\;\;f_{beat}=\frac{1}{4T_{S}};\;\;h=2\Delta fT_{S}=1
\end{equation}
\par\end{center}

\subsubsection{CT Envelope and Bit Error Rate}
\lsec{feasibility:model:ct_envelope_ber}
At the receiver, we assume a simple non-coherent energy detector for demodulation, as normally used in low-power IoT devices~\cite{proakis2008digital}. 
In non-coherent demodulation, an envelope detector compares the energy in the two orthogonal branches, $f_{c}+\Delta f$ and $f_{c}-\Delta f$, to detect zeros and ones.
We note that no phase information is used, unlike the \emph{coherent} demodulators; \ie the phase differences do not deteriorate the demodulation process.
With two concurrent transmitters, the envelope detected in each frequency band fluctuates, as seen in \rfig{Interference-between-two-1-1}. 
Assuming $A_1 \geqslant A_2$, the positive envelope can be expressed as:
\begin{equation}
envelope(t)=(A_{1}-A_{2})+2A_{2}\left|\cos\left(2\pi\frac{f_{beat}}{2}t\right)\right|
\end{equation}

For two concurrent transmitters and non-coherent BFSK, we obtain the analytical expression of the expected BER. 
We assume that both transmissions are received with the same energy ($A_{1}=A_{2}$), 
where $E_{b0}$ is the constant energy of a single transmitter:
\begin{equation}
E_{b}(t)_{2CT, (A_{1}=A_{2})}=4E_{b0}\cos^{2}\left(2\pi\frac{f_{beat}}{2}t\right)
\end{equation}
Note that the theoretical Bit Error Rate (BER) of a single transmitter, which we use as a reference, is the expected performance of non-coherent orthogonal BFSK detection in Additive White Gaussian Noise (AWGN) channels \cite{proakis2008digital}. 
It is given by the following expression:
\begin{center}
\begin{equation}
BER_{BFSK}=\frac{1}{2}\exp\left(-\frac{E_{b}}{2N_{0}}\right)\label{eq:ber_bfsk}
\end{equation}
\par\end{center}

For simplicity, we consider no timing errors at the symbol level ($\Delta t=0$)
and a fairly constant energy during one bit reception ($T_{beat}\gg T_{S}$).
The average BER during one beating period, by using Eq. \ref{eq:ber_bfsk} together with the energy of the beating envelope, is:

\begin{align}
BER_{2CT} & =\frac{1}{T_{beat}}\int_{0}^{T_{beat}}\frac{1}{2}\exp\left(-\frac{E_{b}(t)}{2N_{0}}\right)\mathrm{\,d}t\\
 & =\frac{1}{2}\exp(-E_{b0}/N_{0})\;\;\mathrm{I_{0}}\left(-E_{b0}/N_{0}\right)\nonumber 
\end{align}

where $\mathrm{I_{n}}(z)$ is the modified Bessel function of the first kind.

\review{\rfig{-ber_ebn0-1-1-1-5-1-1} shows the analytical BER comparison between the case of a single transmitter versus two concurrent transmitters. 
Only for noisy receptions ($E_{b}/N_{0}<$ 4 dB), we observe a gain in the performance with two concurrent transmitters instead of a single transmitter.
In low-noise scenarios ($E_{b}/N_{0}>$ 4 dB), we notice a decreased performance with two concurrent transmitters, since the BER curve decreases more slowly. 
Concurrent transmissions are error prone in low-noise environments, but they offer an improved performance when the signal-to-noise ratio is low, which is commonly the case in low- to intermediate-quality links. 
}

\subsubsection{CT Bit Error Rate with Time Shifts and Power Difference}
\begin{figure*}[t]
    \centering
    \begin{subfigure}[t]{0.32\columnwidth}
    \centering
    \includegraphics[width=1.04\columnwidth]{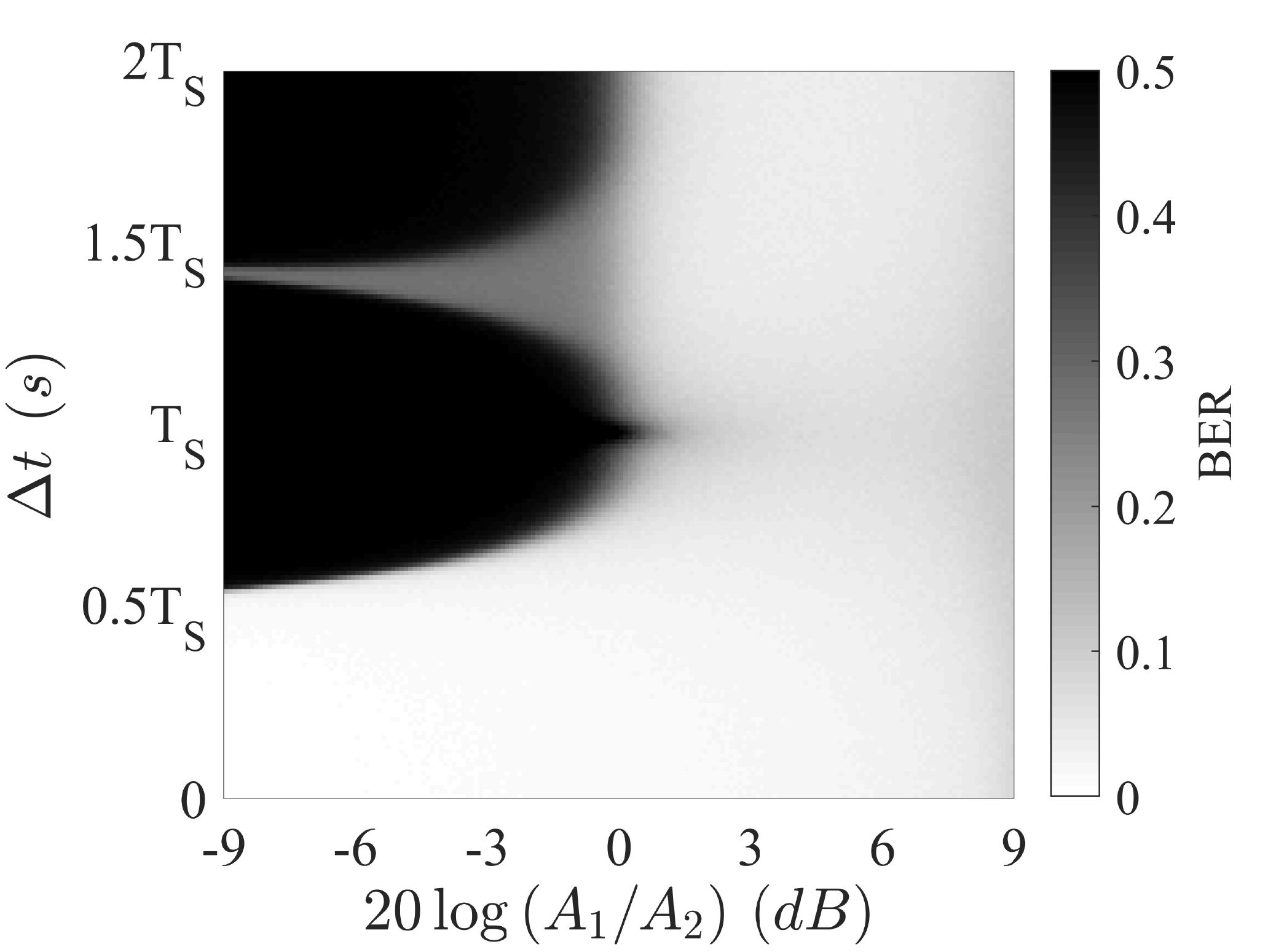}
    \caption{Simulated BER for two concurrent transmitters received with different relative power levels ($A_{1}/A_{2}$) and time displacements ($\Delta t$) in non-coherently received BFSK. AWGN channel with $E_{b}/N_{0}$ = 6 dB. The receiver synchronizes with $y_1(t)$. \label{fig:-ber_ebn0-1-1-1-5-1-2}}
    \end{subfigure}%
    \hfill%
    \begin{subfigure}[t]{0.32\columnwidth}
    \centering
    \includegraphics[width=1.0\columnwidth]{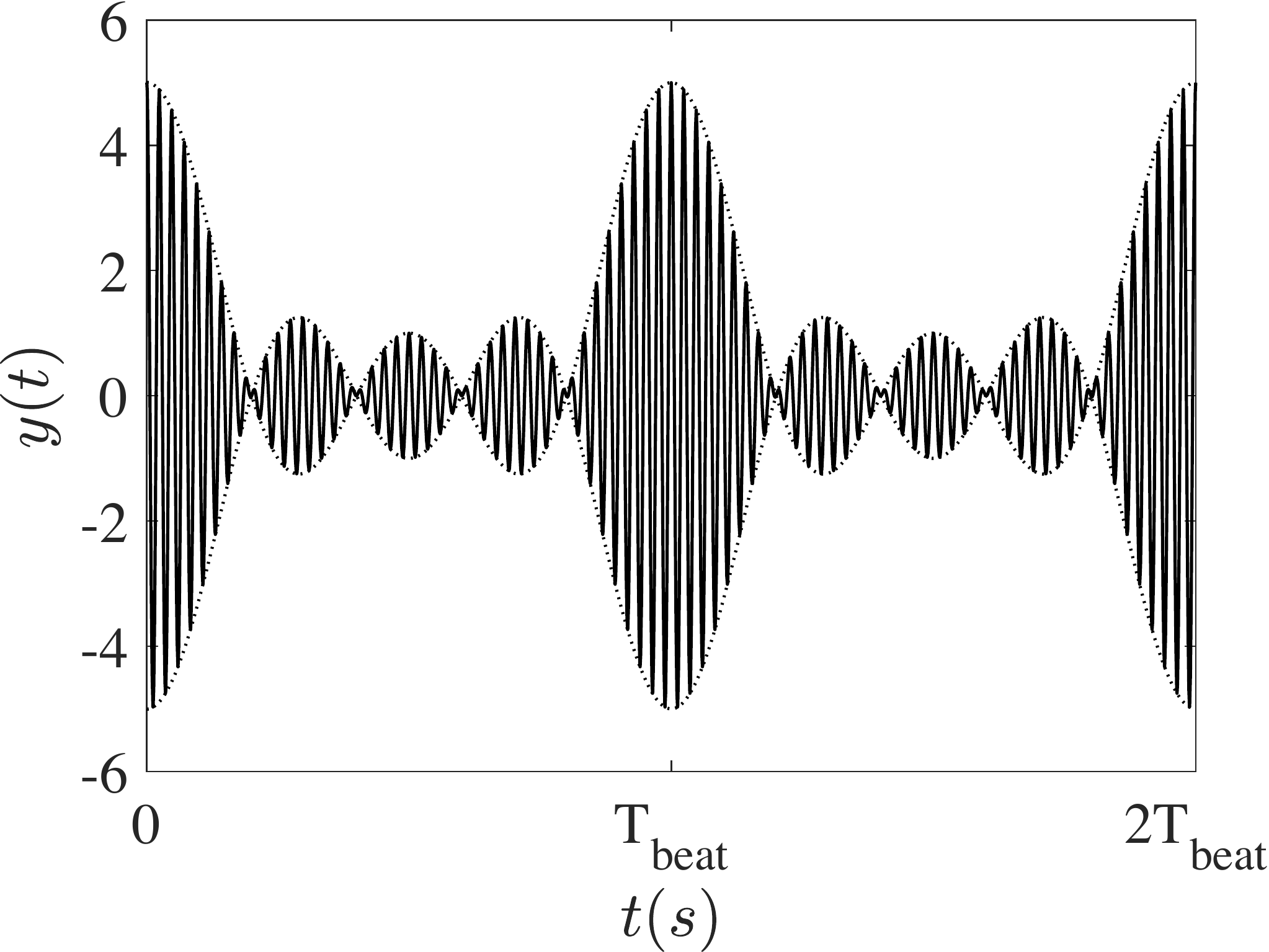}
    \caption{Example of a received waveform composed by the overlap of five concurrent carriers with the same energy and slightly different frequencies: complex envelope.
    \label{fig:-ber_ebn0-1-1-1-3-1-3}}
    \end{subfigure}%
    \hfill%
    \begin{subfigure}[t]{0.32\columnwidth}
    \centering
    \includegraphics[width=1.0\columnwidth]{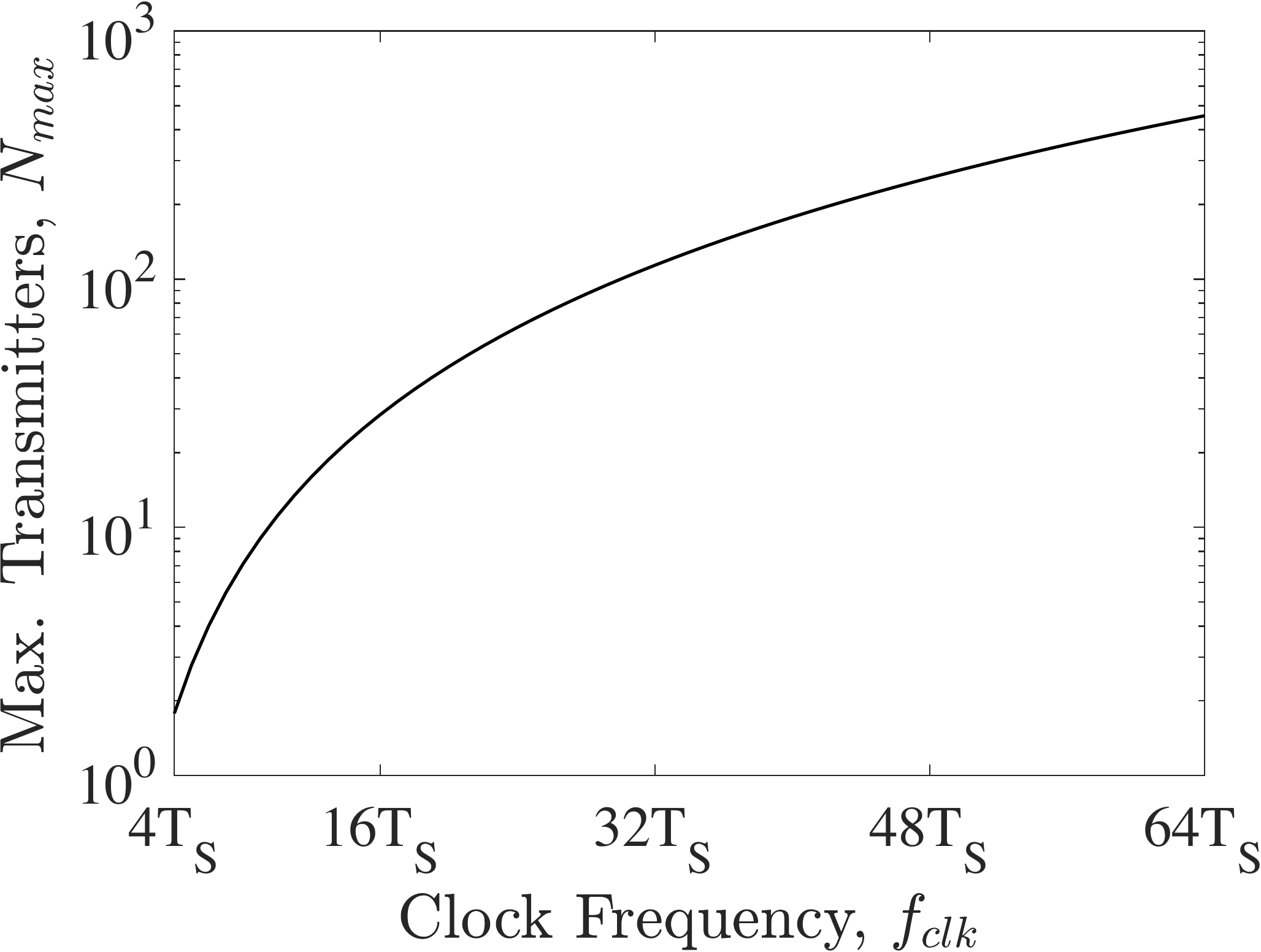}
    \caption{Maximum number of concurrent transmitters depending on the clock frequency of the processor of wireless nodes to keep the ISI low, assuming typical clock jitter and one-tick synchronization \cite{escobar2019}. \label{fig:-ideal-shaping-1-1}}
    \end{subfigure}
    \caption{Scalability of Concurrent Transmissions. [analytic]\label{fig:big-fig-models2}}
\end{figure*}

\lsec{feasibility:model:ct_time_shift_power_delta}
For more complex scenarios, involving different relative power levels and temporal misalignment ($A_1$, $A_2$ and $\Delta{t}$), finding a closed-form expression for the energy envelope is beyond the scope of this work. 
Thus, we use simulations.
\review{\rfig{-ber_ebn0-1-1-1-5-1-2}, shows the Monte Carlo simulation results that we obtain with MATLAB for the following scenarios with different time- and power-deltas}:
\begin{itemize}
\item For $\Delta{t} > T_{S}/2$ and $A_2 > A_1$, the performance is low (high BER) due to inter-symbol interference (ISI). 
In fact, for $\Delta{t} > T_{S}$, \review{different symbols from the two transmissions interfere. 
Thus,} the scenario is equivalent to two transmitters sending different bit-streams.
\item For $A_1 > A_2$, the receiver synchronizes with the strongest signal, and the well-known capture effect for frequency modulations kicks in, decreasing the BER; thus, improving the performance.
\item \review{For $\Delta{t} < T_{S}/2$, both transmitters are properly synchronized for concurrent transmissions. 
In contrast to the capture effect, concurrent transmissions of the same bit-streams} can be decoded with high probability even if the receiver synchronizes with the weakest signal.
\end{itemize}

\subsubsection{Scalability of Concurrent Transmissions}
\lsec{feasibility:model:ct_scalability}
In networks with more than two concurrent transmitters, the amplitude distortion becomes more severe, with complex envelopes hard to describe analytically. 
Multiple peaks and valleys typically appear during the beating period (\rfig{-ber_ebn0-1-1-1-3-1-3}). 
Assuming the $N$ individual transmissions arrive with the same energy, $E_b$, we remark:
\begin{itemize}
\item When all the concurrent transmissions overlap perfectly in phase,
there is a temporal instant of pure constructive interference, 
in which the reception reaches a high energy peak ($N^{2}E_{b}$).
\item When all the transmissions overlap in purely destructive interference,
the transmission fades.
\end{itemize}

In practice, concurrent transmissions interfere with time-varying energy and frequency offsets. 
Thus, we expect complex shapes of the amplitude envelope, with successive depressions and peaks. 
During a packet reception, there are quick successions of high- and low-energy periods. 
The length of these periods depends on the accuracy of the local oscillators. 
The higher the accuracy, and thus the lower the offsets between their carrier frequencies, the longer the resulting beating period. 
This is usually the case with accurate oscillators: the beating period spans the reception of multiple symbols, or even whole packets. 

Scalability is also constrained by the limited clock accuracy of the wireless transceivers. Usually, CT-based protocols \cite{ferrari11glossy, 201311LandsiedelChaos, escobar2019competition}, achieve synchronization at the instruction level by enforcing a constant number of executed instructions or clock ticks after a packet is received in every concurrent transmitter. 
Assuming the temporal displacement between $N$ transmitters follows a normal distribution, with variance, $N\sigma^2$, and using the standard deviation for normal distributions, we can approximate the maximum number of concurrent transmitters, $N_{max}$, \review{that depends on the clock speed of the CPU,} as:
\begin{center}
\begin{equation}
\sigma\simeq\frac{1}{2f_{clock}};\;\;N_{max}=\left(\frac{T_{S}f_{clock}}{3}\right)^{2}\label{sca}
\end{equation}
\par\end{center}

Which means that 95\% of the re-transmissions have an equivalent temporal displacement below half the symbol period. In order to deploy denser networks, either the clock frequency of the digital processor of the wireless mesh nodes, $f_{clock}$, needs to be increased or the communication bitrate needs to be reduced \cite{escobar2019}.

\subsubsection{\review{Impact of the Beating Frequency on the Packet Error Rate}}%
\lsec{feasibility:model:beating_prr_analysis}

In AWGN channels, bit errors are uncorrelated. Therefore, the Packet Error Rate (PER) with only one transmitter, for packets of length $L$, can be expressed as:
\begin{center}
\begin{equation}
PER_{1T} = 1-(1-BER)^L
\end{equation}
\par\end{center}

However, in concurrent transmissions, bit errors tend to appear in bursts during the energy depressions (valleys) of the beating waveform.
In these periods, the concurrent transmissions are interfering destructively.
Similarly, there are periods of constructive interference during the energy peaks of the beating wave.
The beating frequency $f_{beat}$ is equal to the carrier frequency offset (CFO) between the two transmitters~\cite{escobar2019}.
This CFO (and $f_{beat}$) depends on the local oscillator accuracy of the transmitters.
\review{For example, with an extremely accurate oscillator with an accuracy of 10 parts per billion (ppb), the \unit[2.4]{GHz} carrier will have an offset of $\pm10\times10^{-9}\times\unit[2.4\times10^9]{Hz}=\pm\unit[24]{Hz}$. 
This represents a maximum CFO and beating frequency of \unit[48]{Hz}, which has a period of \unit[20.8]{ms}. 
For comparison, a standard Bluetooth iBeacon packet takes \unit[0.368]{ms} air time~\cf\rtab{eval:slotlength}.
However, for low-power devices, we expect a worse accuracy of 1 to 40 parts per million (ppm).
Thus, a worst-case beating frequency of \unit[4.8]{KHz} to \unit[192]{KHz}, \ie beating periods of \unit[0.2]{ms} to \unit[5]{$\mu s$}. 
At the same time, we expect the frequency error to follow a normal distribution. 
Therefore, the actual CFO is statistically better than the worst-case expectations.
}

Based on this discussion, we expect different performance regions with regard to the beating frequency (and period) relative to the packet size: %
\begin{itemize}
    \item \emph{slow beating}: $T_{beat} \gg T_{S}$ and $T_{beat} \geqslant T_{Packet}$, and
    \item \emph{fast beating}: $T_{beat} \gg T_{S}$ and $T_{beat} < T_{Packet}$.
\end{itemize}
\majorreview{To quantify the difference, we perform Monte Carlo simulations in MATLAB. Our results show:
}
\majorreview{\fakeparagraph{Beating with the same power and no time displacement (\rfig{modeling:PER_beating_a})}
In high-noise scenarios ($E_b/N_0 < 10$), two CTs give a performance gain when compared to the single transmitter case, as we saw in the BER analysis~\rsec{feasibility:model:ct_envelope_ber}.
Nevertheless, when the noise is lower, the performance is again worse than sending with only one transmitter. 
As the beating period increases, packets have a better chance to be completely transmitted during a constructive interference phase of the beating waveform. 
However, the destructive regions also get wider. 
The net effect is that the PER tends to slowly decrease when increasing the beating period. The observed performance shares similarities with multipath Rayleigh channels and, similarly to fast fading. This indicates that fast beating may be mitigated by using error-correcting codes and interleaving.
}
\majorreview{\fakeparagraph{Beating with a power delta (\rfig{modeling:PER_beating_b} and \rfig{modeling:PER_beating_d})}
In real-world scenarios, when dense mesh networks are deployed, signals are received with time-varying energy levels. Energy differences between the concurrently received transmissions help diminishing the beating distortion. Already from only 1dB power difference ---easy to achieve in practical deployments---, the PER decreases dramatically, quickly approaching the no beating performance as the power delta is increased. In comparison, when packets with different data overlap in the air, a power delta around 8~dB is required to trigger the capture effect and achieve comparable results. This stresses the unique characteristics of CT with packets with the same data and its exploitable performance.
}
\majorreview{\fakeparagraph{Beating with a time delta (\rfig{modeling:PER_beating_c}, \rfig{modeling:PER_beating_e} and \rfig{modeling:PER_beating_f})}
As widely discussed, CT need to be properly synchronized at the symbol level to operate properly. A time delta of half the symbol period is the hard threshold from which the performance is practically the same as sending packets with different data, and a large power delta (higher than 8~dB) is required to trigger the capture effect and enable the communication. The region in which the performance is optimal is time deltas below a quarter of the symbol period. 
}
\majorreview{\fakeparagraph{Discussion -- Potential performance gain with coding}
Our results show that distortion due to CT shares similarities with fading channels, and uncoded transmissions are fragile in these environments. 
Adding burst error-correcting codes effectively increases the chances of a successful packet reception, since errors appear in bursts during fading-like energy depressions. 
However, this comes with the cost of the added complexity in the communication system and longer packet periods, thus, a higher energy expenditure for the same payload.
Coding techniques, like the DSSS used in IEEE 802.15.4 and the FEC convolutional codes used in the coded PHY modes of Bluetooth 5, greatly decrease the PER in the presence of concurrent transmissions.
We validate this experimentally in~\rsec{feasibility:txPower}.
While for uncoded transmissions a slow beating regime is clearly preferred, fast beating offers better chances of error reduction through coding. 
The reason is that time diversity can be exploited within a packet reception, and techniques like interleaving combined with FEC are very effective. 
On the contrary, deep and wide fades during very slow beatings create error bursts too long to be practically recovered, which limits the potential gain of applying error-correcting techniques. 
}

\majorreview{
Our model is an abstraction. It gives insights into challenges of CT with (uncoded) BLE and explains the expected behaviour. 
However, it is not intended to predict exact thresholds since, for example, it does not model phase locking or automatic gain-control on the receiver side. 
Therefore, it is important to validate our conclusions from this analytic study using experimental results. 
In the next section, we evaluate the performance of CT on commercial Bluetooth transceivers in a controlled scenario.
}

\begin{figure*}[t]
  \centering
    \begin{subfigure}[t]{0.32\columnwidth}
    \centering
    \includegraphics[width=1.0\columnwidth]{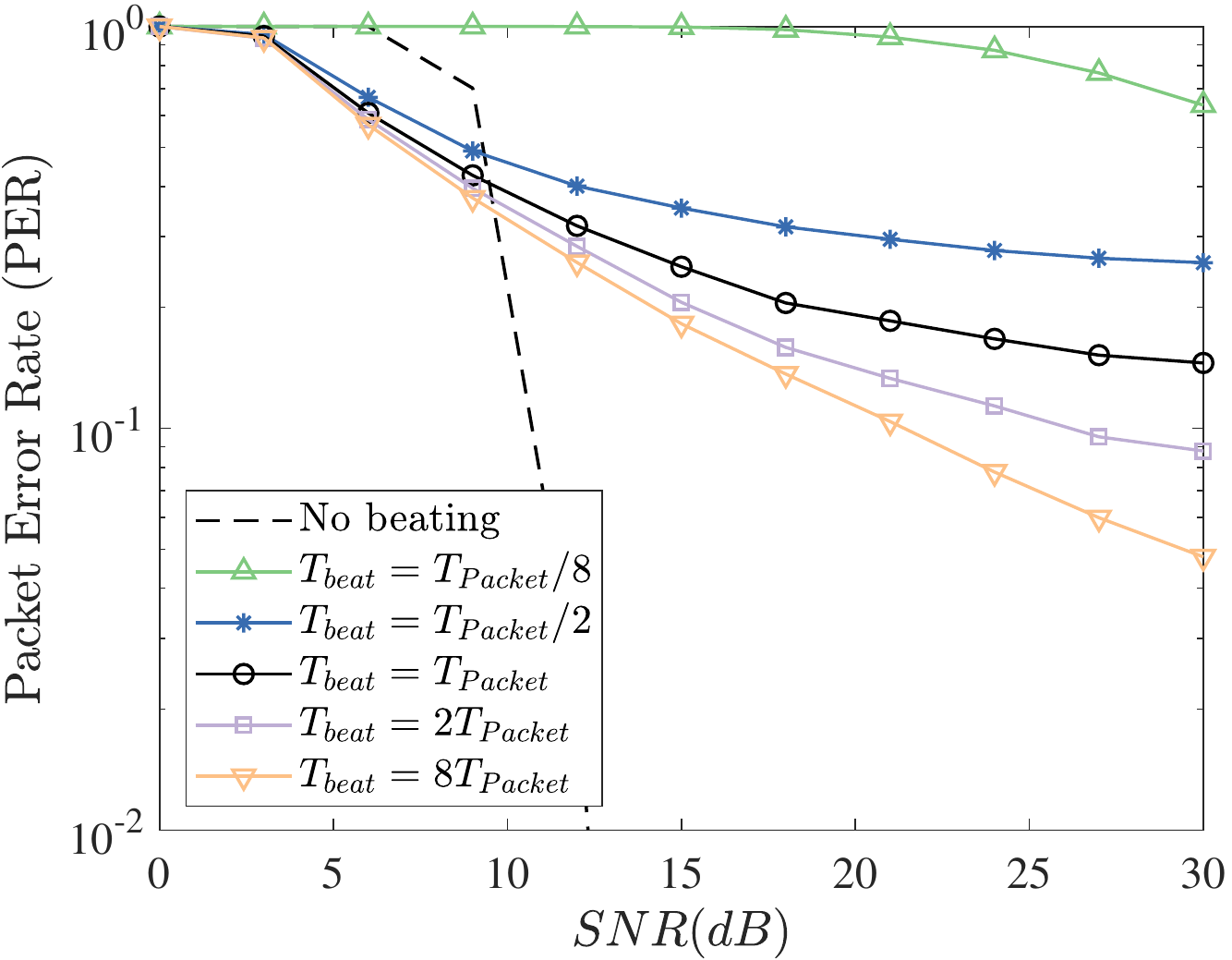}
    \caption{$\Delta P$=0dB, $\Delta t$=0, Same Data \lfig{modeling:PER_beating_a}}
    \end{subfigure}%
    \hfill%
    \begin{subfigure}[t]{0.32\columnwidth}
    \centering
    \includegraphics[width=1.0\columnwidth]{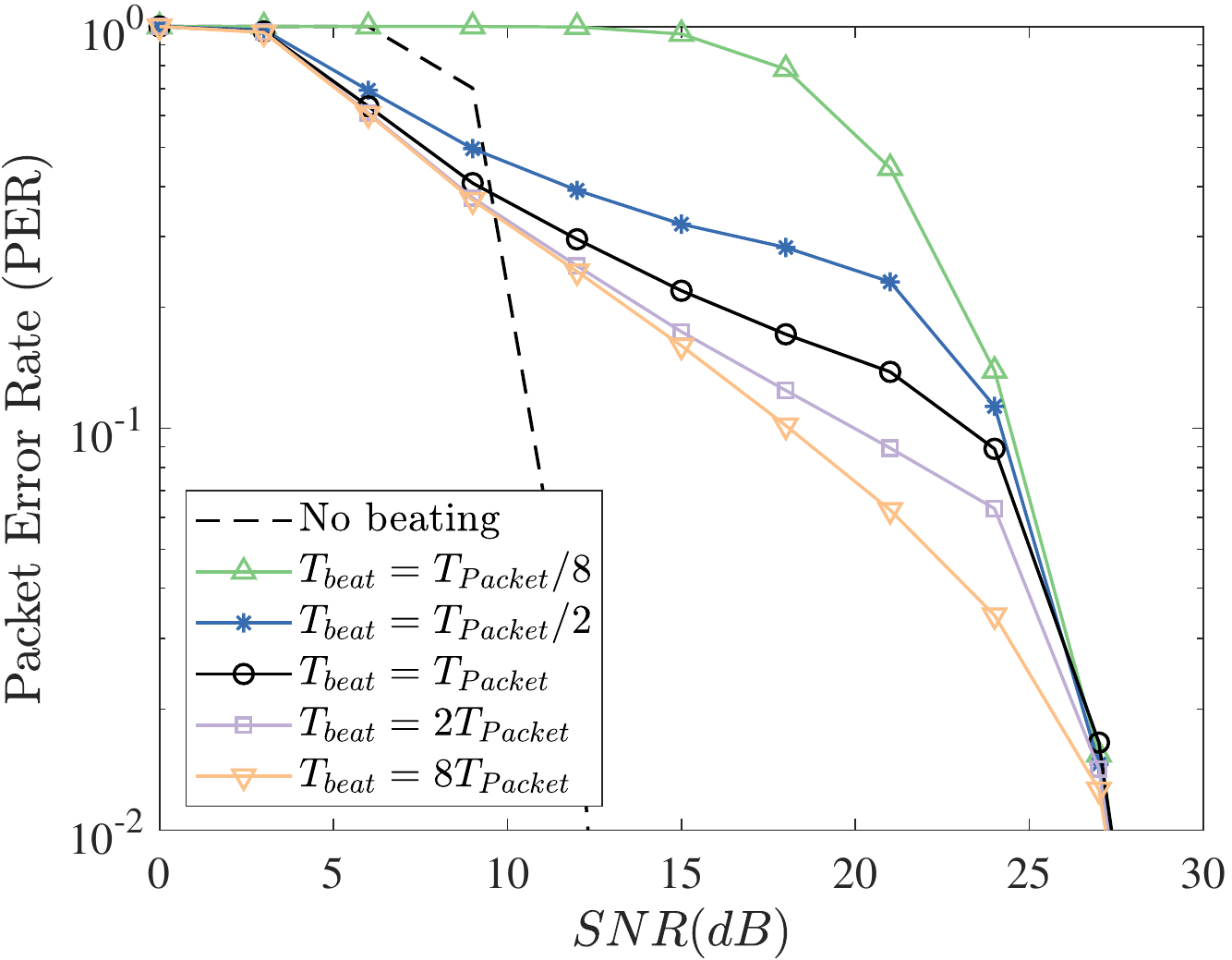}
    \caption{$\Delta P$=1dB, $\Delta t$=0, Same Data \lfig{modeling:PER_beating_b}}
    \end{subfigure}
    \hfill
    \begin{subfigure}[t]{0.32\columnwidth}
    \centering
    \includegraphics[width=1.0\columnwidth]{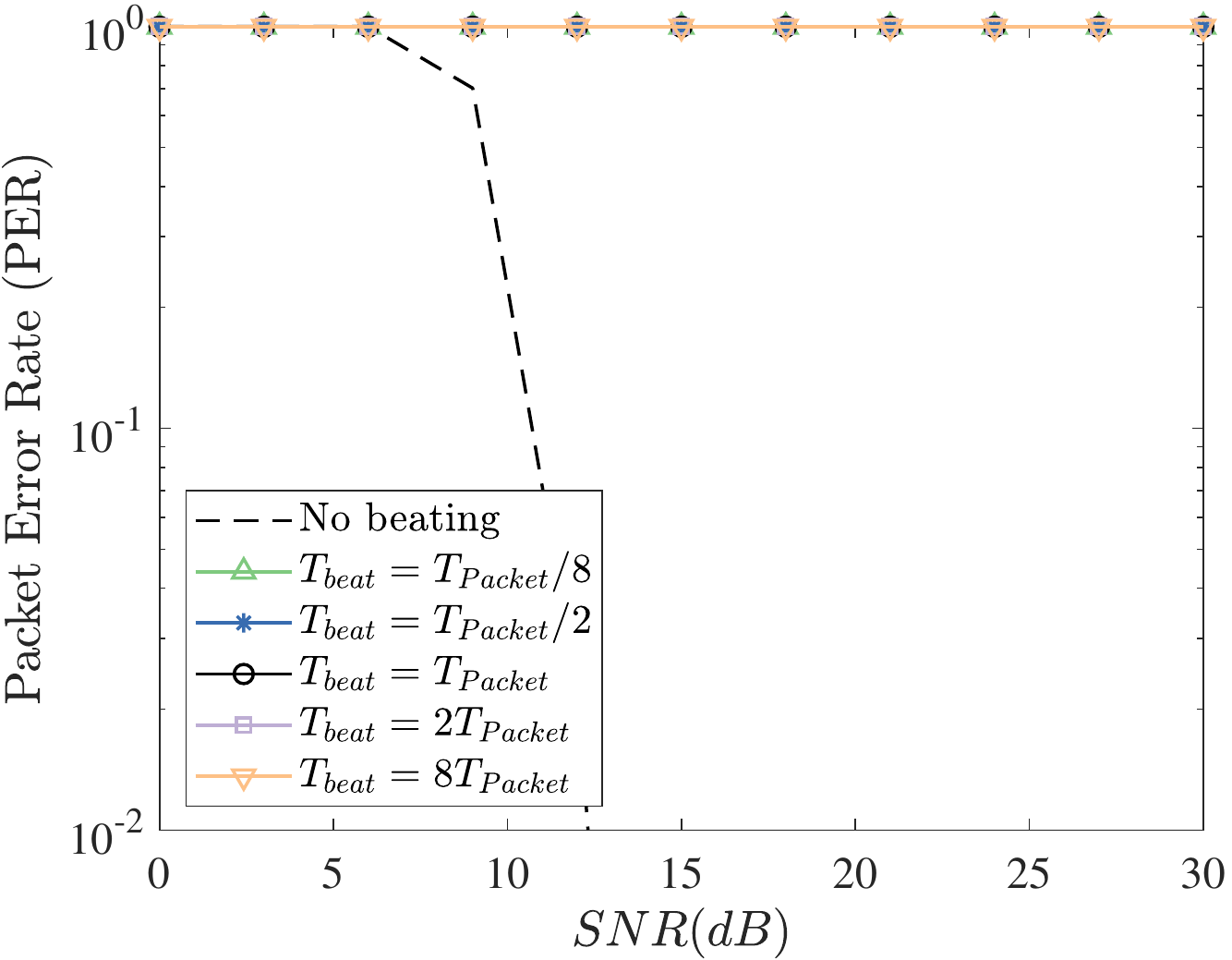}
    \caption{$\Delta P$=0dB, $\Delta t$=$T_S$/2, Same Data
    \lfig{modeling:PER_beating_c}}
    \end{subfigure}\\%
    \vspace{1em}
        \begin{subfigure}[t]{0.32\columnwidth}
    \centering
    \includegraphics[width=1.0\columnwidth]{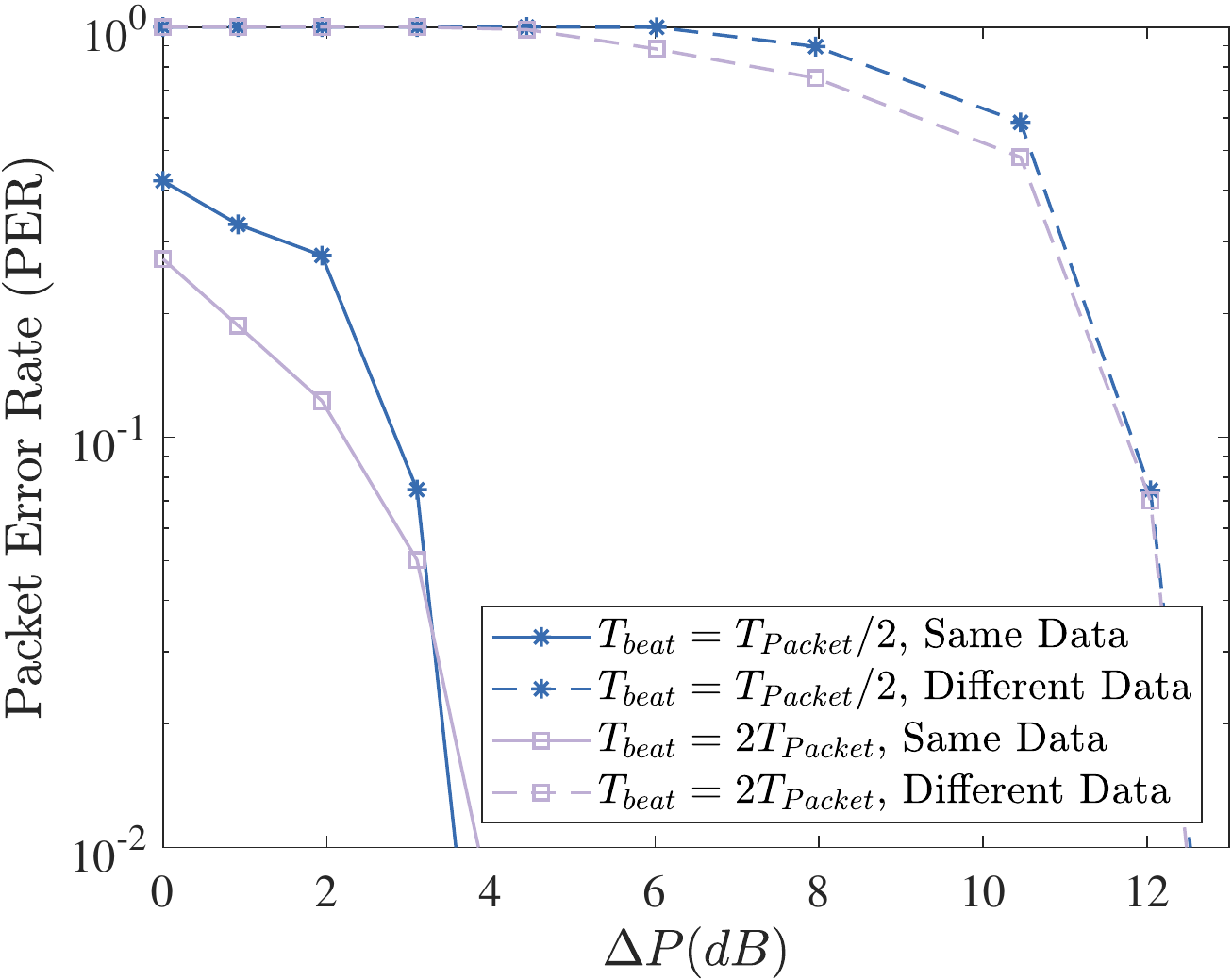}
    \caption{SNR=12 dB, $\Delta t$=0
     \lfig{modeling:PER_beating_d}}
    \end{subfigure}%
    \hfill
    \begin{subfigure}[t]{0.32\columnwidth}
    \centering
    \includegraphics[width=1.0\columnwidth]{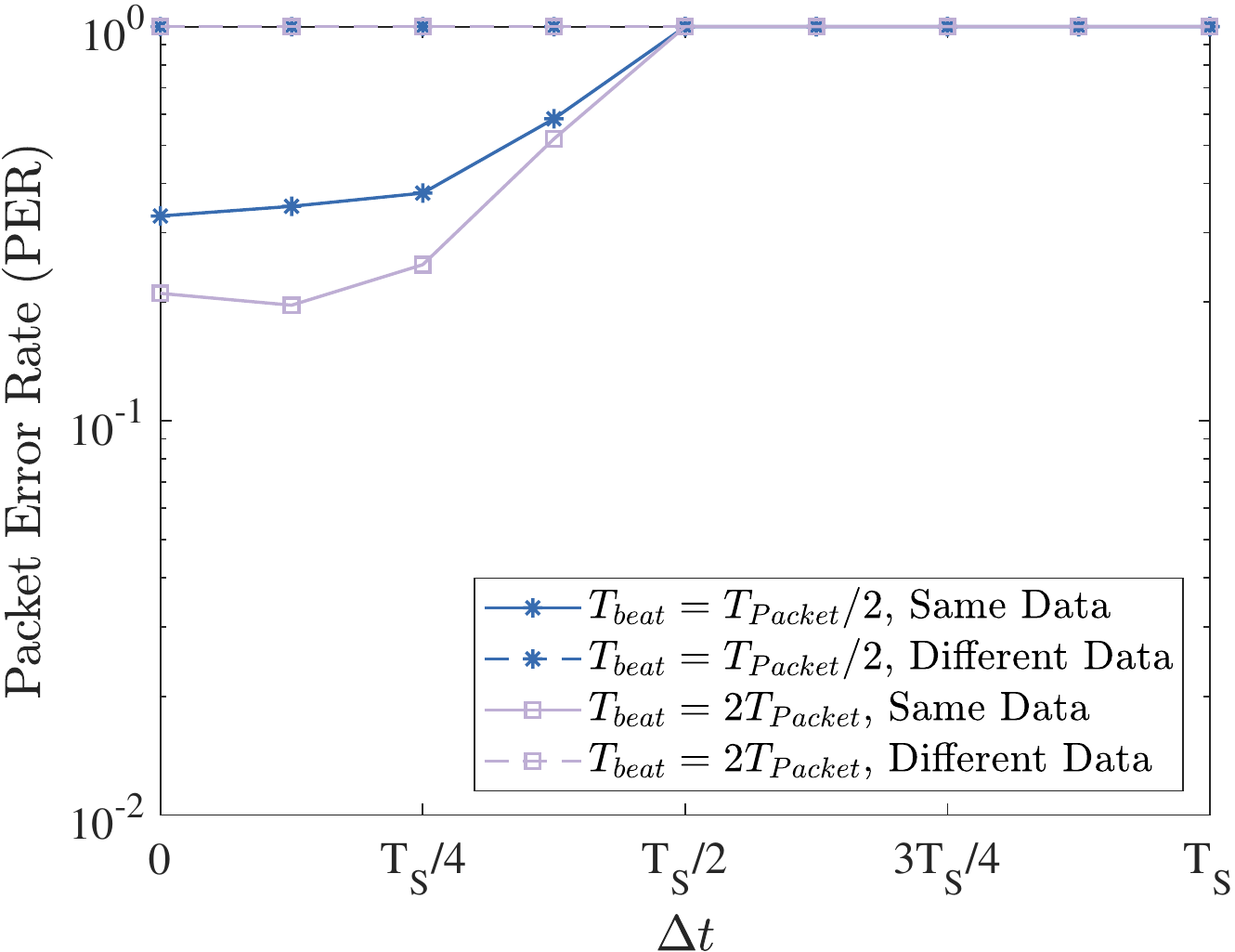}
    \caption{SNR=12 dB, $\Delta P$=0 dB
    \lfig{modeling:PER_beating_e}}
    \end{subfigure}%
    \hfill
    \begin{subfigure}[t]{0.32\columnwidth}
    \centering
    \includegraphics[width=1.0\columnwidth]{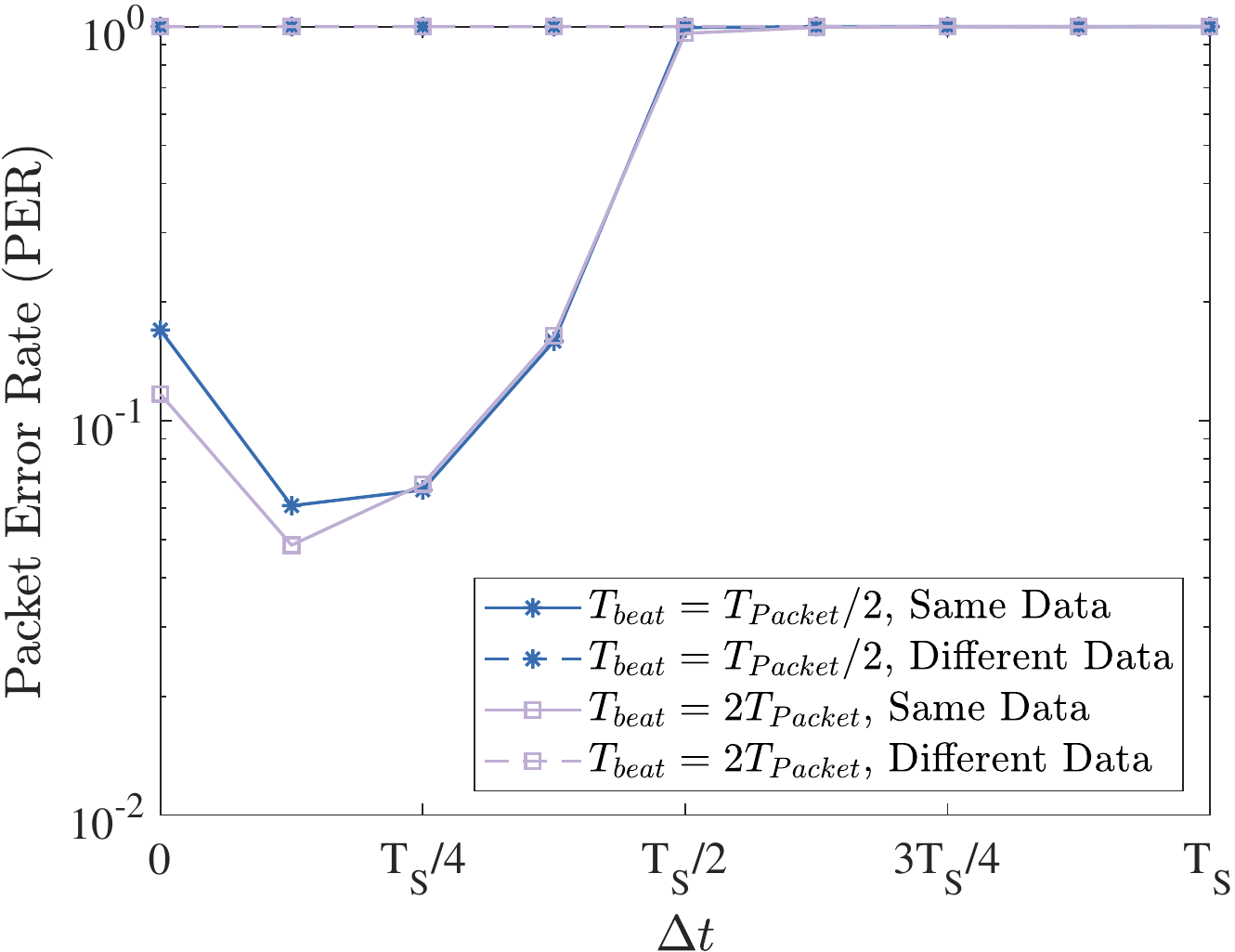}
    \caption{SNR=12 dB, $\Delta P$=2 dB 
    \lfig{modeling:PER_beating_f}}
    \end{subfigure}%
    
\caption{PER for non-coherent BFSK and uncoded transmissions in AWGN channels with one transmitter (no beating) and two CT for different noise levels (SNR) beating periods, $T_{beat}$, power deltas, $\Delta P = 20\log(A_1/A_2)$, and time deltas $\Delta t$.  Considering packets of a given air-time, $T_{Packet}$, and length, $L = 128\;bit$.\label{fig:big-fig-models3}}
\end{figure*}

\subsection{CT over Bluetooth: Controlled Experimental Study}
\lsec{feasibility:CT_experimental}
The results of our numerical analysis indicate the feasibility of CT over Bluetooth.
Before devising and implementing a full system for concurrent transmissions in Bluetooth in Section \rsec{Design}, we complement the numerical analysis with a series of controlled and reproducible experiments.

\fakeparagraph{Objectives}
In this section, we validate our analysis in a controlled environment using off-the-shelf Bluetooth 5 modules (nRF52840) and a Software Defined Radio (SDR).
We show the feasibility of CT over Bluetooth by answering four questions:
(i) What are the carrier beating patterns resulting from concurrent transmissions, and how do they affect the performance of CT?
(ii) How reliable is a Bluetooth CT link depending on the difference in the received signal strength of two concurrent transmitters?
(iii) How does timing accuracy affect the reliability of CT?
(iv) How does CT in the Bluetooth PHY perform when sending same vs. different data?

\begin{figure}
\centering
\begin{minipage}{0.66\textwidth}
\centering
    \begin{subfigure}[t]{0.48\columnwidth}%
        \centering%
        \includegraphics[width=1\columnwidth,page=2,clip]{pics/wired-nodes.pdf}%
        \caption{Setup for measuring carrier beating with CT\lfig{cfo_setup}}%
    \end{subfigure}%
    \hfill%
    \begin{subfigure}[t]{0.48\columnwidth}%
        \centering%
        \includegraphics[width=1\columnwidth,page=1,clip]{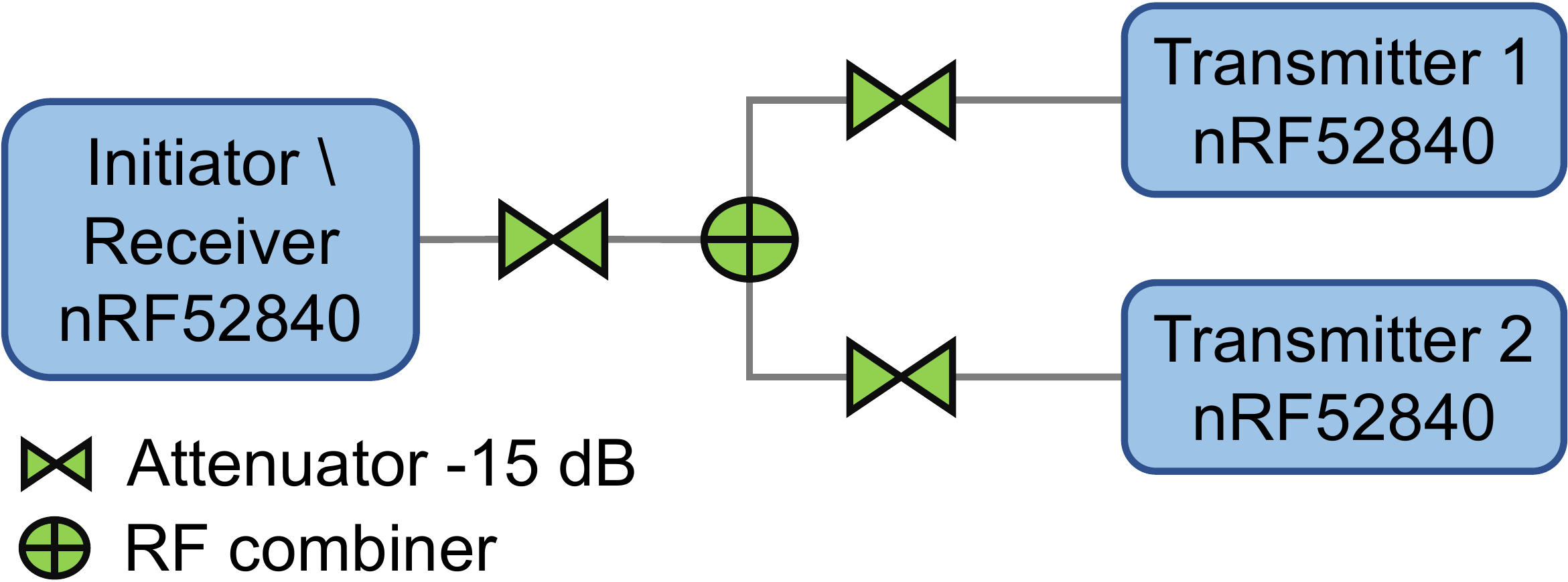}%
        \caption{Setup for measuring Packet Reception Rate PRR\lfig{prr_setup}}%
    \end{subfigure}%
    \caption[Evaluation Setup of the CT Feasibility Study]{Feasibility of CT over Bluetooth PHY: \capt{micro-evaluation setup of two concurrent transmitters and one receiver connected via coaxial cables and attenuators through their antenna connectors\lfig{experiment_setup}    \lfig{wirednodes}.}}%
\end{minipage}
\hfill
\begin{minipage}{0.32\textwidth}
    \centering
    \captionsetup{type=table} %
    \caption{Parametrization of the Bluetooth transmitters deviations\ltab{eval_cfo}}%
    \footnotesize%
    \begin{widetable}{1\linewidth}{lcc}%
    \toprule%
    \bf{Transmitter} & \bf{$CFO_{avg}$} & \bf{Magnitude}\\%
    {\unit{\#}} & {[\unit{KHz}]} & {[\unit{mV}]}\\%
    \midrule%
    1 & -8.58 & 0.195\\%
    2 & -6.75 & 0.21\\%
    3 & -18.25 & 0.19\\%
    4 & -18.23 & 0.20\\%
    \bottomrule%
    \end{widetable}%
\end{minipage}
\end{figure}

\begin{figure*}[t]
  \centering
    \begin{subfigure}[t]{0.24\columnwidth}
    \centering
    \includegraphics[width=1\columnwidth]{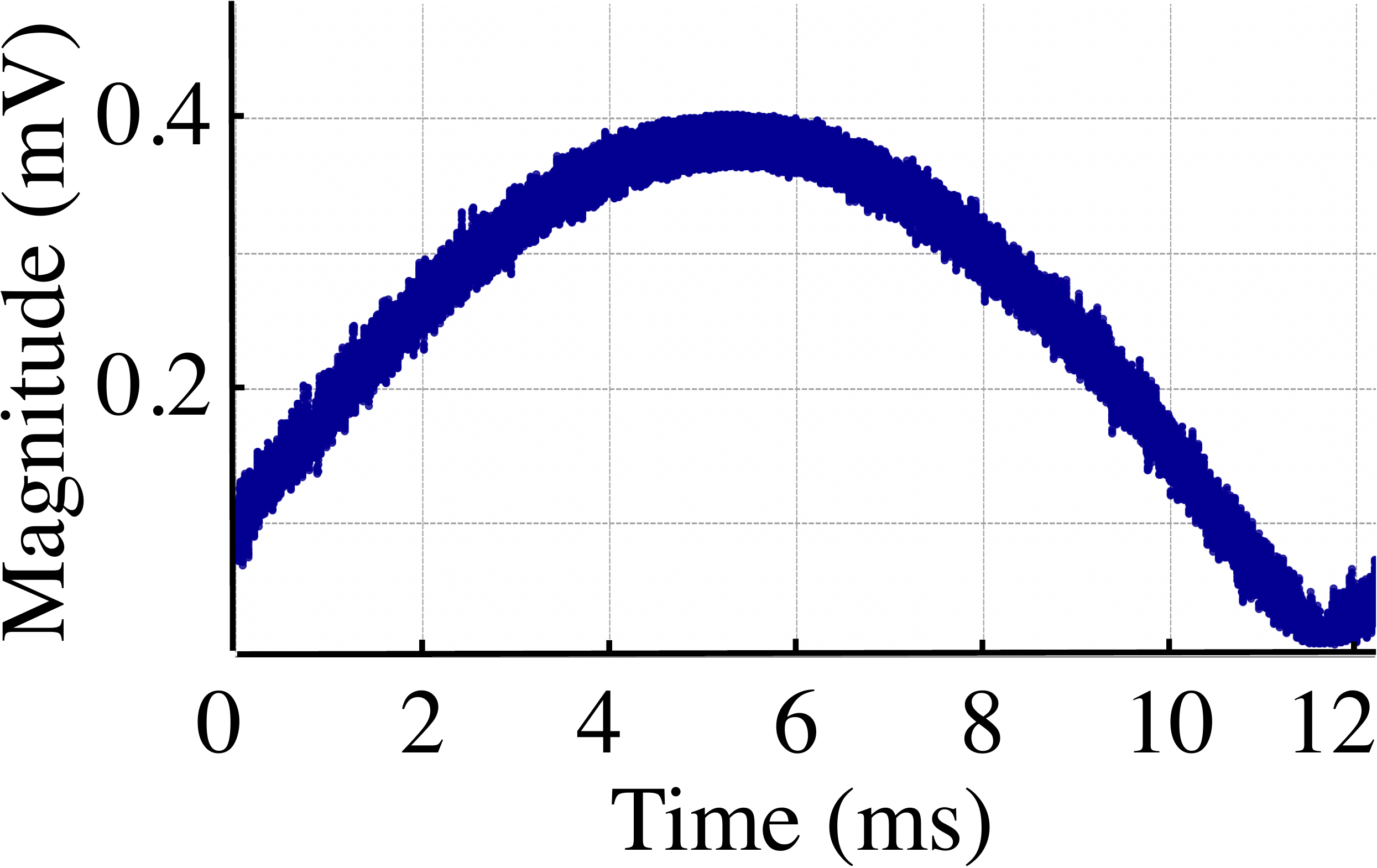}
    \caption{Slow beating (transmitters 3, 4): $f_{beat}\approx80~\unit{Hz}$ \lfig{sdr-beating-wide}}
    \end{subfigure}%
    \hfill%
    \begin{subfigure}[t]{0.24\columnwidth}
    \centering
    \includegraphics[width=1\columnwidth]{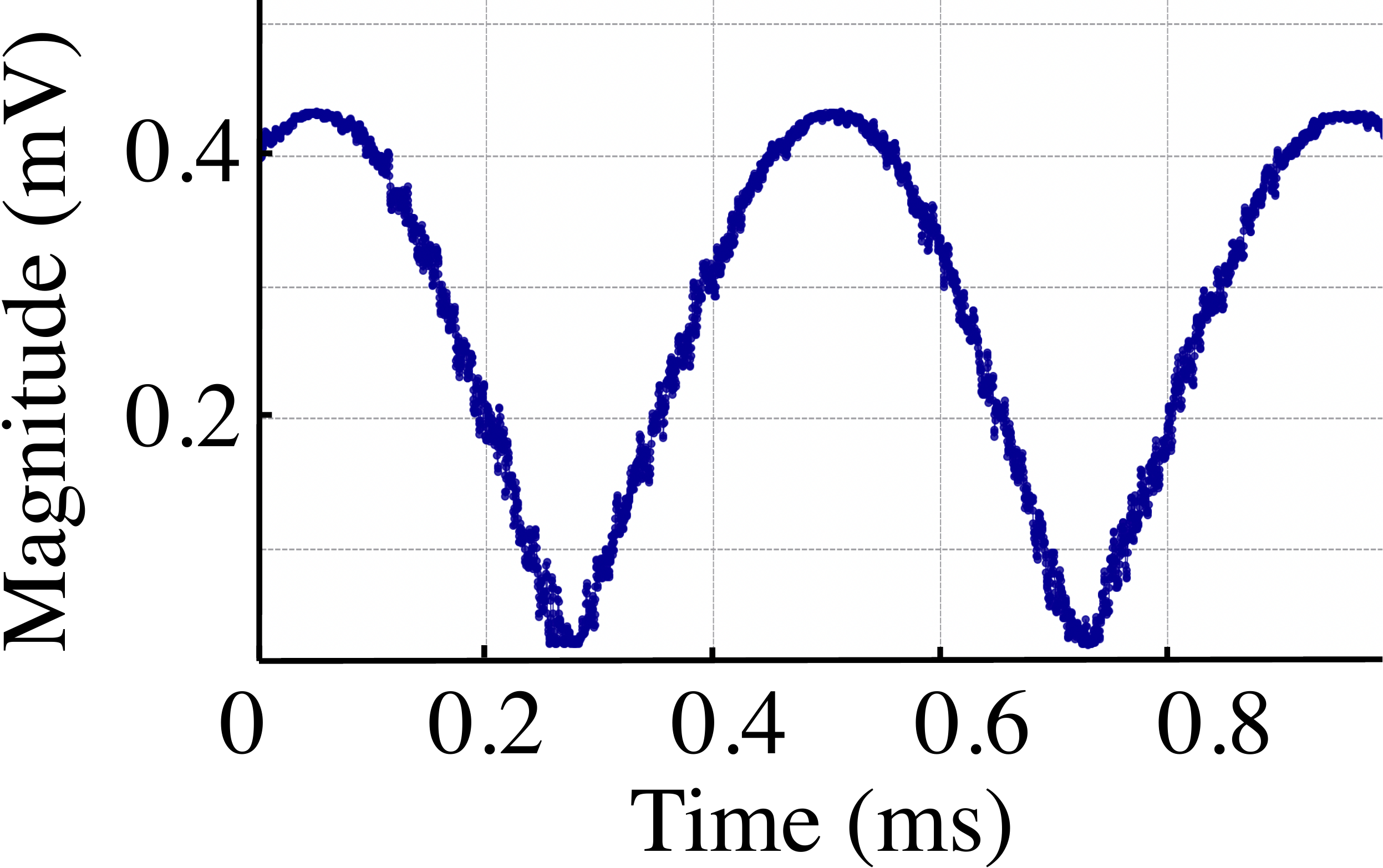}
    \caption{Fast beating (transmitters 1, 2): $f_{beat}\approx2.1~\unit{KHz}$ \lfig{beating2K}}
    \end{subfigure}%
    \hfill%
    \begin{subfigure}[t]{0.24\columnwidth}
    \centering
    \includegraphics[width=1\columnwidth]{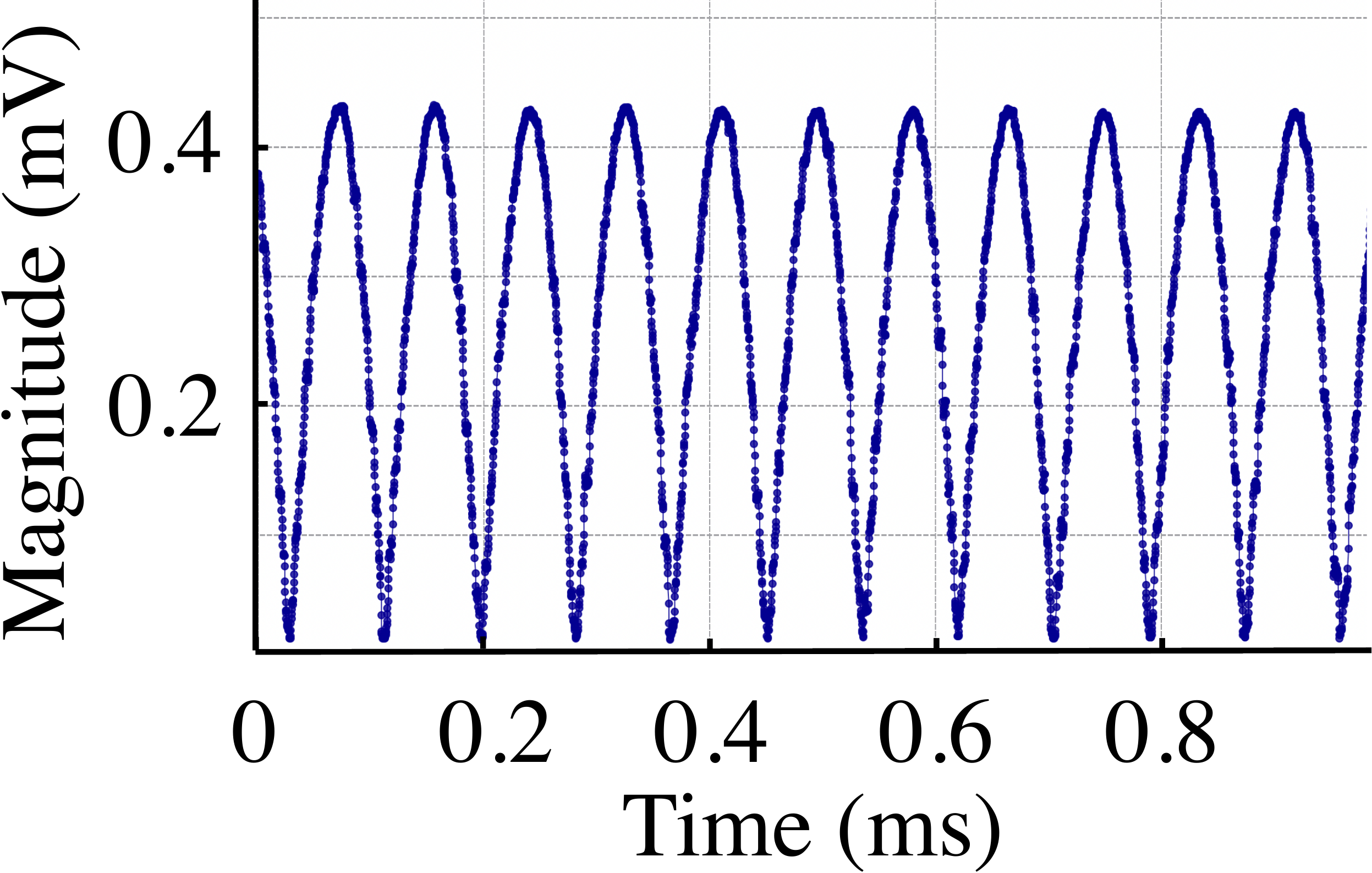}
    \caption{Fast beating (transmitters 2, 4): $f_{beat}\approx12~\unit{KHz}$\lfig{beating12K}}
    \end{subfigure}%
    \hfill%
    \begin{subfigure}[t]{0.24\columnwidth}
    \centering
    \includegraphics[width=1\columnwidth]{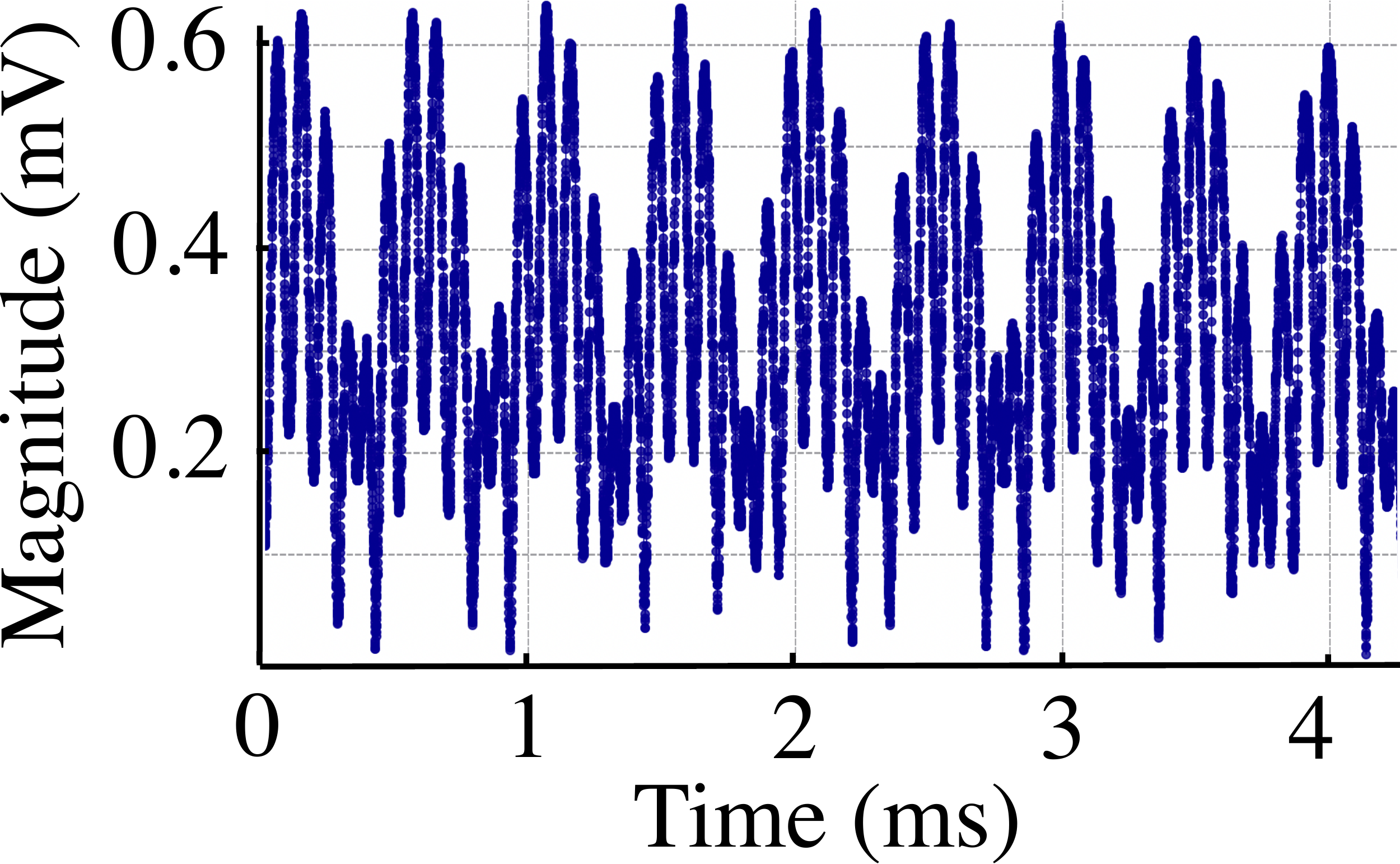}
    \caption{Beating with three carriers (transmitters 2, 3, 4): complex envelope\lfig{beating3carriers}}      
    \end{subfigure}%
\caption{Capturing the Envelope of the Beating Carrier in CT reception. The beating frequency is equal to the frequency difference of the transmitters. Slow beating results in a lower distortion. [experiment] \lfig{beating-pattern-sdr}}
\end{figure*}

\fakeparagraph{Setup} \review{For simplicity and without loss of generality, we focus in our feasibility study on the case of two concurrent transmitters and one receiver.
To enable reproducible results, we employ a symmetric and controlled communication-channel free from external interference: 
all nodes in this feasibility study are connected via equal-length coaxial cables and equal attenuators through their antenna connectors.
Thus, the received power from every concurrent transmitter can be easily controlled, avoiding the typical distortions of wireless channels, such as multi-path fading. 
This setup is common in related work, \eg, Wilhelm et.~al. ~\cite{wilhelm14ct}.}

We evaluate the performance of concurrent transmissions over Bluetooth using a setup of three nodes equipped with nRF52840 SoCs (see \rtab{eval:nodes} and \rfig{prr_setup}) capable of Bluetooth~5 communication:
(i) an initiator node that starts periodic rounds by transmitting a packet, then switches to receive mode, and (ii) two CT nodes that transmit concurrently after hearing the first packet.
We send iBeacon packets \review{with a PDU of 38 bytes which corresponds to frames of 46 and 47 octets in the \unit[1~and 2]{Mbps} modes, respectively, as shown in~\rfig{bg:blepacket}}.
We test both cases of sending the same data and different data.
Each experiment is run until at least 2000 packets are sent.

\subsubsection{Carrier Beating with Concurrent Transmissions}
\lsec{feasibility:carrier_beating_eval}

We measure the actual operating frequency of the transmitters, as compared to their nominal frequency. 
Then, we capture the different carrier beating patterns resulting from concurrent transmissions by different transmitter pairs.
We use a wired setup with SDR to receive the combined signal and up to four nRF52840 nodes, as shown in \rfig{cfo_setup}. 

\fakeparagraph{Carrier Frequency Offset}
We configure the transceivers in the test mode to transmit an unmodulated carrier in the same channel and with the same transmission power. We use the Bluetooth channel 37, which has a nominal frequency of \unit[2.402]{GHz}. 
However, we expect that the actual frequency deviates due to oscillator inaccuracies and operating factors, such as temperature. 
We measure the Carrier Frequency Offset (CFO) on four Bluetooth boards in \rtab{eval_cfo}. 
We also note slight transmission power deviations between the boards.

\fakeparagraph{Carrier Beating}
As expected from the models in \rsec{feasibility:CT_model}, concurrent transmissions result in signal distortion in the form of beating.
A very wide beating pattern when the frequency difference of the two transmissions is small, as shown in \rfig{sdr-beating-wide} and \rfig{beating2K}.
It shall be noted that a standard Bluetooth advertisement packet of \unit[46]{octets} lasts for \unit[368]{\textmu{s}} when transmitted using the legacy \unit[1]{Mbps} mode. 
This is a typical situation of \emph{slow} beating. 
It is very likely for the packet to be transmitted during a period that does not encompasses the potentially destructive deep fade of the valley.

However, when the frequency offset of the two transmitters is large, this typically results in a \emph{fast} beating pattern, as in \rfig{beating12K}.
Fast beating deteriorates the quality of the \emph{uncoded} concurrent transmission link, as most of the packets have to survive several valleys to be properly decoded.
Finally, \rfig{beating3carriers} shows the beating pattern with three concurrent transmitters, resulting in complex shapes of the beating envelope. 
The combined signal has a magnitude at least as strong as one of the transmitters for a larger portion of the beating period, which we expect to have a positive impact on the resulting link quality.

\subsubsection{CT Performance vs. Carrier Frequency Offset (CFO)}
\lsec{feasibility:cfo_eval}

We evaluate the performance of two concurrent transmissions over Bluetooth using the setup depicted in \rfig{prr_setup}: 
(i) an initiator node that starts periodic rounds by transmitting a packet, then switches to receive mode, and (ii) two nodes that transmit concurrently for 8 times after receiving and synchronizing on the first packet.
This simple strategy gives a synchronization error of 0 to \unit[0.25]{\textmu{s}} during the 8 packets round as measured on the initiator node.

We run the experiments on the different Bluetooth~5 modes, sending at least 4000 packets. 
We confirmed that 4000 packets give a representative measurement by running some of the experiments for a longer time and collecting 30000 packets.

\fakeparagraph{Objective}
We compare the PER of different pairs of concurrent transmitters, which have different CFOs, see \rtab{eval_cfo}.
This in turn results in various beating patterns (slow vs. fast)
, as shown in \rsec{feasibility:carrier_beating_eval}.
The goal is to measure how the different beating patterns affect PER, as predicted by our analysis in \rsec{feasibility:model:beating_prr_analysis}.
Also, we evaluate the performance of the coded versus the uncoded modes under beating.

    \begin{table}[t]
    \centering%
    \caption{Comparing PER of concurrent transmissions with different CFO and Bluetooth modes. \review{The ratio $T_{packet}/T_{beat}$ represents the average number of beats a single packet endures.} We notice that (i) slow beating ($T_{packet}/T_{beat}\leqslant 1$) gives a better performance in uncoded modes, and (ii) the coded modes help recovering the packets under beating. [experiment]\ltab{eval_prr}}%
    \footnotesize%
    \begin{widetable}{1\linewidth}{lc|cc|cc|cc}%
    \toprule%
           & \bf Tx Pairs          & \multicolumn{2}{c|}{\bf{(1, 4)}}   & \multicolumn{2}{c|}{\bf{(1, 2)}} & \multicolumn{2}{c}{\bf{(3, 4)}}\\%
           & CFO (\unit{KHz}) & \multicolumn{2}{c|}{\bf{9.645}}    & \multicolumn{2}{c|}{\bf{1.83}} & \multicolumn{2}{c}{\bf{0.025}}\\%
          & $T_{beat}$ (\unit{ms})  & \multicolumn{2}{c|}{\bf{0.10} (fast beating)}     & \multicolumn{2}{c|}{\bf{0.55} (slower beating)} & \multicolumn{2}{c}{\bf{40.00} (slow beating)}\\%
    \midrule%
    \textbf{Mode} (\textnormal\unit{Mbps}) & $T_{packet}$ (\unit{ms})    & $\frac{T_{packet}}{T_{beat}}$ & \bf PER & $\frac{T_{packet}}{T_{beat}}$ & \bf PER & $\frac{T_{packet}}{T_{beat}}$ & \bf PER \\%
    \midrule%
    uncoded  2 & 0.18 & 1.8 & 73.56\% & 0.33 & 38.81\% & 0.0045 & 24.86\%\\
    uncoded   1 & 0.36 & 3.6 & 93.96\% & 0.65 & 76.88\% & 0.0090 & 8.66\%\\
    coded   0.5 & 0.958 & 9.58 & 16.57\% & 1.74 & 47.45\% & 0.0240 & 6.79\%\\
    coded   0.125 & 2.944 & 29.44 & 0.86\% & 5.35 & 9.84\% & 0.0736 & 3.87\%\\
    \bottomrule%
    \end{widetable}%

    \end{table}%
    
\fakeparagraph{Results}
\rtab{eval_prr} lists the PER for the scenarios.
First, we note that in the case of a single transmitter, all the packets shall be received due to the absence of significant external noise.
However, concurrent transmissions never achieve perfect reliability in our settings, as it is a worst-case scenario: both transmitters are received with similar energy, which results in a deep fading. 
Thus, we can safely conclude that synchronous concurrent transmissions are not \emph{constructive}, but, even in the worst case, they are not completely \emph{destructive} either.

We summarize the results: 
(i) slow beating resulting from low CFO gives better performance in uncoded modes, as our simulation results predict in \rsec{feasibility:model:beating_prr_analysis}, and
(ii) the coded modes help recovering the packets under all beating conditions. They are particularly effective under fast beating conditions since they potentially have enough diversity and correct bits to perform the FEC. Note that the coded \unit[125]{Kbps} mode performs consistently with over 90\% reliability.
(iii) The uncoded \unit[2]{Mbps} mode performs generally better than the legacy mode \unit[1]{Mbps} due to the shorter packet duration, which cause it to experience a slower relative beating. It is, however, more sensitive to synchronization errors, and
(iv) the concurrent transmissions link quality (reflected by the PER) depends heavily on the ratio of the beating carrier to the packet period ($T_{beat}/T_{packet}$) in both uncoded modes.

\begin{figure}[tb]%
  \centering
  \small

  \begin{subfigure}[t]{0.47\textwidth}
    \centering\includegraphics[width=1\textwidth]{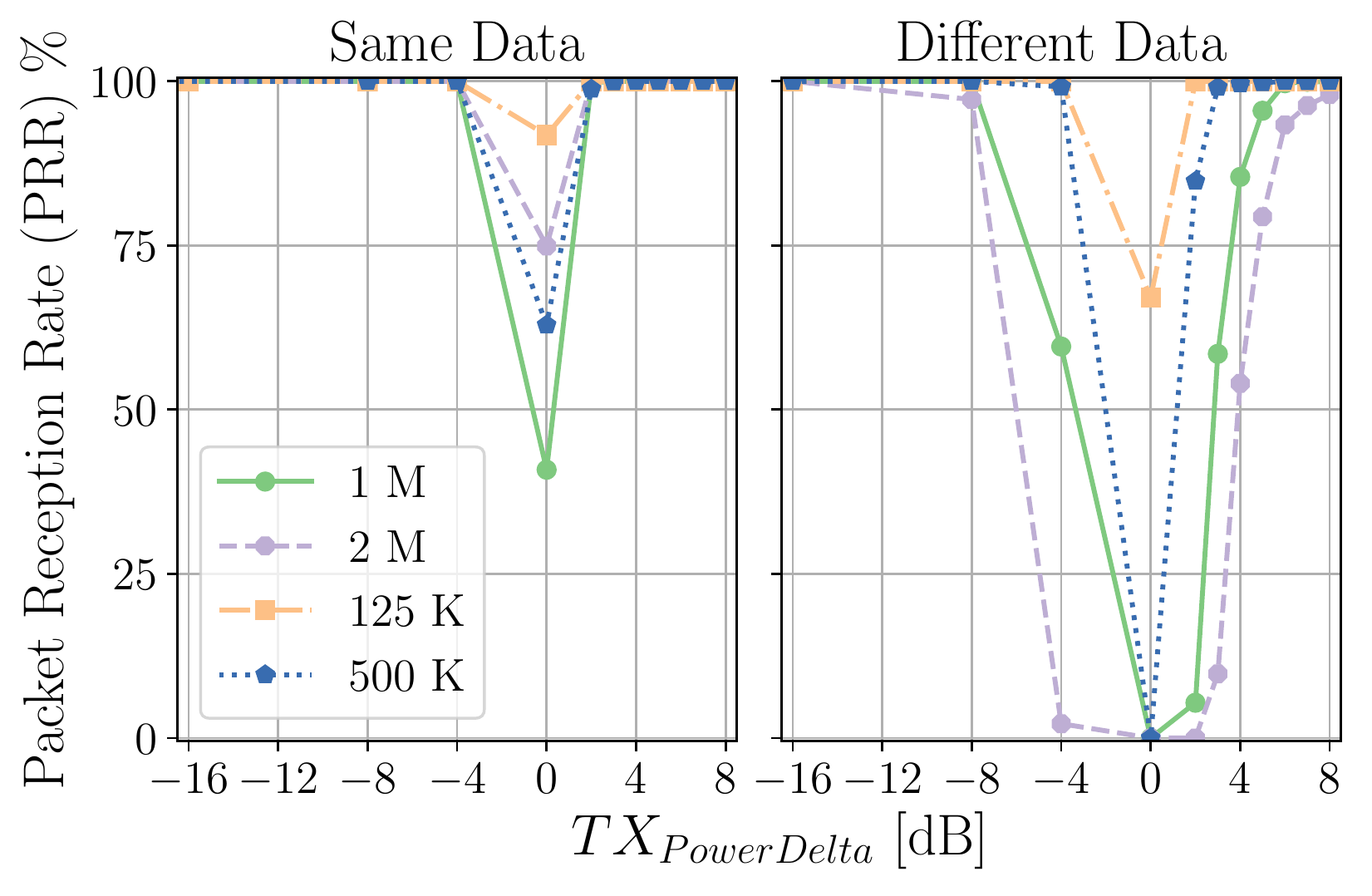}
	\caption{\lfig{ble_ct_0}TX power delta: \capt{CT of the same data is feasible even when the two signals have the same signal strength. However, the capture of different data suffers greatly when power delta is less than 8~dB.}}
  \end{subfigure}%
  \hfill
  \begin{subfigure}[t]{0.47\textwidth}
    \centering\includegraphics[width=1\textwidth]{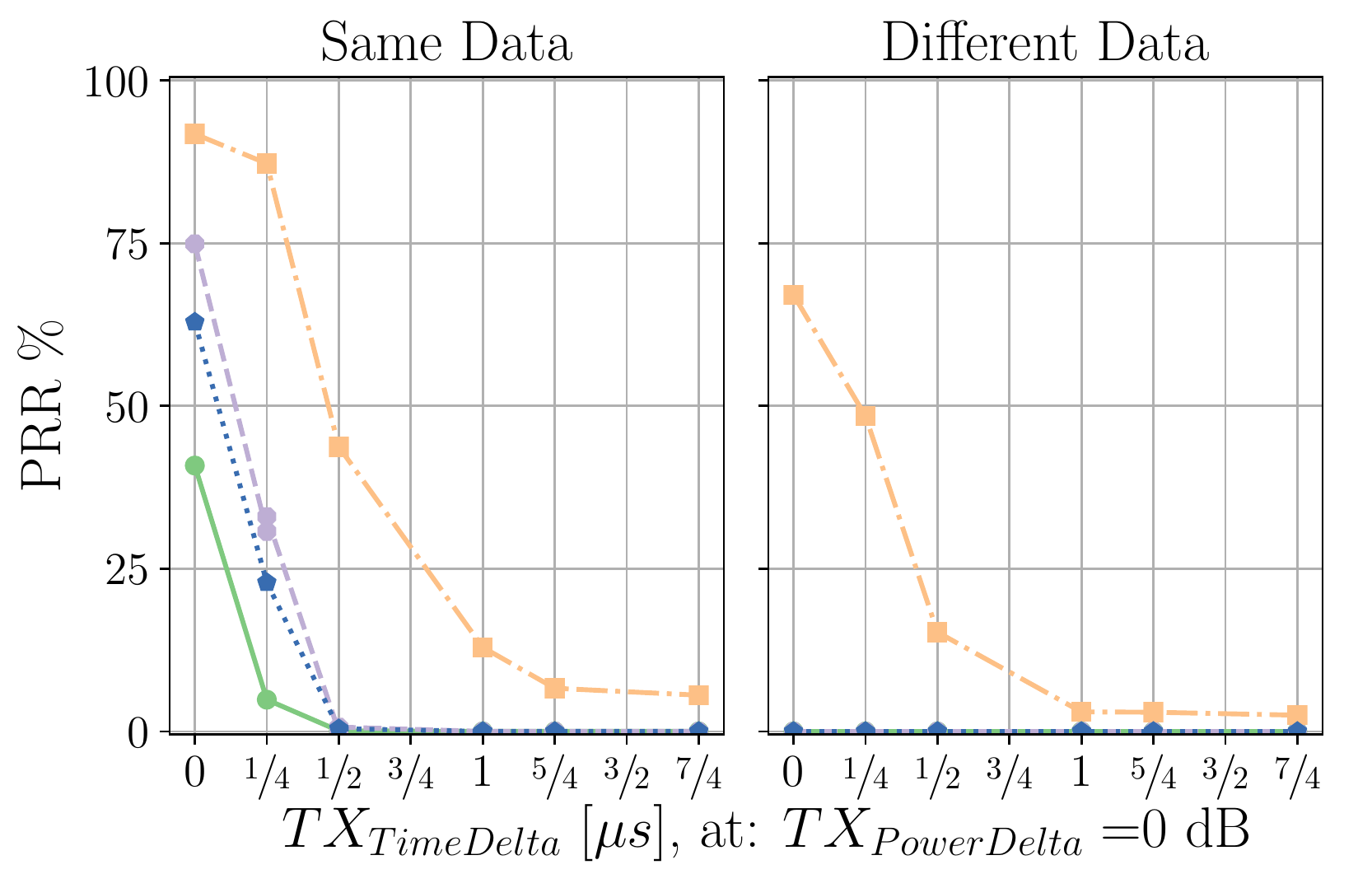}
	\caption{\lfig{ble_ct_td_p0}TX time delta without power delta: \capt{CT performance drops significantly \review{when one of the concurrent transmissions is delayed}.}}
  \end{subfigure}\\%
  \vspace{1em}
  \begin{subfigure}[t]{0.32\textwidth}
    \centering\includegraphics[width=1\textwidth]{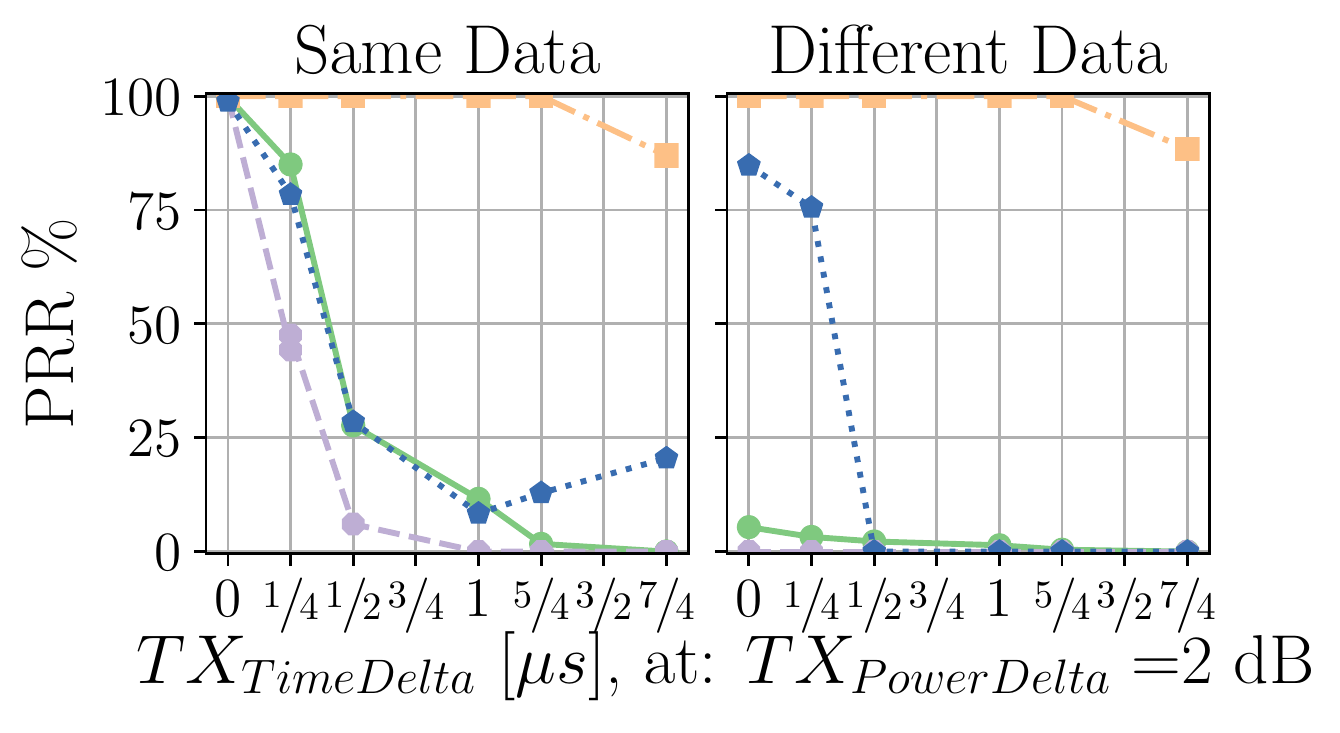}
	\caption{\lfig{ble_ct_td_p2}TX time delta with 2~dB power delta: \capt{CT performance drops significantly \review{when one of the concurrent transmissions is delayed}.}}
  \end{subfigure}
  \hfill
  \begin{subfigure}[t]{0.32\textwidth}
    \centering\includegraphics[width=1\textwidth]{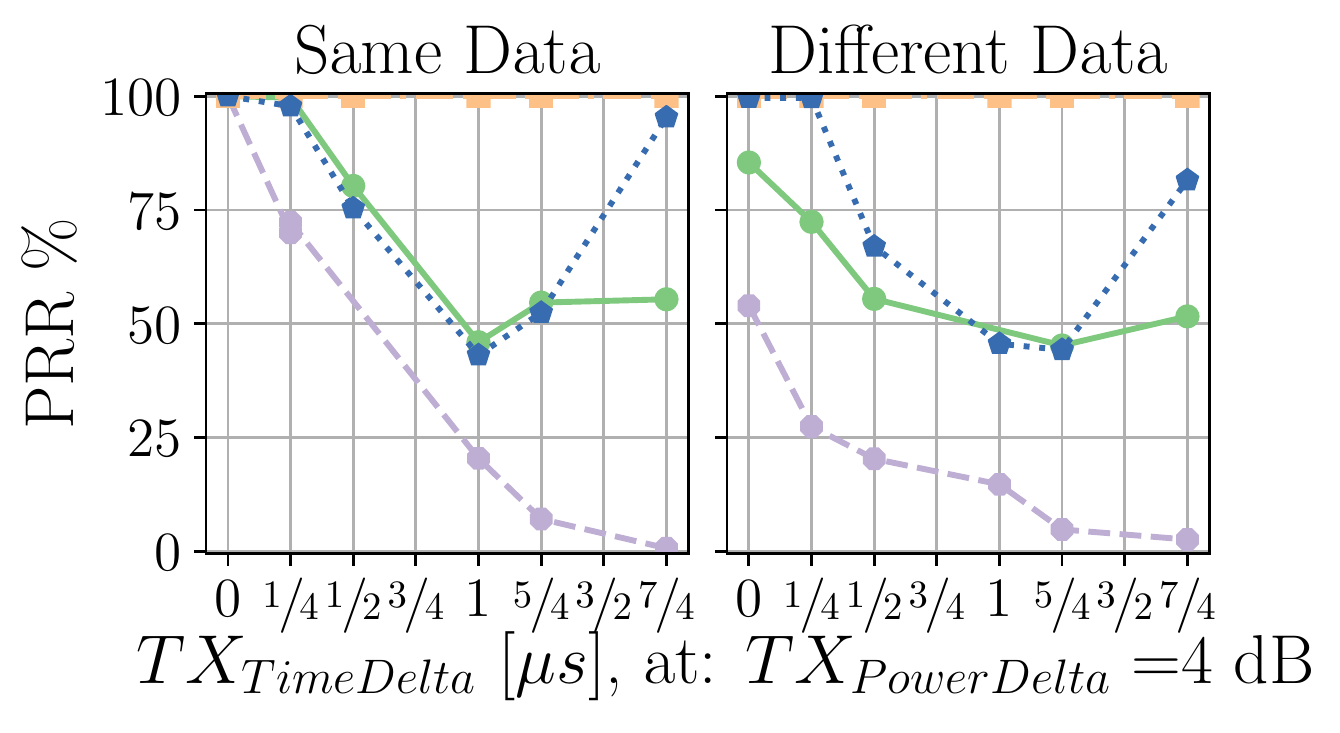}
	\caption{\lfig{ble_ct_td_p4}TX time delta with 4~dB power delta: \capt{CT performance drops significantly when one transmitter is delayed, then recovers when the delay crosses the symbol boundaries ($1\mu s$ for \unit[1]{Mbps}) due to power-capture.}}
  \end{subfigure}
    \hfill
    \begin{subfigure}[t]{0.32\textwidth}
    \centering\includegraphics[width=1\textwidth]{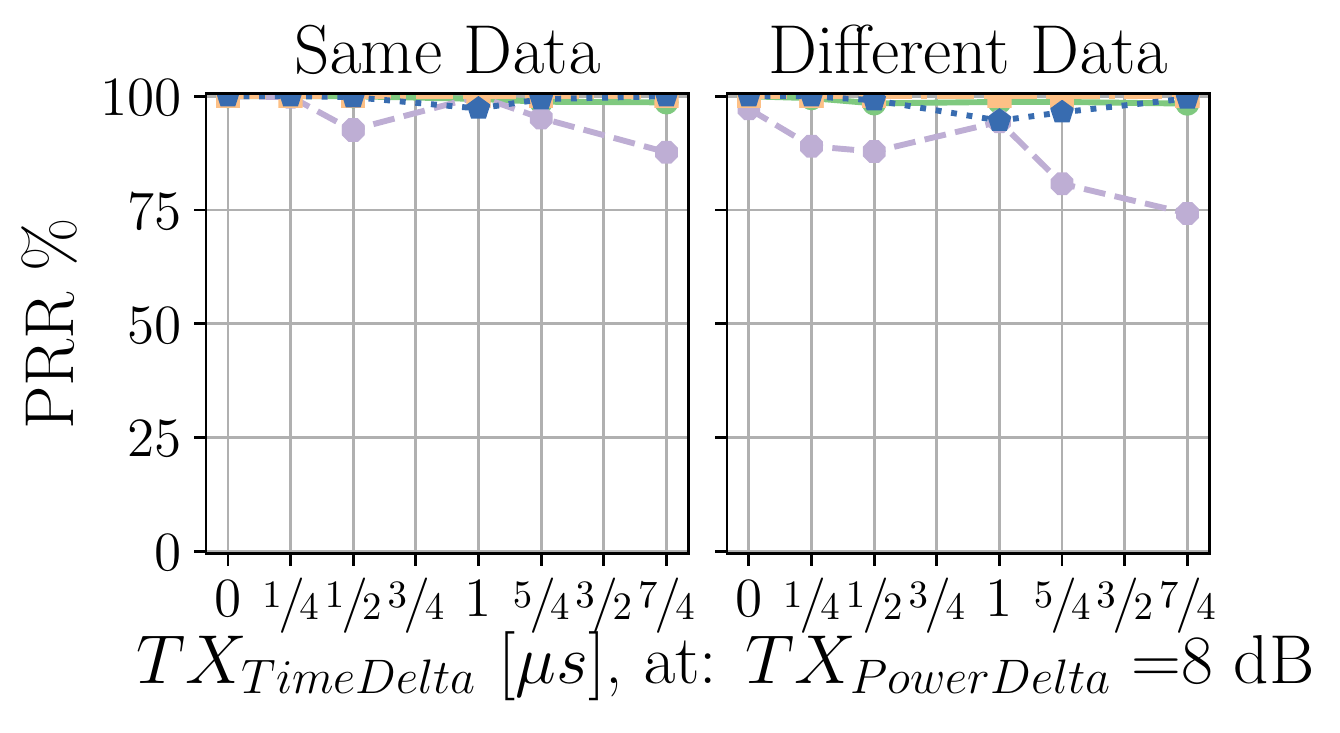}
	\caption{\lfig{ble_ct_td_p8}TX time delta with $8$~dB power delta: \capt{CT performance is penalized lightly when time delta is introduced between the transmissions, as it is operating in the power-capture zone.}}
  \end{subfigure}
\caption[Micro-evaluation of CT over Bluetooth]{Micro-evaluation of CT over Bluetooth PHY: \capt{effect of power delta and time delta when transmitting identical or independent payloads.} [experiment]}
\lfig{ble_ct}
\end{figure}
\subsubsection{CT Performance vs. TX Power Delta}
\lsec{feasibility:txPower}
We fix the transmission power of one CT node to 0~dB and vary the transmission power of the second to sweep all the factory calibrated TX power settings: [-40, -20, -16, -8, -4, 0, 2, 3, 4, 5, 6, 7, 8]~dB.
We cross check at the receiver (initiator) and confirm that the received signals have a matching power delta as the configuration.
We repeat this experiment on the four modes of Bluetooth 5; namely, \unit[2]{Mbps}, \unit[1]{Mbps}, \unit[500]{Kbps} and \unit[125]{Kbps}.

\rfig{ble_ct_0} shows the results of the experiments, which summarize in the following takeaways:
\begin{enumerate}[(i)]
    \item CT of the same data is feasible over all the Bluetooth 5 modes regardless of the power delta.
    \item While the long range mode \unit[125]{Kbps} with FEC 1:8 has the best performance, the other modes perform well once there is a difference of at least \unit[2]{dB} in the CT signal strength.
    \item In the case of different data, capture is only practical when there is a power delta of roughly 6~dB or more.
\end{enumerate}
For these reasons, we base our design on CT of the same data.
It should be noted, however, that the performance of concurrent transmissions over Bluetooth PHY is considerably weaker than over 802.15.4 (as reported in \eg~\cite{ferrari11glossy}) and it is greatly affected by the setting.
Nevertheless, we show in this paper that we can utilize it to build efficient end-to-end flooding.

\subsubsection{CT Performance vs. TX Time Delta}
\lsec{feasibility:txTimeDelta}
We inject a constant delay in the transmission time of one CT node, and vary it to [0, 4, ... 28] clock ticks; \ie [0, 0.25, ... 1.75] $\mu s$, and fix the transmission power of both nodes. 
Note that one clock tick is $1/16=0.0625~\mu s$, and the symbol period of the \unit[1]{Mbps} PHY is equal to $1~\mu s$, which is the same for the Bluetooth modes (\unit[1]{Mbps}, \unit[500]{Kbps}, \unit[125]{Kbps}), while the \unit[2]{Mbps} mode has a symbol period of $0.5~\mu s$.
We repeat the test with different TX power deltas (as we did in the previous section: fix one node to TX power of 0~dB and change the other) to study the combined effect of signal power and transmission delay.

\rfig{ble_ct_td_p0}--~\rfig{ble_ct_td_p8} show the results of the experiments.
We distinguish the following phenomena:
\begin{itemize}
    \item \emph{Destructive interference: 0~--~2~dB}. The first take away of this experiment is that the performance of CT of the same data drops significantly with TX time delta when we operate in the 0~--~2~dB power delta zone as Figures~\emph{\rfig{ble_ct_td_p0}~--~\rfig{ble_ct_td_p2}} show.
    The reason is that the two signals interfere destructively when the symbols are misaligned.
    
    \item \emph{Power capture for the coded \unit[125]{Kbps} mode: 0~--~2~dB}. We notice in the case of 0~dB tx power and different data that only the high fidelity \unit[125]{Kbps} mode survives up to time delta of 8 ticks, which equals half of PHY symbol.
    On the other hand, having as little as 2~dB power delta makes the time delta effect on performance insignificant for the \unit[125]{Kbps} mode.
    
    \item \emph{Slightly destructive: 4~dB and half a symbol delay}.
    We notice that the CT performance drops with the time delta up to $0.5 \mu s$ (half a symbol for \unit[500]{Kbps} and \unit[1]{Mbps}). It partially starts recovering after crossing the symbol boundary.
    The \unit[2]{Mbps} mode exhibits a performance drop similar to \unit[1]{Mbps} ($\approx 60\%$) but at $0.25 \mu s$ (half a symbol at \unit[2]{Mbps}), and does not recover, as \rfig{ble_ct_td_p4} shows.
    
    \item \emph{Power capture at, for example, 8~dB}.
    We notice that the time delta effect on CT performance is almost negligible except for the \unit[2]{Mbps} mode where we see a drop of PRR to 80~--~90\%.
    Thus, we conclude that at this power difference we mainly have capture. 
\end{itemize}

\subsection{Discussion and Practical Implications}
\lsec{feasibility:conclusion}
\majorreview{\fakeparagraph{Practical Implications}
Both our numerical and experimental results reveal key trends of CT behaviour over Bluetooth that give guidelines for designing a mesh protocol that utilizes CT over Bluetooth:
(i) CT does not cause constructive interference in commercial transceivers, but it is not totally destructive either;
(ii) the performance of CT depends on several factors with varying degrees of difficulty to control:
a. time synchronization accuracy: it is feasible to have clock-tick synchronization accuracy, see \cite{lcn17brachmann}; 
b. signal power delta seen at the receiver: this is more difficult to control, as it depends on the relative location of the receiver and the environment, and 
c. carrier frequency offset: this requires custom hardware; 
(iii) CT of the same data is feasible for all Bluetooth~5 modes: 
a. it can tolerate a time delta of a couple of MCU clock ticks, \ie in the order of half a microsecond, and 
b. it achieves a relatively high reception rate with as little as $2~dB$ power delta. 
We note that such a small power delta is common in practical indoor deployments, as it merely represents a few centimeters distance between two nodes \cite{rssindoor12} and even links quality fluctuate over time at higher power levels \cite{watteyne12challenges}.
(iv) CT of different data needs a relatively high ($6$ to $8~dB$) signal power difference to be practical, which makes it more difficult to use over Bluetooth, except for the \unit[125]{Kbps} coded mode.
}
\majorreview{\fakeparagraph{Model vs. Reality Gap} 
We would like to highlight that} the results of our experimental study and analytic models discussed in \rsec{feasibility:possibilitiesAndChallenges} differ slightly:
For example, our results indicate Bluetooth CT cannot tolerate more than $\tau/4 = 0.25~\mu s$ time delta as opposed to the expected $\tau/2 = 0.5~\mu s$ in the case of $0~dB$ power delta.
Moreover, we see that CT of different data is successful with \majorreview{$6$~to $8~dB$} Tx power delta as opposed to the expected \majorreview{$8~dB$} for uncoded modes (1~--~\unit[2]{Mbps}).
\review{Resolving these mismatches requires a more detailed model of CT over Bluetooth, which is beyond the scope of this paper.
Rather, our focus in this paper is to show that CT is a viable design choice.%
}
\majorreview{\fakeparagraph{Conclusion}} Our main conclusion from the analysis and experiments is that CT of the same data over Bluetooth is feasible for all Bluetooth 5 modes.
Therefore, we focus on data dissemination in this paper, and utilize CT of the same data to build \name: a reliable end-to-end flooding protocol, as we show next.

\section{\name}
\lsec{Design}

\begin{figure}[tb]
\begin{subfigure}[t]{0.49\textwidth}
	\centering
		\includegraphics[width=1\textwidth]{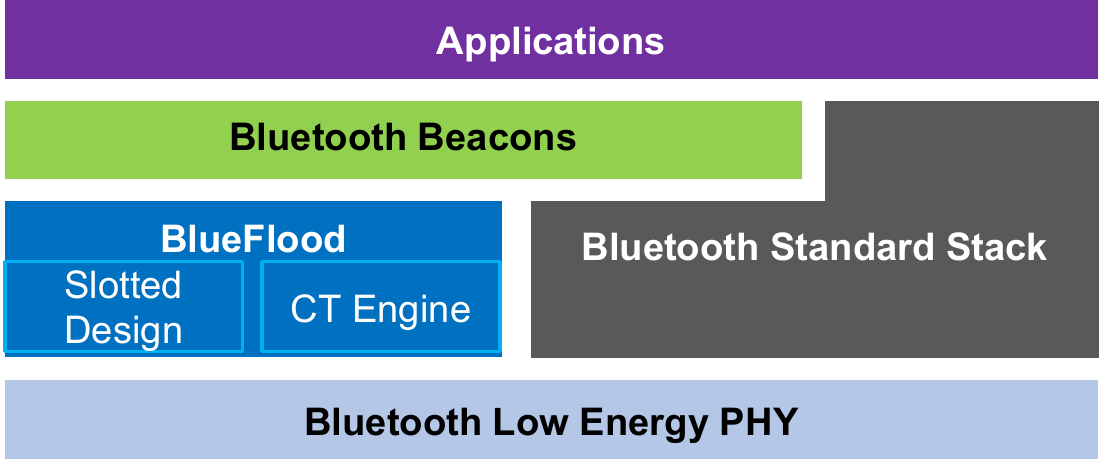}
		\caption[System Architecture of \name]{System architecture: \capt{\name operates over the Bluetooth PHY, but it is transparent to the application which interfaces with standard Bluetooth beacons.}}
		\lfig{arch}
	\end{subfigure}
\hfill
	\begin{subfigure}[t]{0.49\textwidth}
	\centering
		\includegraphics[width=0.95\columnwidth]{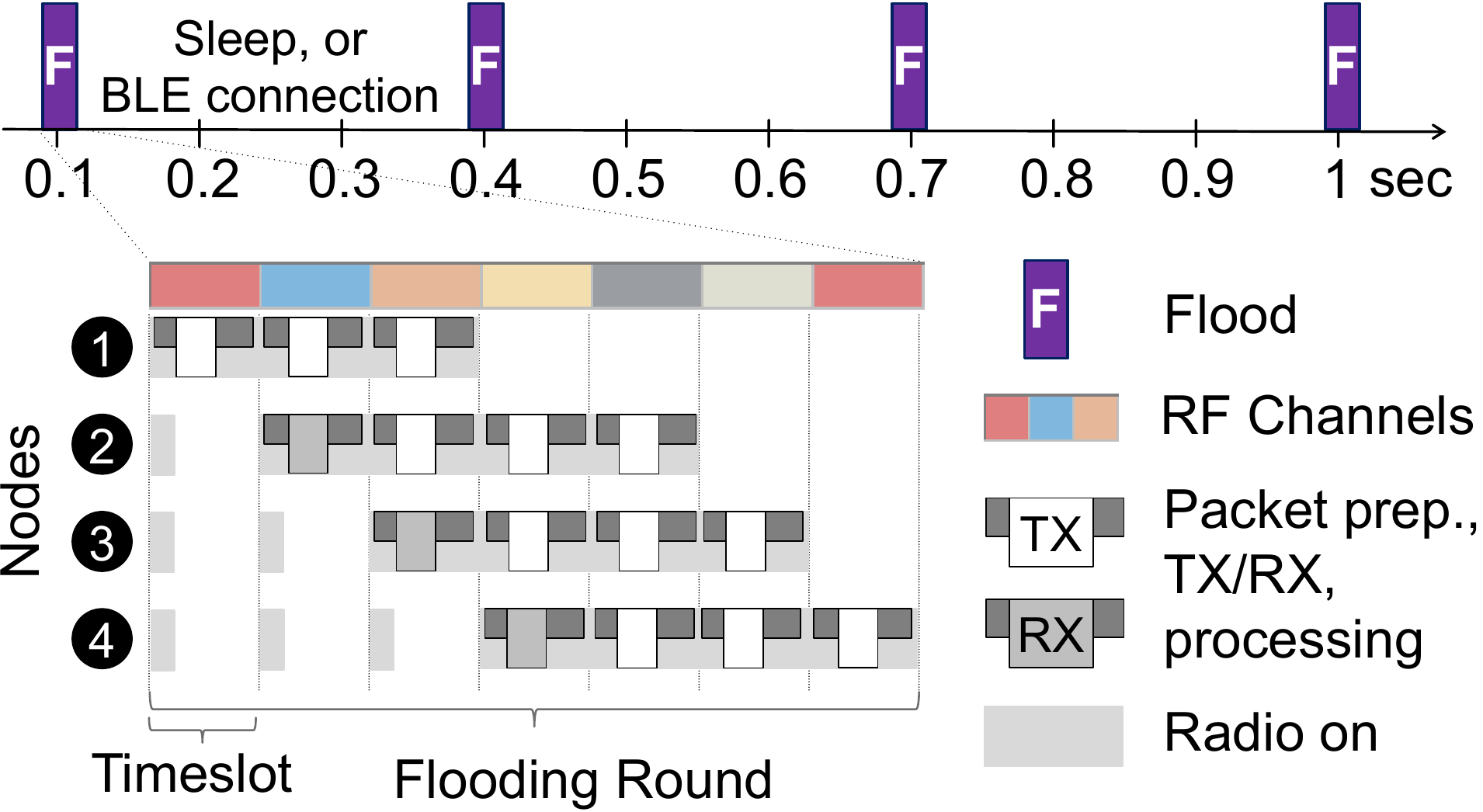}
		\caption[Overview of \name Operation]{Overview of \name operation: \capt{synchronous flooding that utilizes CT with RX / 3~TX transmission policy. Every timeslot accommodates one packet transmission or reception, handling, and channel hopping.}}
		\lfig{bleglossy}
	\end{subfigure}
	\caption{\name System Architecture and Operation Details.}
\end{figure}

In this section, we motivate the use of concurrent flooding and how we tackle its challenges for Bluetooth, then we introduce the design of \name.

\fakeparagraph{Motivation}
We seek to design a low-power protocol for multi-hop data dissemination that can be received with unmodified devices.
Thus, in \name, a backbone of \name-enabled devices floods Bluetooth-compliant advertisements through concurrent transmissions, which are received by off-the-shelf Bluetooth devices. 
Based on our insights from the feasibility study of concurrent transmissions in Bluetooth, see \rsec{feasibility}, we adopt a design that borrows from Glossy and related protocols.

\fakeparagraph{Challenges and \majorreview{Opportunities}}
As explained in \rsec{feasibility:possibilitiesAndChallenges}, concurrent transmissions are challenging over Bluetooth.
Mainly, (i) concurrent transmissions need to be synchronized down to \unit[250]{ns}, and (ii) the CT links are fragile in case of a near zero power delta.
However, since the link quality stays above 30\% in the worst case in \rfig{ble_ct_0}, we argue that CT stays a viable strategy.
Besides, the link quality improves drastically once the CT is composed of signals \majorreview{with a power difference as little as \unit[1]{dB}.
We benefit from the nature of RF propagation in real deployments where the signals undergo reflections and multi-path propagation and so, in practice, two signals rarely have a power difference of less than \unit[1]{dB}. 
Literature in WSN and localization notes that \cite{rssindoor12, watteyne12challenges}: 
(i) RSS and channel reliability vary over time and frequency in the order of several dB, and (ii)~even little differences in distance to a receiver cause a few decibels difference in RSS.
Note that a channel-hopping slotted protocol benefits from these phenomena and} the frequency diversity over 40~channels in Bluetooth helps surviving external interference.
Moreover, the different Bluetooth~5 modes give an interesting reliability--energy trade-off and widen the design space.
Plus, the modern SoCs simplify the realization of the required tight synchronization, as we show later. 

\fakeparagraph{Overview}
\lsec{Design:overview}
We build \name, a synchronous flooding protocol that utilizes CT of the same data, as depicted in \rfig{bleglossy}.
We 
design our protocol to be a round-based and time-slotted protocol.
Thus,
we schedule individual communication rounds on a network-wide scale. 
A round is further split into time slots in which nodes either transmit, listen or sleep, according to a transmission policy. 

From a system integration perspective, we design \name to be transparent to the application.
In our example, applications interact with a standard Bluetooth beacon library without having to know about the existence of \name, see \rfig{arch}.
As a result, \name distributes Bluetooth beacons on network scale instead of the traditional one-hop announcements, enabling the application scenarios discussed in \rsec{Introduction}.

Next, we discuss the logical components of \name: time-slotted design, synchronization, transmission policy and frequency agility.
Later, we discuss the design simplifications on modern SoCs.

\subsection{Design Elements}
\lsec{Design:Components}
In this section we discuss the design elements of \name.
We take inspirations from Glossy, $A^2$ and the winners of the EWSN dependability competition 2016, 2017 and 2018 \cite{ewsn17Comp,Lim17Comp,bigbangbus18comp}.

\subsubsection{Time-Slotted Design}
Each slot fits one packet transmission or reception, and processing.
Within each slot, a node transmits, receives or sleeps according to the selected transmission policy. 
The default transmission policy in \name is to concurrently transmit a packet $N$ times, \ie in the $N$ slots following the reception of a packet, before completing the round and entering a deep-sleep mode until the beginning of the next round, see~\rfig{bleglossy}. 

\fakeparagraph{Power-saving}
To save power, during each slot, a node turns the radio off as soon as the transmission or reception has ended or in case it fails to detect a valid packet at the beginning of the slot, \ie once the guard time expired.
The combination of CT with these simple power-saving techniques allows \name to provide a backbone of energy-efficient flooding devices.
This is in contrast to Bluetooth Mesh, which restricts the forwarding task to mains-powered devices.

\subsubsection{Frequency Agility}
Glossy and related CT approaches see their performance degrade in presence of interference~\cite{ferrari12lwb, he16arpeggio}.
We address this by employing Bluetooth frequency agility over the 40 available channels.
Thus, in \name, nodes switch to a new channel to transmit or receive in each timeslot following a network-wide schedule.
The round and slot numbers are used to index this hopping sequence. 
Once the node is synchronized, it has the same view of the slot and round numbers as the rest of the network; thus, it does not need to start each round on the same channel.
This is similar to the channel-hopping of TSCH~\cite{stdieee802154-2015} and Bluetooth~\cite{bluetooth5}. 
It has proven its robustness even under strong interference in the EWSN dependability competitions~\cite{Lim17Comp, Escobar17Comp, Nahas17Comp}.

\subsubsection{Transmission Policy}
\majorreview{Each round has a single initiator and in the beginning of a round, other nodes wake up aiming to receive, as in Glossy.}
Since we only require the reception of one valid packet per round to keep the synchronization, we utilize a transmission policy that follows the pattern: one valid Rx, then N consecutive Tx; \ie we wait for the first valid packet then transmit N times in a row.
This has a lower overhead of $N+1$ slots instead of $2 \times N$ for the original Glossy transmission policy (N times Rx--Tx).
\name eliminates the need to listen to repeated packets. 
Thus, it needs half the slots plus one to do N transmissions.

\subsubsection{Synchronization}
A key requirement is to keep the nodes tightly synchronized for a complete round within the bounds of 250~ns to successfully achieve CT.
\fakeparagraph{Round synchronization}
We require each node to receive a single valid packet during each round, which we then use for the per-round synchronization based on the radio-registered timestamp.
\majorreview{A node waits for a constant number of slots each round until a successful reception before joining the round or going back to sleep.
This threshold can be a function of the diameter of the network and the number of transmissions $N$.
For example, $Wait_{slots} = N + 2 \times TestbedDiameter$.}

\fakeparagraph{Scanning for Networks}
When a node wants to join the network, it listens on one frequency for $2 \times C$ periods, where $C$ is the number of channels.
Until it receives a valid packet, it hops to a random channel and repeats. 
Upon receiving a valid packet, it uses the slot number to synchronize to the beginning of the round.
\fakeparagraph{Re-synchronization} 
If a node does not receive a packet for a configurable number of rounds consecutively, it assumes it lost synchronization.
Subsequently, it switches to the scanning mode.

\fakeparagraph{Power Budget}
\lsec{design:txpolicypower}
With the aforementioned transmission policy, in each slot, a node stays on for receive guard time ($Rx_{GuardTime}$) waiting to receive a valid packet. 
Once it receives a valid packet, it transmits it N times. 
This strategy gives an average power budget $P_{Avg}$ as a function of Tx and Rx power $P_{Tx}, P_{Rx}$, and average radio time $R_{Avg}$ per node per successful round of:
\begin{equation}
\label{equ:avgPower}
P_{Avg}=(Avg_{HopCount} \times Rx_{GuardTime} + AirTime) \times P_{Rx} + N \times AirTime \times P_{Tx}
\end{equation}
\begin{equation}
\label{equ:avgRadioTime}
R_{Avg}=(Avg_{HopCount} \times Rx_{GuardTime}) + (N+1) \times AirTime
\end{equation}
\majorreview{We can estimate the radio duty-cycle $D_C$, taking into account failed rounds, with probability of failure $PER$ and the conservative assumption of receiving full-length corrupted packets in each slot:
\begin{equation}
\label{equ:pFail}
D_C=(R_{Avg} \times (1-PER) + Wait_{slots} \times AirTime \times PER) / RoundPeriod
\end{equation}}
\fakeparagraph{Bluetooth Modes Trade-off}
The fastest mode \unit[2]{Mbps} has the shortest radio air-time. 
Thus, it has the lowest energy budget, but a lower reliability and shorter range than the coded \unit[125]{Kbps} mode which has up to $2-4\times$ longer range in comparison.
In the same time, the coded \unit[125]{Kbps} mode has 1:8 FEC, which means $8-16\times$ longer air-time and higher energy budget than the \unit[1]{Mbps} and \unit[2]{Mbps} modes, respectively.
In other words, N transmissions in the \unit[125]{Kbps} mode cost as much as $16\times N$ transmissions in the \unit[2]{Mbps} mode.
In our evaluation in \rsec{evaluation}, we show how the different transmission modes impact the reliability and \review{latency of multi-hop dissemination using CT}. 

\subsubsection{Bluetooth Compatibility and Packet Structure}
To keep receive compatibility with off-the-shelf devices; \eg smartphones, we utilize the standard Bluetooth beacons; \eg non-connectable undirected advertisements, to flood events.
In particular, we use iBeacons (see \rfig{bg:blepacket}) and override the major and minor numbers to designate the round and slot numbers, respectively.
\review{We note that Bluetooth~5 supports longer advertising packets with payloads up to \unit[255]{bytes}. 
This makes it easier to support a wider range of applications. 
However, it is interesting to explore the performance implications of packet size in Bluetooth under CT. 
We expect to get a higher loss rate with longer packets due to the higher probability of colliding with an external interfering signal and due to the increased timing jitters with longer slot sizes. 
We highlight this in our evaluation later in~\rsec{evaluation:packetsize}.}

\subsection{Simplified Design on Modern SoCs}
Modern SoCs commonly integrate MCU and radio and provide a memory-mapped packet buffer.
Moreover, some provide configurable triggering of peripherals based on HW events to eliminate SW delays of processing interrupts.
For example, on the nRF51 and 52 series, it is possible to control the radio by scheduling a HW timer that \review{directly triggers a radio operation, \eg transmit, receive or turn off, at a specific moment, without MCU interaction or further code execution.}
In the same time, it achieves both timely and synchronous HW events by timing the radio and the peripherals with a 16~MHz clock derived from the common high resolution 64~MHz CPU clock~\cite{nrf52840}.

In \name, we utilize both the direct wiring of events and the high resolution clock to strongly simplify our design and implementation when compared to Glossy. 
Practically, it allows us to avoid many of the SW complexities the original design of Glossy deals with to achieve the tight timing requirements on older-generation systems such as TelosB motes.
For example, due to these limitations of the platform, the implementation of Glossy: (i) relies on a radio-driven execution model, (ii) builds on a complex management of execution timing to minimize the packet transfer delay between the radio and the MCU, and (iii) relies on a Virtual High-resolution Timer (VHT)~\cite{vht} for synchronization. 
In our experience, this makes Glossy and protocols building on Glossy such as, for example, LWB~\cite{ferrari12lwb}, Chaos~\cite{201311LandsiedelChaos}, and Crystal~\cite{Crystal} hard to manage and difficult to port to new platforms. 
We note that Glossy was later ported to several SoC platforms such as the CC2538~\cite{lwbSocCC2538} and the subGHz CC430 SoC~\cite{lwbSocSubGhz}.
To our best knowledge, the synchronous transmission kernel of these ports stays complex due to the lack of the ability to wire hardware events on these platforms.

\section{\name Evaluation}
\lsec{evaluation}

In this section we describe our implementation briefly and evaluate \name performance in a multihop mesh scenario.

\subsection{Evaluation Setup}
\lsec{evaluation:setup}

We present our \name implementation, the scenario, the metrics and the evaluation testbeds.

\fakeparagraph{Implementation} 
We implement \name in C for the Contiki OS~\cite{dunkels04contiki} targeting Nordic Semiconductor nRF nodes equipped with an on-SoC Bluetooth radio.
\rtab{eval:nodes} lists supported platforms and their specifications.
\majorreview{
Please note that we optimized the code base we use in this evaluation when compared to the original publication \cite{alnahas2019blueflood}. 
Most notably, we shorten the slot sizes by about 50\% , improve synchronization, and fix bugs that led to occasional packet losses.
As a consequence, we improve the performance and the experimental results are not directly comparable. 
}

\fakeparagraph{Scenario}
The evaluation scenario is a connection-less multihop dissemination.
We use standard Bluetooth channels; as a result, we run \name with co-existing Bluetooth and WiFi traffic.
For the single channel experiments, we use the Bluetooth advertising channel~37.
Unless otherwise mentioned, dissemination rounds repeat at a 0.2~s period.
We run each experiment until we get more than 3000 rounds.

\fakeparagraph{Configuration}
Depending on the Bluetooth mode, the slot size varies between 0.4~and 3.7~ms.
\majorreview{
We use a guard time, \ie the time we wake up to receive before we actually expect a transmission, of 0.032~ms, independent of the Bluetooth mode. 
}

\fakeparagraph{Transmission Policy}
\majorreview{We use a constant number of transmission $N_{Tx}=3$, but with a custom policy for the initiator.
The initiator transmits for $N_{Tx} + 2 \times TestbedDiameter$, while other nodes transmit for $N_{Tx}$ after a successful reception, then go back to sleep.
A node waits for a valid packet for at most $N_{Tx} + 2 \times TestbedDiameter$ slots before going to sleep.}

\fakeparagraph{Goals} We evaluate \name performance on two testbeds (described next) and test reception on a smartphone. %
Moreover, we evaluate how the different parameters affect the performance.
Namely, we look at the effects of different transmission powers, numbers of retransmissions and packet sizes.

\fakeparagraph{Metrics}
We focus on the following performance metrics:
\review{
\begin{itemize}
\item {\em End-to-End Packet Error Rate (PER)}: is the ratio of failed deliveries, which gives an indication of the protocol's reliability. 
We consider a round failed when one or more nodes do not receive the disseminated value;
\item {\em Hop count}: is the average number of time slots until each node receives the disseminated value. 
It is affected by the testbed deployment geography, interference (internal from CT and external from coexisting networks), and the protocol transmission strategy;
\item {\em Latency}: is the average duration (of a round) until each node receives the disseminated value. It is a function of hop count: $Latency = hopCount \times slotSize$; and
\item {\em Active slots}: is the total number of slots the protocol is active during a round.
It serves as a proxy for the maximum energy consumed during a round.
\end{itemize}
}
\begin{table}[tb]
	\centering
	\caption[\name Slot-length to Send a Single iBeacon Frame]{\name slot length needed to send a single iBeacon (38 bytes) for the different Bluetooth modes.
	\capt{Air time: is the air time for the packet and represents the relative power budget for each mode. The radio slot is longer than the air time as we need to setup the radio and to compensate for the various SW delays.}
	\ltab{eval:slotlength}}
	\footnotesize%
	\begin{widetable}{\linewidth}{lcccccc}
	\toprule
	\emph{Radio} & \emph{Mode and Bitrate} & \emph{PHY symbols} & \emph{Air time} [ms]& \emph{Radio slot} [ms]& \emph{Guard} [ms]& \emph{Slot } [ms]\\
	\cmidrule(r){1-1}
	\cmidrule(lr){2-2}
	\cmidrule(lr){3-3}
	\cmidrule(lr){3-3}
	\cmidrule(lr){4-4}
	\cmidrule(lr){5-5}
	\cmidrule(lr){6-6}
	\cmidrule(l){7-7}
	Bluetooth & uncoded \unit[2]{Mbps} & 376 & 0.188 & 0.358 & 0.032 & 0.4016 \\
	Bluetooth & uncoded \unit[1]{Mbps} & 368 & 0.368 & 0.522 & 0.032 & 0.5705 \\
	Bluetooth & coded \unit[500]{Kbps} & 1134 & 1.134 & 1.25 & 0.032 & 1.362 \\
	Bluetooth & coded \unit[125]{Kbps} & 3408 & 3.408 & 3.603 & 0.032 & 3.715 \\
	IEEE 802.15.4 & \unit[250]{Kbps} & 90  & 1.440 & 1.768 & 0.032 & 1.867 \\
	\bottomrule
	\end{widetable}
	\end{table}

\begin{figure}
\centering
\begin{minipage}{0.465\textwidth}
\centering
\captionsetup{type=table} %
\caption{\ltab{eval:nodes}Supported platforms}
\footnotesize%
\begin{widetable}{\linewidth}{lcccc}
\toprule
\emph{SoC} & \emph{CPU} & \emph{RAM} & \emph{Firmware} & \emph{Bluetooth} \\
\emph{nRF} & \emph{freq.} & \emph{size} & \emph{storage} & \emph{modes}\\
\cmidrule(r){1-1}
\cmidrule(lr){2-2}
\cmidrule(lr){3-3}
\cmidrule(lr){4-4}
\cmidrule(lr){5-5}
\emph{Version} & \emph{[MHz]} & \emph{[KB]} & \emph{[KB]} & \emph{[bps]}\\
\midrule
51822 & Cortex M0 - 16 & 16 & 128 & \unit[1]{M} \\
52832 & Cortex M4 - 64 & 64 & 512 & 1-\unit[2]{M} \\
52840 & Cortex M4 - 64 & 64 & 512 & 125-\unit[500]{K}, 1-\unit[2]{M} \\
\bottomrule
\end{widetable}
\end{minipage}
\hfill
\begin{minipage}{0.255\textwidth}
\centering
\captionsetup{type=table} %
\caption[\name evaluation testbeds]{Testbeds. \capt{
They are subject to interference from users and network deployments.
}}
\footnotesize%
\ltab{evaluation:testbeds}
\begin{widetable}{\linewidth}{lcc}
\toprule
\emph{Testbed} & \emph{Nodes} & \emph{Diameter} \\
\midrule
D-Cube              & 48 & 5 \\
Kiel     & 20 & 3 \\
\bottomrule
\end{widetable}
\end{minipage}
\hfill
\begin{minipage}{0.25\textwidth}
\captionsetup{type=figure} %
	\centering
		\includegraphics[width=.85\textwidth]{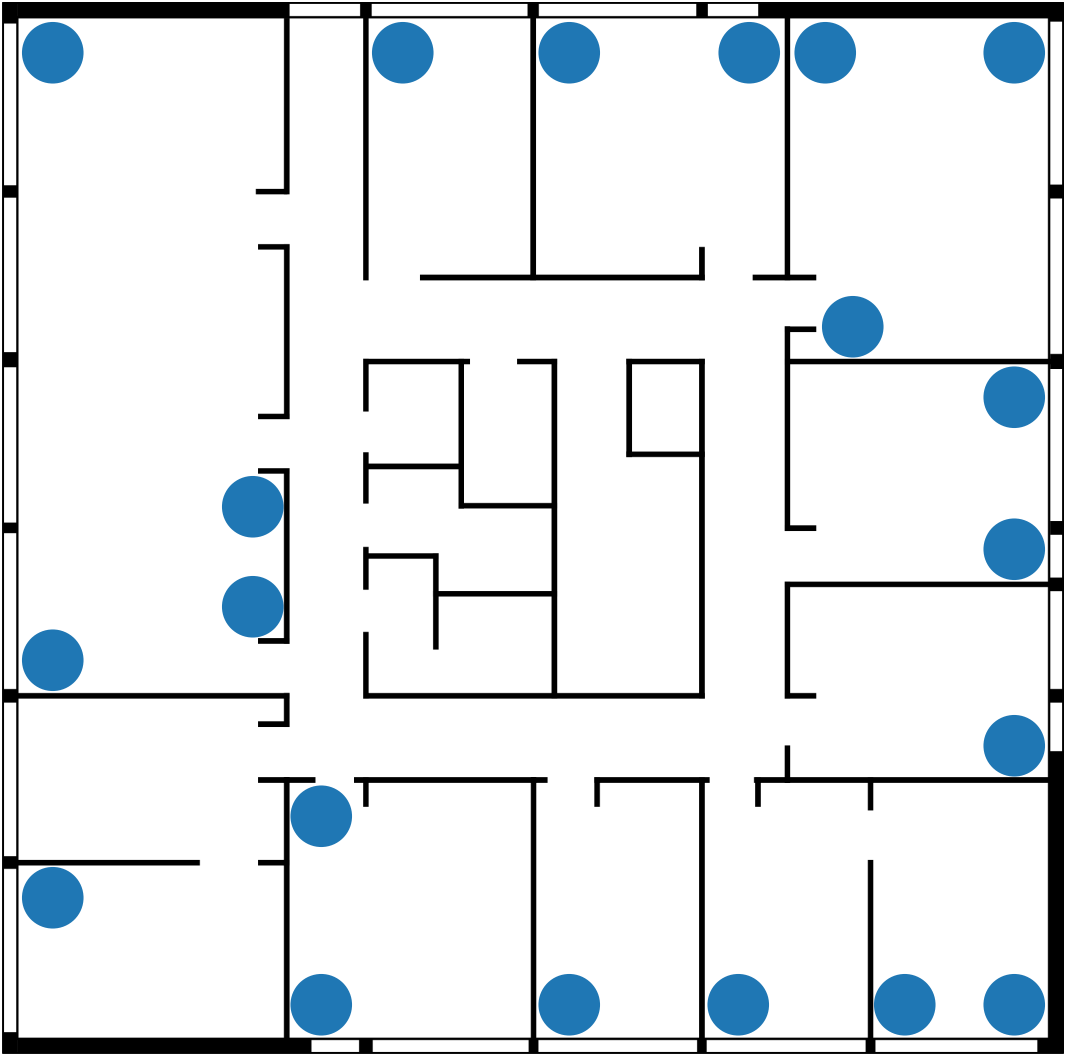}
		\caption[Kiel Testbed]{\lfig{kiel_testbed}Kiel testbed. %
		}
\end{minipage}
\hfill
\end{figure}

\fakeparagraph{Testbeds}
We run our evaluation on two testbeds with 20 to 48 nodes, deployed in university buildings at Kiel University (see~\rfig{kiel_testbed}) and D-Cube~\cite{ewsn17Comp} at TU Graz, respectively. 
The properties of these testbeds are summarized in~\rtab{evaluation:testbeds}.
\majorreview{Note that diameter is an approximation of the maximum shortest-path length between any pair of nodes.}
The testbeds suffer from uncontrolled interference from co-located WiFi networks and Bluetooth devices.
\finalreview{Note that the original publication \cite{alnahas2019blueflood} had a limited evaluation on a small testbed of 8 nodes.}

\subsection{Transmission Power}
\lsec{evaluation:txpower}
We evaluate the performance of \name for different transmission powers.
We use the three Bluetooth advertisement channels (37, 38, 39) to send iBeacon packets; \ie 38-byte packets with 30 bytes payload and 46 bytes = 368 symbols on air including PHY headers on the \unit[1]{Mbps} PHY.
We vary the Tx power and repeat the experiments using the four Bluetooth modes, and the IEEE~802.15.4 mode, for comparison. \majorreview{Note that all nodes use the same configured transmit power.}

\rfig{eval:txpow_diss} summarizes the results.
The end-to-end loss rate stays below 1\% for all modes and all transmission powers, except for the \unit[2]{Mbps} mode at 0~dB in the D-Cube testbed. 
\review{The reason is the shorter transmission range of the \unit[2]{Mbps} mode at the same transmission power, due to decreased \emph{RF processing gain}: the ratio of information transmitted per bandwidth. 
In this case, the channel bandwidth of \unit[2]{MHz} is the same, but the bitrate is doubled; thus, the processing gain and, subsequently, the range will be lower~\cite{rfbook2009rouphael, demystifyingLpwan16}.}

\review{We notice a trend: the higher the transmission power, the lower the loss rate. 
The reason is the increased transmission range; thus, reduced hop count.
This shows that in real deployments, CT does not cause destructive interference, due to multipath fading: the CT signals reach receivers with different powers. 
Therefore, they \emph{escape} the destructive zone of same-power CT.
Moreover, the uncoded modes \unit[1 and 2]{Mbps} exhibit a higher loss rate; thus, a lower reliability, but a faster operation, when compared to the coded modes and 802.15.4.}

We take the 4~dB configuration on D-Cube as an example: it leads to a 2--3~hops network, depending on the transmission mode; \ie it takes 2--3~slots to get the packet.
In this setting, the end-to-end reliability is greater than $99.9\%$.
In the same time, the \unit[2]{Mbps} mode offers about 8~times less latency as compared to the \unit[125]{Kbps} mode which has close to $99.999\%$ PDR, but up to 8~ms in latency.
Overall, \name offers an attractive low-latency power-saving alternative -- something that Bluetooth~Mesh can fundamentally not achieve, as relay nodes must be always-on.

\fakeparagraph{Estimated Duty-cycle}
For the \unit[2]{Mbps} mode, we use \equref{equ:avgRadioTime} with $N_{Tx} = 3$, an average hop count of 2.5~hops, and air time, guard time and total slot length of 0.188~ms, 0.032~ms, 0.4~ms, respectively.
We get $R_{Avg}=0.832$~ms average radio time per node per round and $Latency=2.5\times0.4=1$~ms on average.
\majorreview{With $PER=0.001$, $Wait_{slots}=13$, and a round period of 200~ms, we get an estimate of the radio duty-cycle with \equref{equ:pFail}: $D_C\approx0.42\%$.
For rounds that repeat every second, we extrapolate an average radio duty-cycle of $D_C\approx0.08\%$.}

\begin{figure}[tb]
	\centering
		\includegraphics[width=\textwidth]{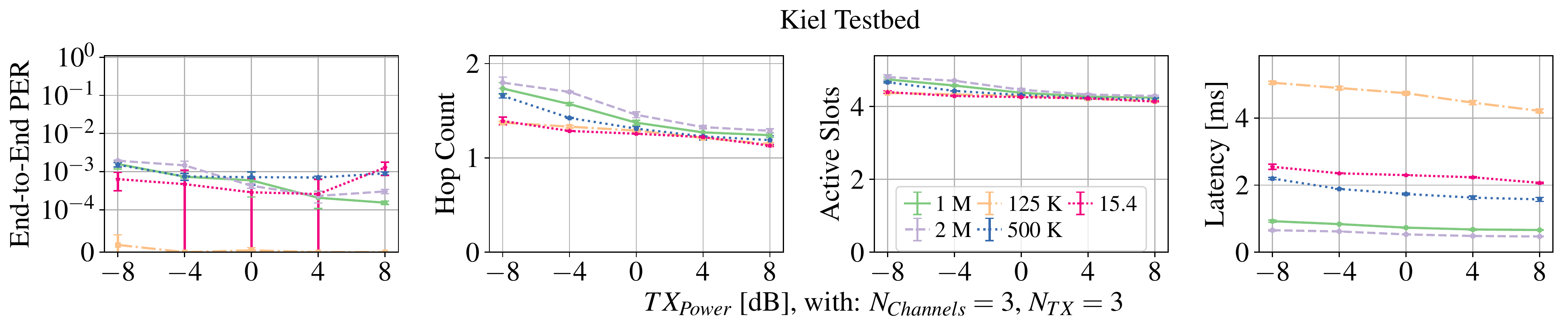}
		\includegraphics[width=\textwidth]{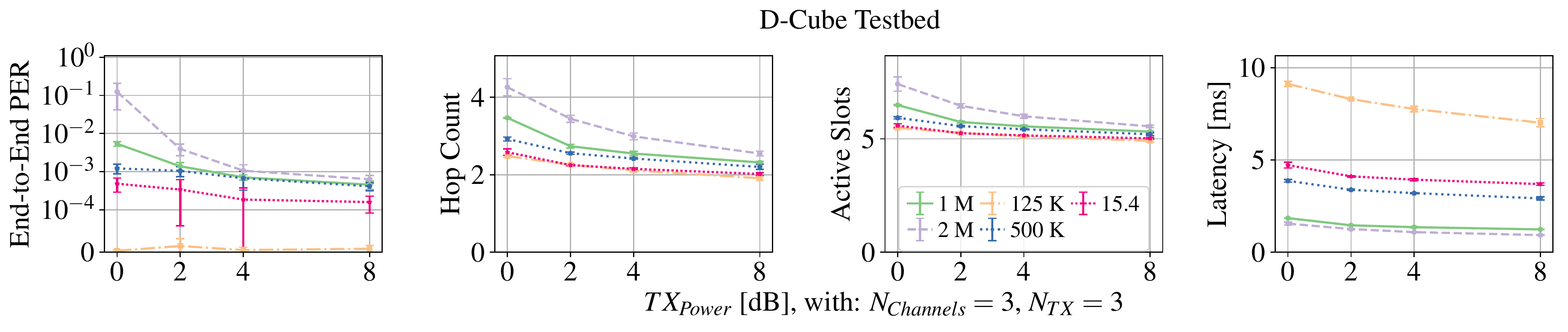}
\caption[\name Performance vs. Transmission Power]{\lfig{eval:txpow_diss}\name dissemination at different transmission powers, over 3 channels: \capt{while all modes have a low loss rate on average of less than 1 per 100, the \unit[2]{Mbps} mode is particularly interesting as it is ~8 times faster than \unit[125]{Kbps}. The upper plot is for Kiel testbed, and the lower is for D-Cube.}}
\end{figure}

\begin{figure}[tb]
	\centering
	\begin{subfigure}[t]{0.49\columnwidth}
	\centering
		\includegraphics[width=0.49\textwidth]{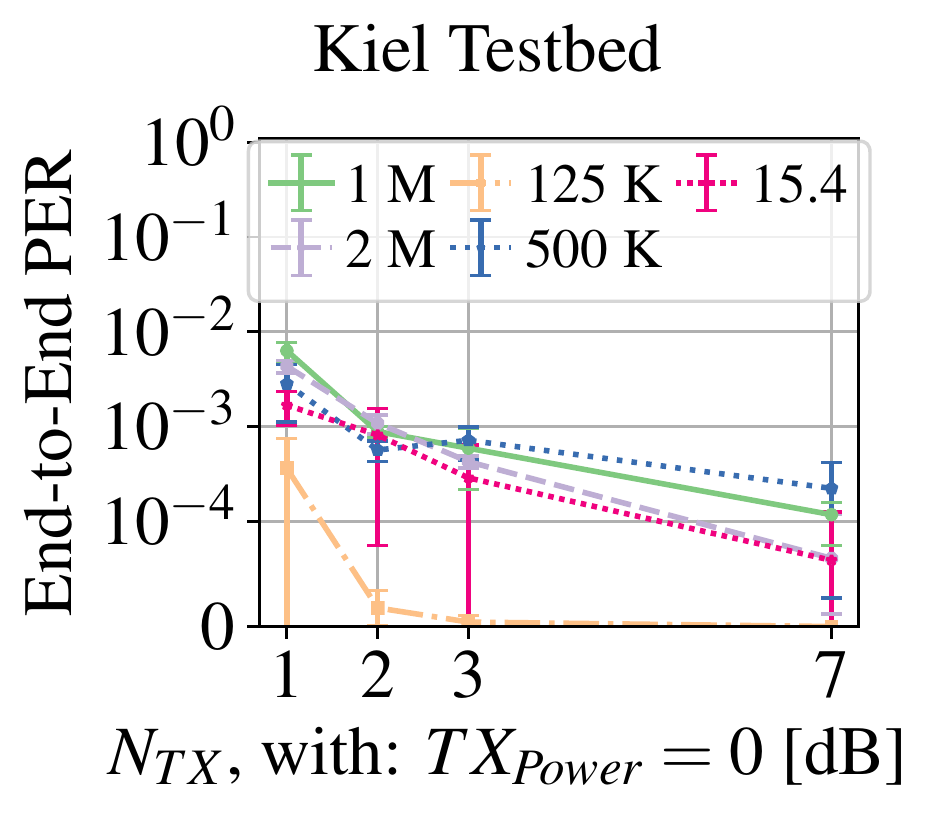}
		\includegraphics[width=0.49\textwidth]{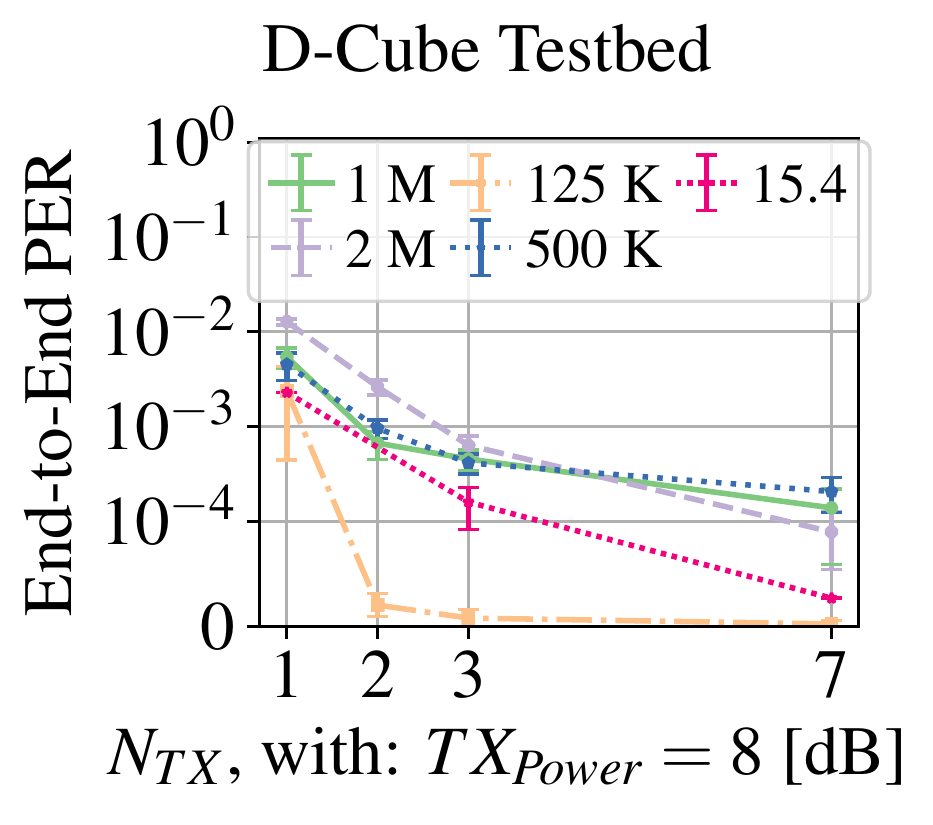}
		\caption[\name Performance vs. Number of Repetitions]{\lfig{eval:ntx}\name performance with longer repetition bursts. \capt{End-to-end reliability improves with more retransmissions.}}
	\end{subfigure}
	\hfill
	\begin{subfigure}[t]{0.48\columnwidth}
	\centering
		\includegraphics[width=1\textwidth]{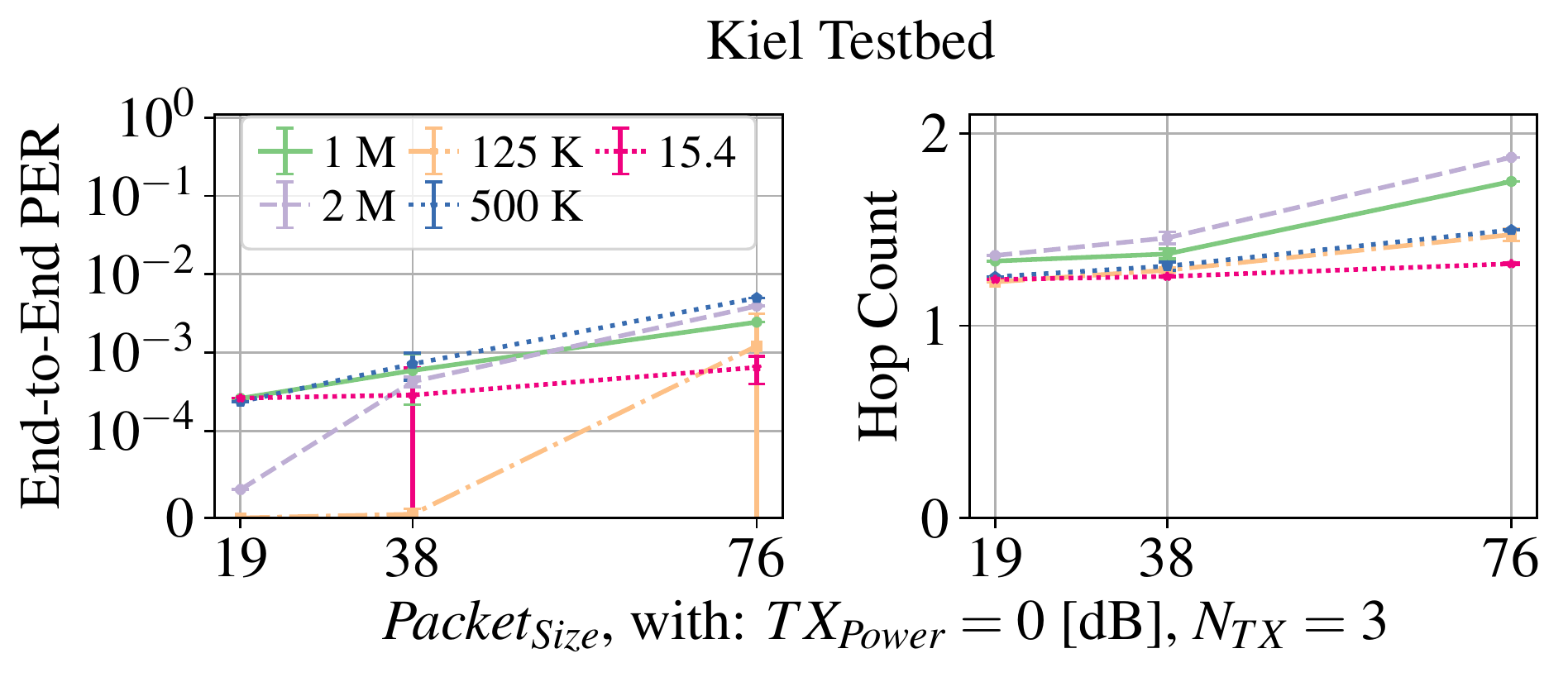}
		\caption[\name Performance vs. Packet Size]{\lfig{eval:packetsize}\name performance with different packet sizes. \capt{The longer the packet, the higher the losses.}}
	\end{subfigure}%
	\caption[\name Evaluation: $N_{Tx}$ and Packet size]{\lfig{evaluationall}\name Evaluation on Kiel Testbed: number of Tx repetitions and packet size.}
	\end{figure}%

\subsection{Repetitions: Number of Transmissions}
\lsec{evaluation:ntx}

We evaluate the performance of \name for different numbers of transmissions $N_{Tx}$.
We use 3~channels to send iBeacon packets with 0~dB, and 8~dB transmission power at Kiel and D-Cube testbeds, respectively. 
We vary the number of transmissions in [1, 2, 3, 7]. %

\rfig{eval:ntx} shows the results.
We notice that the end-to-end loss rate decreases with increasing number of transmissions.
The reason is that repeated transmissions improve the end-to-end packet delivery ratio exponentially: $PDR = 1-(1-PRR)^N$.
This leads to an end-to-end $PDR > 99\%$ for all modes, at the expense of energy.
On the other hand, we see an interesting energy trade-off for the different modes:
The \unit[2]{Mbps} mode reliability with 7~Tx is better than 1~Tx with \unit[125]{Kbps}.
Yet, the \unit[2]{Mbps} mode costs about 16~times less energy to send one packet; \ie the cost of sending the whole round with the packet repeated 7~times is less than that for sending one packet in the mode \unit[125]{Kbps}.
Note that \name covers the whole network even with $N_{Tx}=1$ at the selected Tx powers for these two testbeds. 
However, we notice a similar trend with other Tx powers, but some nodes become weakly connected when $N_{Tx} < 3$.

\subsection{Packet Size}
\lsec{evaluation:packetsize}

We evaluate the performance of \name when sending larger packets.
We use 3~channels to send iBeacons with 0~dB Tx power at Kiel testbed.
We vary the size of the packet in [19, 38, 76] bytes. %
Note that packets larger than 38 bytes are not compatible with iBeacons although we use the same format with longer payload, but they are still Bluetooth~5 compliant. 
Besides, we use a round period of \unit[400]{ms} for disseminating \unit[76]{bytes}, to accommodate the doubled slot length and logging.

\rfig{eval:packetsize} summarizes the results.
We notice that the end-to-end loss rate increases with the larger packet size.
With a larger packet, the probability of corruption due to both interference and fading increases, as does the packet air time.
The protection of FEC helps retaining a reliability close to $99.9\%$ for the \unit[125]{Kbps} mode.
\review{The most affected are the \unit[2]{Mbps}, then \unit[1]{Mbps} modes, which show the relative fragility of CT over Bluetooth for packets larger than standard beacons of \unit[38]{bytes}.
The increased hop count for the larger packet size indicates that the internal interference of CT decreases the packet reception probability, as predicted by our numerical results. 
Further, we note three interesting artifacts: 
(a) we notice the relative ineffectiveness of the \unit[500]{Kbps} mode in all the evaluation scenarios so far,
(b) we notice the effectiveness of the 802.15.4 mode for the larger packets of 76~bytes, and
(c) we notice that the \unit[2]{Mbps} has a 10~times lower loss rate for the 19~bytes packets, when compared to the other modes, except the \unit[125]{Kbps} mode.
The reason is the shorter airtime, which allows the packets to escape the carrier beating.
}
\subsection{Compatibility with Unmodified Phones}
\lsec{evaluation:smartphones}

Next, we evaluate, whether CT beacons can be received by off-the-shelf smartphones. 
\majorreview{
This is important to evaluate as on the BLE nodes we control the complete software stack and the SoC at low-level details, while on the smartphone we cannot achieve low-level control of the BLE SoC and the BLE protocol stack of the operating system.
Thus, we want to verify if off-the-shelf BLE devices receive \name communication without custom software or hardware changes. 
}

We run \name and test the reception of the CT of iBeacons from our testbed at Kiel using an unmodified Samsung Galaxy~S9.
We run \name using $N_{Tx}=3$ and $Tx_{power}=-8~dB$ on channel 37 in the legacy \unit[1]{Mbps} mode.
We install a Bluetooth beacon scanner application and enable the scanning mode. 
We place the phone in several locations in the testbed, and it is able to correctly decode our beacons.
\majorreview{
We do not evaluate the reliability of reception because Bluetooth beacon scanning in general is asynchronous and is meant to receive from different advertisers, not only a single source.
Moreover, due to the tight timing requirements of CT and our lack of control over the BLE stack on the phone, the phone cannot participate in the flood, but it receives it.
}

\section{Conclusion}
\lsec{Conclusion}

This paper models and evaluates concurrent transmissions over Bluetooth PHY.
We argue that the recent approaches to concurrent transmissions based on Glossy, \review{are key enablers for efficient multi-hop communications over Bluetooth.}
We present \name: a network stack based on concurrent transmissions to provide low power, low latency and reliable flooding and data dissemination to Bluetooth mesh networks that are battery operated.

\review{Our analytic models and experimental evaluation show that:
(i) CT over Bluetooth is feasible, though fragile, when used in the uncoded modes, due to beating;
(ii) the coding employed in the \unit[125]{Kbps} mode improves the reliability in both cases of same or different data CT;}
(iii) %
despite this fragility of CT over Bluetooth PHY, it is a viable communication strategy for network-wide dissemination;
(iv) \name achieves data dissemination with high reliability, low power and low latency;
(v) the choice of the transmissions mode provides a trade-off between reliability, energy, and latency; and
(vi) \name floods can be received on unmodified phones.

\begin{acks}
We would like to thank Coen Roest for his master thesis work on Enabling the Chaos Networking Primitive on Bluetooth LE.
This work was supported by 
the Swedish Research Council VR through the ChaosNet project,
the Swedish Foundation for Strategic Research SSF through the LoWi project, 
and Sweden's innovation agency VINNOVA.
\end{acks}

\balance
\bibliographystyle{ACM-Reference-Format}
\bibliography{paper}

\end{document}